\input harvmac.tex
\input epsf
\noblackbox

\def\figin{\epsfcheck\figin}\def\figins{\epsfcheck\figins}
\def\epsfcheck{\ifx\epsfbox\UnDeFiNeD
\message{(NO epsf.tex, FIGURES WILL BE IGNORED)}
\gdef\figin##1{\vskip2in}\gdef\figins##1{\hskip.5in}
\else\message{(FIGURES WILL BE INCLUDED)}%
\gdef\figin##1{##1}\gdef\figins##1{##1}\fi}
\def\DefWarn#1{}
\def\figinsert{\goodbreak\topinsert}
\def\ifig#1#2#3#4{\DefWarn#1\xdef#1{fig.~\the\figno}
\writedef{#1\leftbracket fig.\noexpand~\the\figno}%
\figinsert\figin{\centerline{\epsfxsize=#3mm \epsfbox{#2}}}
\bigskip\medskip\centerline{\vbox{\baselineskip12pt
\advance\hsize by -1truein\noindent\footnotefont{\sl Fig.~\the\figno:}\sl\ #4}}
\bigskip\endinsert\noindent\global\advance\figno by1}


\def\encadremath#1{\vbox{\hrule\hbox{\vrule\kern8pt\vbox{\kern8pt
 \hbox{$\displaystyle #1$}\kern8pt}
 \kern8pt\vrule}\hrule}}
 %
 %
 

 \font\cmss=cmss10
 \font\cmsss=cmss10 at 7pt
 \def\rlx{\relax\leavevmode}
 \def\inbar{\vrule height1.5ex width.4pt depth0pt}
 \def\IC{\relax\,\hbox{$\inbar\kern-.3em{\rm C}$}}
 \def\IN{\relax{\rm I\kern-.18em N}}
 \def\IP{\relax{\rm I\kern-.18em P}}

\def\ZZ{\rlx\leavevmode\ifmmode\mathchoice{\hbox{\cmss Z\kern-.4em Z}}
  {\hbox{\cmss Z\kern-.4em Z}}{\lower.9pt\hbox{\cmsss Z\kern-.36em Z}}
  {\lower1.2pt\hbox{\cmsss Z\kern-.36em Z}}\else{\cmss Z\kern-.4em Z}\fi}
 \def\IZ{\relax\ifmmode\mathchoice
 {\hbox{\cmss Z\kern-.4em Z}}{\hbox{\cmss Z\kern-.4em Z}}
 {\lower.9pt\hbox{\cmsss Z\kern-.4em Z}}
 {\lower1.2pt\hbox{\cmsss Z\kern-.4em Z}}\else{\cmss Z\kern-.4em Z}\fi}
 \def\IZ{\relax\ifmmode\mathchoice
 {\hbox{\cmss Z\kern-.4em Z}}{\hbox{\cmss Z\kern-.4em Z}}
 {\lower.9pt\hbox{\cmsss Z\kern-.4em Z}}
 {\lower1.2pt\hbox{\cmsss Z\kern-.4em Z}}\else{\cmss Z\kern-.4em Z}\fi}

 \def\narrowplus{\kern -.04truein + \kern -.03truein}
 \def\narrowminus{- \kern -.04truein}
 \def\narrowminussub{\kern -.02truein - \kern -.01truein}

 \def\m{{\mu}}

 \def\a{{\alpha}}
 
 \def\frac#1#2{{#1\over #2}}
 
 \def\G{{\Gamma}}
 
 \def\g{{\gamma}}

 \def\p{\partial}

\def\ts{\tilde s}
\def\tu{\tilde u}

 \def\IZ{\relax\ifmmode\mathchoice
 {\hbox{\cmss Z\kern-.4em Z}}{\hbox{\cmss Z\kern-.4em Z}}
 {\lower.9pt\hbox{\cmsss Z\kern-.4em Z}}
 {\lower1.2pt\hbox{\cmsss Z\kern-.4em Z}}\else{\cmss Z\kern-.4em Z}\fi}
 \def\IB{\relax{\rm I\kern-.18em B}}
 \def\IC{{\relax\hbox{$\inbar\kern-.3em{\rm C}$}}}
 \def\Ic{{\relax\hbox{$\inbar\kern-.22em{\rm c}$}}}
 \def\ID{\relax{\rm I\kern-.18em D}}
 \def\IE{\relax{\rm I\kern-.18em E}}
 \def\IF{\relax{\rm I\kern-.18em F}}
 \def\IG{\relax\hbox{$\inbar\kern-.3em{\rm G}$}}
 \def\IGa{\relax\hbox{${\rm I}\kern-.18em\Gamma$}}
 \def\IH{\relax{\rm I\kern-.18em H}}
 \def\II{\relax{\rm I\kern-.18em I}}
 \def\IK{\relax{\rm I\kern-.18em K}}
 \def\IP{\relax{\rm I\kern-.18em P}}
\def\Tr{{\rm Tr}}
 \font\cmss=cmss10 \font\cmsss=cmss10 at 7pt
 \def\IR{\relax{\rm I\kern-.18em R}}

 %

 %
 %
 \def\eqnn#1{\xdef
#1{(\secsym\the\meqno)}\writedef{#1\leftbracket#1}%
 \global\advance\meqno by1\wrlabeL#1}
 \def\eqna#1{\xdef
#1##1{\hbox{$(\secsym\the\meqno##1)$}}

\writedef{#1\numbersign1\leftbracket#1{\numbersign1}}%
 \global\advance\meqno by1\wrlabeL{#1$\{\}$}}
 \def\eqn#1#2{\xdef
#1{(\secsym\the\meqno)}\writedef{#1\leftbracket#1}%
 \global\advance\meqno by1$$#2\eqno#1\eqlabeL#1$$}

\newdimen\tableauside\tableauside=1.0ex
\newdimen\tableaurule\tableaurule=0.4pt
\newdimen\tableaustep
\def\phantomhrule#1{\hbox{\vbox to0pt{\hrule height\tableaurule width#1\vss}}}
\def\phantomvrule#1{\vbox{\hbox to0pt{\vrule width\tableaurule height#1\hss}}}
\def\sqr{\vbox{%
  \phantomhrule\tableaustep
  \hbox{\phantomvrule\tableaustep\kern\tableaustep\phantomvrule\tableaustep}%
  \hbox{\vbox{\phantomhrule\tableauside}\kern-\tableaurule}}}
\def\squares#1{\hbox{\count0=#1\noindent\loop\sqr
  \advance\count0 by-1 \ifnum\count0>0\repeat}}
\def\tableau#1{\vcenter{\offinterlineskip
  \tableaustep=\tableauside\advance\tableaustep by-\tableaurule
  \kern\normallineskip\hbox
    {\kern\normallineskip\vbox
      {\gettableau#1 0 }%
     \kern\normallineskip\kern\tableaurule}%
  \kern\normallineskip\kern\tableaurule}}
\def\gettableau#1 {\ifnum#1=0\let\next=\null\else
  \squares{#1}\let\next=\gettableau\fi\next}

\tableauside=1.0ex
\tableaurule=0.4pt

\def\IE{\relax{\rm I\kern-.18em E}}
\def\IP{\relax{\rm I\kern-.18em P}}

\def\cC{{\cal C}} 
\def\cF{{\cal F}} 
\def\cH{{\cal H}}

\def\cW{{\cal W}}

\def\tt{{\widetilde{t}}}
\def\tp{{\widetilde{p}}}
\def\tx{{\widetilde{x}}}
\def\tf{{\widetilde{\phi}}}
\def\f{{\phi}}
\def\hf{{{1\over 2}}}
\def\Z{{\bf Z}}
\def\R{{\bf R}}
\def\C{{\bf C}}
\def\F{{\Phi}}
\def\tu{{\tilde u}}
\def\CF{{\cal F}}

\lref\Dijkreview{
R.~Dijkgraaf,
``Intersection theory, integrable hierarchies and topological field theory,''
in J. Frohlich et al. (eds), {\it New symmetry principles in
quantum field theory}, Plenum Press,
p. 95, arXiv:hep-th/9201003.
}
\lref\dvI{
R.~Dijkgraaf and C.~Vafa,
``Matrix models, topological strings, and supersymmetric gauge theories,''
Nucl.\ Phys.\ B {\bf 644}, 3 (2002)
[arXiv:hep-th/0206255].
}
\lref\dvII{
R.~Dijkgraaf and C.~Vafa,
``On geometry and matrix models,''
Nucl.\ Phys.\ B {\bf 644}, 21 (2002)
[arXiv:hep-th/0207106].
}
\lref\goshv{
D.~Ghoshal and C.~Vafa,
``$c = 1$ string as the topological theory of the conifold,''
Nucl.\ Phys.\ B {\bf 453}, 121 (1995)
[arXiv:hep-th/9506122].
}

\lref\lowm{
I.~Low and A.~V.~Manohar,
``Spontaneously broken spacetime symmetries and Goldstone's theorem,''
Phys.\ Rev.\ Lett.\  {\bf 88}, 101602 (2002)
[arXiv:hep-th/0110285].
}

\lref\wittenanomaly{
E.~Witten,
``Quantum background independence in string theory,''
[arXiv:hep-th/9306122].
}

\lref\kostovcftt{
I.~K.~Kostov,
``Conformal field theory techniques in random matrix models,''
arXiv:hep-th/9907060.
}

\lref\ckr{S.~Chiantese, A.~Klemm, I.~Runkel,
``Higher order loop equations for A(r) and D(r) quiver matrix models,''
arXiv:hep-th/0311258.
}

\lref\akmvII{
M.~Aganagic, A.~Klemm, M.~Mari\~no and C.~Vafa,
``The topological vertex,''
arXiv:hep-th/0305132.
}
\lref\dmp{
R.~Dijkgraaf, G.~W.~Moore and R.~Plesser,
``The Partition function of 2-D string theory,''
Nucl.\ Phys.\ B {\bf 394}, 356 (1993)
[arXiv:hep-th/9208031].
}
\lref\ty{E.~Calabi,
``M\'etriques K\"ahl\'eriennes et Fibr\'es Holomorphes,''
Ann. scient. \'Ec. Norm. Sup. (1979) 294,
G.~Tian and S.-T.~Yau, ``Existence of K\"ahler-Einstein Metrics
on complete K\"ahler manifolds and their
applications to algebraic geometry,''
Mathematical Aspect of String Theory, 1986 at UCSD,
Ed. S.T. Yau, World Scientific (1987),
``Complete K\"ahler manifolds with zero Ricci curvature, I,''
Journal of AMS {\bf 3} (1990),
``Complete K\"ahler manifolds with zero Ricci curvature, II,'' Inv. Math. {\bf 106} (1991).}

\lref\AV{
M.~Aganagic and C.~Vafa,
``Mirror symmetry, D-branes and counting holomorphic discs,''
arXiv:hep-th/0012041.
}

\lref\van{
C.~Vafa,
``Brane/anti-brane systems and U(N$|$M) supergroup,''
arXiv:hep-th/0101218.
}

\lref\fivebrane{E.~Witten,
``Five-brane effective action in M-theory,''
J.\ Geom.\ Phys.\  {\bf 22}, 103 (1997)
[arXiv:hep-th/9610234].
}

\lref\no{
N.~Nekrasov and A.~Okounkov,
``Seiberg-Witten theory and random partitions,''
arXiv:hep-th/0306238.
}

\lref\civ{
F.~Cachazo, K.~A.~Intriligator and C.~Vafa,
``A large N duality via a geometric transition,''
Nucl.\ Phys.\ B {\bf 603}, 3 (2001)
[arXiv:hep-th/0103067].
}

\lref\dvv{
R.~Dijkgraaf, H.~Verlinde and E.~Verlinde,
``Loop Equations And Virasoro Constraints In Nonperturbative 2-D Quantum Gravity,''
Nucl.\ Phys.\ B {\bf 348}, 435 (1991).
}

\lref\wal{
M.~Fukuma, H.~Kawai and R.~Nakayama,
``Continuum Schwinger-Dyson Equations And Universal
Structures In Two-Dimensional Quantum Gravity,''
Int.\ J.\ Mod.\ Phys.\ A {\bf 6}, 1385 (1991),
``Infinite Dimensional Grassmannian Structure Of Two-Dimensional Quantum Gravity,''
Commun.\ Math.\ Phys.\  {\bf 143}, 371 (1992).
}

\lref\ckv{
F.~Cachazo, S.~Katz and C.~Vafa,
``Geometric transitions and N = 1 quiver theories,''
arXiv:hep-th/0108120.
}

\lref\iz{C.~Itzykson and J.~B.~Zuber,
``The Planar Approximation. 2,''
J.\ Math.\ Phys.\  {\bf 21}, 411 (1980).
}

\lref\cfikv{
F.~Cachazo, B.~Fiol, K.~A.~Intriligator, S.~Katz and C.~Vafa,
``A geometric unification of dualities,''
Nucl.\ Phys.\ B {\bf 628}, 3 (2002)
[arXiv:hep-th/0110028].
}

\lref\minmat{
S.~Kharchev, A.~Marshakov, A.~Mironov, A.~Morozov and S.~Pakuliak,
``Conformal matrix models as an alternative to conventional multimatrix models,''
Nucl.\ Phys.\ B {\bf 404}, 717 (1993)
[arXiv:hep-th/9208044]. A.~Morozov,
``Integrability And Matrix Models,''
Phys.\ Usp.\  {\bf 37}, 1 (1994)
[arXiv:hep-th/9303139].
}

\lref\GrossAW{
D.~J.~Gross and A.~A.~Migdal,
``A Nonperturbative Treatment Of Two-Dimensional Quantum Gravity,''
Nucl.\ Phys.\ B {\bf 340}, 333 (1990).
}

\lref\bk{E.~Br\'ezin and V.~A.~Kazakov,
``Exactly Solvable Field Theories Of Closed Strings,''
Phys.\ Lett.\ B {\bf 236}, 144 (1990).
}

\lref\ovknot{H.~Ooguri and C.~Vafa,
``Knot invariants and topological strings,''
Nucl.\ Phys.\ B {\bf 577}, 419 (2000)
[arXiv:hep-th/9912123].
}

\lref\DouglasDD{
M.~R.~Douglas,
``Strings In Less Than One-Dimension And The Generalized KdV Hierarchies,''
Phys.\ Lett.\ B {\bf 238}, 176 (1990).
}

\lref\WittenHR{
E.~Witten,
``Two-Dimensional Gravity And Intersection Theory On Moduli Space,''
Surveys Diff.\ Geom.\  {\bf 1}, 243 (1991).
}

\lref\ds{M.~R.~Douglas and S.~H.~Shenker,
``Strings In Less Than One-Dimension,''
Nucl.\ Phys.\ B {\bf 335}, 635 (1990).
}

\lref\mvw{
E.~J.~Martinec,
``Criticality, Catastrophes And Compactifications,'' in Brink, L. (ed.) et al., {\it
Physics and mathematics of strings}, World Scientific, p. 389.
C.~Vafa and N.~P.~Warner,
``Catastrophes And The Classification Of Conformal Theories,''
Phys.\ Lett.\ B {\bf 218}, 51 (1989).
}

\lref\kknmm{
S.~Y.~Alexandrov, V.~A.~Kazakov and I.~K.~Kostov,
``2D string theory as normal matrix model,''
Nucl.\ Phys.\ B {\bf 667}, 90 (2003)
[arXiv:hep-th/0302106].
}

\lref\gvI{R.~Gopakumar and C.~Vafa,
``On the gauge theory/geometry correspondence,''
Adv.\ Theor.\ Math.\ Phys.\  {\bf 3}, 1415 (1999)
[arXiv:hep-th/9811131].
}

\lref\kekeli{K.~Li,
``Topological Gravity With Minimal Matter,''
Nucl.\ Phys.\ B {\bf 354}, 711 (1991).
``Recursion Relations In Topological Gravity With Minimal Matter,''
Nucl.\ Phys.\ B {\bf 354}, 725 (1991).
}

\lref\dvvts{R.~Dijkgraaf, H.~Verlinde and E.~Verlinde,
``Topological Strings In $d<1$,''
Nucl.\ Phys.\ B {\bf 352}, 59 (1991).
}

\lref\OVblackhole{
H.~Ooguri and C.~Vafa,
``Two-Dimensional Black Hole and Singularities of CY Manifolds,''
Nucl.\ Phys.\ B {\bf 463}, 55 (1996)
[arXiv:hep-th/9511164].
}

\lref\im{
C.~Imbimbo and S.~Mukhi,
``The Topological matrix model of $c = 1$ string,''
Nucl.\ Phys.\ B {\bf 449}, 553 (1995)
[arXiv:hep-th/9505127].
}

\lref\witteng{
E.~Witten,
``Ground ring of two-dimensional string theory,''
Nucl.\ Phys.\ B {\bf 373}, 187 (1992)
[arXiv:hep-th/9108004].
}

\lref\llz{C.-C. M. Liu, K. Liu and J. Zhou, ``A proof of a conjecture of
Mari\~no-Vafa on Hodge integrals,'' math.AG/0306434; ``A formula of two-partition
Hodge integrals,'' math.AG/0310272.}

\lref\res{
J.~McGreevy and H.~Verlinde,
``Strings from tachyons: The $c = 1$ matrix reloaded,''
arXiv:hep-th/0304224.
T.~Takayanagi and N.~Toumbas,
``A matrix model dual of type 0B string theory in two dimensions,''
JHEP {\bf 0307}, 064 (2003)
[arXiv:hep-th/0307083].
M.~R.~Douglas, I.~R.~Klebanov, D.~Kutasov, J.~Maldacena, E.~Martinec and N.~Seiberg,
``A new hat for the $c = 1$ matrix model,''
arXiv:hep-th/0307195.
}

\lref\nikitab{
A.~S.~Losev, A.~Marshakov and N.~A.~Nekrasov,
``Small instantons, little strings and free fermions,''
arXiv:hep-th/0302191.
}
\lref\nikitaa{
N.~A.~Nekrasov,
``Seiberg-Witten prepotential from instanton counting,''
arXiv:hep-th/0206161.
}

\lref\kostov{I.~K.~Kostov,
``String equation for string theory on a circle,''
Nucl.\ Phys.\ B {\bf 624}, 146 (2002)
[arXiv:hep-th/0107247].
S.~Y.~Alexandrov, V.~A.~Kazakov and I.~K.~Kostov,
``Time-dependent backgrounds of 2D string theory,''
Nucl.\ Phys.\ B {\bf 640}, 119 (2002)
[arXiv:hep-th/0205079].
I.~K.~Kostov, ``Integrable flows in $c = 1$ string theory,''
J.\ Phys.\ A {\bf 36}, 3153 (2003)
[arXiv:hep-th/0208034].
}

\lref\HV{K.~Hori and C.~Vafa,``Mirror symmetry,'' hep-th/0002222.
}

\lref\bcov{
M.~Bershadsky, S.~Cecotti, H.~Ooguri and C.~Vafa,
``Kodaira-Spencer theory of gravity and exact results for quantum string amplitudes,''
hep-th/9309140,
Commun.\ Math.\ Phys.\  {\bf 165} (1994) 311.}


\lref\amv{M.~Aganagic, M.~Mari\~no and C.~Vafa,
``All loop topological string amplitudes from Chern-Simons theory,''
hep-th/0206164.}

\lref\macdonald{I.G. Macdonald, {\it Symmetric functions and Hall polynomials},
Oxford University Press, 1995.}
\lref\solitons{T. Miwa,
M. Jimbo and E. Date, {\it Solitons}, Cambridge University Press, 2000.}
\lref\kl{V.~G.~Kac and J.~W.~van de Leur,
``The n-Component KP Hierarchy And Representation Theory,''
J.\ Math.\ Phys.\  {\bf 44}, 3245 (2003)
[arXiv:hep-th/9308137].}

\lref\ehy{T.~Eguchi and S.~K.~Yang,
``The Topological ${\bf CP}^1$ model and the large $N$ matrix integral,''
Mod.\ Phys.\ Lett.\ A {\bf 9}, 2893 (1994)
[arXiv:hep-th/9407134].
T.~Eguchi, K.~Hori and S.~K.~Yang,
``Topological sigma models and large $N$ matrix integral,''
Int.\ J.\ Mod.\ Phys.\ A {\bf 10}, 4203 (1995)
[arXiv:hep-th/9503017].
}

\lref\op{A. Okounkov and R. Pandharipande, ``The equivariant Gromov-Witten
theory of ${\bf P}^1$,'' math.AG/0207233.}

\lref\ik{
A.~Iqbal and A.~K.~Kashani-Poor,
``$SU(N)$ geometries and topological string amplitudes,''
arXiv:hep-th/0306032.
}

\lref\df{D.~E.~Diaconescu and B.~Florea,
``Localization and gluing of topological amplitudes,''
arXiv:hep-th/0309143.
}

\lref\hiv{
T.~J.~Hollowood, A.~Iqbal and C.~Vafa,
``Matrix models, geometric engineering and elliptic genera,''
arXiv:hep-th/0310272.
}

\lref\zhou{
J. Zhou, ``Hodge integrals and integrable hierarchies,'' math.AG/0310408.}

\lref\zhouII{
J. Zhou, ``Curve counting and instanton counting,'' math.AG/0311237.}

\lref\zhouIII{
J. Zhou, ``A conjecture on Hodge integrals,'' math.AG/0310282.}

\lref\ek{T.~Eguchi and H.~Kanno,
``Topological strings and Nekrasov's formulas,''
arXiv:hep-th/0310235.
}
\lref\akv{
M.~Aganagic, A.~Klemm and C.~Vafa,
``Disk instantons, mirror symmetry and the duality web,''
Z.\ Naturforsch.\ A {\bf 57}, 1 (2002)
[arXiv:hep-th/0105045].
}
\lref\faddeev{L.~D.~Faddeev and R.~M.~Kashaev,
``Quantum Dilogarithm,''
Mod.\ Phys.\ Lett.\ A {\bf 9}, 427 (1994)
[arXiv:hep-th/9310070].
}
\lref\orv{A.~Okounkov, N.~Reshetikhin and C.~Vafa,
``Quantum Calabi-Yau and classical crystals,''
arXiv:hep-th/0309208.
}
\def\Title#1#2{\nopagenumbers\abstractfont\hsize=\hstitle\rightline{#1}%
\vskip .5in\centerline{\titlefont #2}\abstractfont\vskip .5in\pageno=0}

\Title
{\vbox{
 \baselineskip12pt
\hbox{hep-th/0312085}
\hbox{UW/PT-03-33}
\hbox{ITFA-2003-58}
\hbox{CERN-TH/2003-290}
\hbox{MAD-TH-03-5}
\hbox{HUTP-03/A083}
}}
{\vbox{
 \centerline{Topological Strings and Integrable Hierarchies}
 }}
\centerline{Mina Aganagic,$^{a}$ Robbert Dijkgraaf,$^{b}$
 Albrecht Klemm,$^{c}$ Marcos Mari\~no,$^{d}$
and Cumrun Vafa$^{e}$}
\bigskip
\centerline{$^a$ Department of Physics, University of Washington at Seattle}
\centerline{Seattle, WA 98195-1560, USA}
\centerline{$^{b}$ Institute for Theoretical Physics \& Korteweg-de Vries
Institute for Mathematics}
\centerline{University of Amsterdam, 1018 XE Amsterdam, The Netherlands}
\centerline{$^c$ Physics Department, University of Wisconsin at Madison}
\centerline{Madison, WI 53706-1390, USA}
\centerline{$^d$ Theory Division, CERN, Geneva 23, CH-1211 Switzerland}
\centerline{$^{e}$ Jefferson Physical Laboratory, Harvard University}
\centerline{Cambridge, MA 02138, USA}

\smallskip
\centerline{\bf Abstract}

We consider the topological B-model on local Calabi-Yau geometries.
We show how one can solve for the amplitudes by using $\cW$-algebra
symmetries which encodes the symmetries of holomorphic
diffeomorphisms of the Calabi-Yau.
In the highly effective fermionic/brane formulation this
leads to a free fermion description of the amplitudes.
Furthermore we argue that topological strings on Calabi-Yau geometries
provide a unifying picture connecting non-critical (super)strings,
integrable hierarchies, and various matrix models.  In particular we
show how the ordinary matrix model, the double scaling limit of matrix
models, and Kontsevich-like matrix model are all related and arise
from studying branes in specific local Calabi-Yau three-folds.  We
also show how A-model topological string on ${\bf P}^1$ and local
toric threefolds (and in particular the
topological vertex) can be
realized and solved as B-model topological string amplitudes on a Calabi-Yau manifold.

\Date{December 2003}

\newsec{Introduction}

Topological strings on Calabi-Yau threefolds have served as a unifying
theme of many aspects of string theory.  They are a rather simple
class of theories which relate to many different aspects of string
theory, including computing F-terms for superstring compactifications
to four dimensions and equivalence with non-critical strings.  Many
deep phenomena in string theory have a simpler and better understood
description in the context of topological strings, in particular large $N$
transitions that encode the connections between gauge theory and
geometry.

The aim of this paper is to study the topological B-model on a special
class of non-compact Calabi-Yau threefolds and develop various
techniques to solve it completely.  In the process of doing this we
end up unifying a number of different areas of string theory.  In
particular we will see that the topological B-model on CY backgrounds
is the right language to understand various properties of non-critical
strings (see \Dijkreview\ for a review of non-critical bosonic
strings).  For example it has been known that non-critical bosonic
strings have two different matrix model descriptions: a double scaling
limit of a matrix model, in which the string world-sheets emerge
through the 't Hooft ribbon diagrams as triangulations, as well as a
finite $N$ matrix model, introduced by Kontsevich, in which the matrix
diagrams can be considered as open string field theory diagrams that
triangulate moduli space.  It has also been known that Virasoro (or
more generally $\cW$-algebra) constraints essentially characterize the
amplitudes. We will see that all these viewpoints are naturally
understood and unified into a single setting through the topological
B-model on Calabi-Yau threefolds.

We will be mainly considering Calabi-Yau threefolds which are
non-compact and can be viewed as a hypersurface
\eqn\nte{zw -H(p,x)=0,}
where $z,w \in {\bf C}$ and we consider cases where $x,p \in {\bf C}$
or ${\bf C}^*$.
One has a well defined
string theory of B-model on the corresponding Calabi-Yau geometry.  We
can then study B-model in the target space of this geometry.  This
will give the Kodaira-Spencer theory of gravity \bcov .  The
observables of this theory (the modes of the `tachyon field' in terms
of the two-dimensional non-critical string) correspond to variations
of the complex structure at infinity.  In each patch this is described
by the modes of a chiral boson $\f(x)$, defined by
$$
\del\f(x) = p(x).
$$
Quantum consistency of the Kodaira-Spencer theory of variation of
complex structure gives rise to Virasoro or more generally
W-constraints. These identities in turn completely fix the string
amplitudes.  The basic idea is very simple:   The symmetries
of the B-model involve holomorphic diffeormorphisms which preserve
the equation of Calabi-Yau \nte\ as well as the holomorphic 3-form
$$\Omega ={dz\over z} \wedge dx\wedge dp$$
(this is analogous to the symmetries of $c=1$ strings \witteng ).
 If we turn off $H(p,x)\rightarrow 0$
the symmetries of the B-model will be enhanced.  In fact arbitrary
symplectic diffeomorphisms of the $(x,p)$ plane (which by
definition preserve $dx\wedge dp$) is the symmetry of this theory
as it now preserves \nte .  This is generated by arbitrary
functions $f(x,p)$ by the symplectic flow. In the quantum theory
this is the $\cW$-algebra symmetry.  Shifting the background back
to $H\not=0$ breaks some of this symmetry. However, the broken
generators of this symmetry corresponds to ``Goldstone Bosons''
which get identified with the ``tachyon fields'' which deform the
complex structure of the B-model.  These symmetries are strong
enough, as we shall see in this paper, to completely fix the
amplitudes.

The connection of the B-model to matrix model comes from two different
directions, both related to the B-branes:  One is in the context
of compact branes leading to large $N$ transitions; the other
is for non-compact branes, leading to relation
to Kontsevich like matrix model with source terms determining
the position of the non-compact brane.

{}For compact branes,
as has been argued before \refs{\dvI ,\dvII}\ the topological
B-model for the class of CY given by \nte\ is a large $N$ dual to a matrix model,
where $p(x)$ is identified with the spectral density of the matrix,
$x$ with the eigenvalue, and $H(p,x)=0$ is the large $N$ limit of the
loop equations.   The way this arises is by considering Calabi-Yau geometries
with compact B-branes wrapped over blown up ${\bf P}^1$ geometries and using
the response of the complex structure to the B-brane, which leads to a flux
for the holomorphic 3-form $\Omega$ over the three-cycle surrounding it.
 In particular if one is interested in having functions of the form
$$H(p,x)=p^r+x^s+lower \ order \ terms$$
then one can take an $A_{r-1}$ quiver matrix theory (with adjoint
matrices at the $r-1$ nodes and bifundamental for edges) and tune the
potential for the adjoints suitably. The exact finite N-loop equations
for the matrix quiver theories can be derived using conformal field theory
techniques \refs{\kostovcftt,\dvII,\ckr}. In the large $N$ limit, with suitable tuning of
these parameters, one can end up with a function with the lower terms
vanishing.  This gets identified with non-critical bosonic string with the background
corresponding to the $(r,s)$ minimal model.  Note that this double
scaling limit is not necessary
to obtain a smooth string theory.  It is only necessary if one wants
to get a specific non-critical bosonic string. The topological string is
a smooth string theory and does
make sense for arbitrary $H(p,x)$. The quadratic case where
$$H(x,p)=p^2-x^2-\mu$$
corresponds to the $c=1$ non-critical bosonic string at self-dual radius
\goshv .

On the other hand the variations of the complex structure at infinity
can also be induced by addition of non-compact branes to the geometry
\akmvII .  This is the topological string analog of
the back-reaction of the branes on gravity.  This in particular
implies that we should be able to get the amplitudes of the closed
string theory upon variation of the complex structure at infinity, by
introducing open string sectors corresponding to branes and
integrating this sector out.  The open string field theory on these
branes turns out to be a Kontsevich-like matrix model whose classical
action can be read off from the Calabi-Yau geometry:
$$W={1\over g_s} \Tr\left[\int P(X)dX -\Lambda X\right]$$
with $P(X)$ obtained by solving $H(p,x)=0$ and $P=\Lambda$ give the
classical position of the branes. The case $(r,s)=(1,2)$ gives the
usual Kontsevich model with action
$$
W = {1\over g_s}\Tr\left[{1\over 3} X^3 -\Lambda X\right].
$$
{}From this point of view it is natural to compute the change in the
closed string partition function, as a result of the back reaction to
the presence of $N$ branes, as a rank $N$ matrix integral. This
explains why a closed string theory can be written both in terms of a
large $N$ and a finite $N$ matrix model, as it was the case for pure
topological gravity.

Another interesting example we will study is given by the case where
$x,p$ are cylindrical variables and
$$H(p,x)=e^p+e^x+1. $$
This is the B-model mirror \HV\ relevant for the topological vertex studied in
\akmvII .  We will show that the full topological vertex can be computed
directly from the perspective of the target space gravity.

In all these models one can also use the brane perspective for formulating
and solving the $\cW$ identities.
Non-compact branes turn out to be fermions of the KS theory \akmvII\ in
terms of which the $\cW$ constraints are bilinear relations.  This
perspective not only gives a natural choice for certain
normal ordering ambiguities, but also leads to a very simple
solution:  In terms of fermions the theory is free (i.e. the corresponding
state is an element of KP hierarchy).\foot{This generalizes the situation
for $c=1$ non-critical bosonic string where the amplitudes only involve
phase multiplication of fermions.}
However these are {\it not}
ordinary free fermions:
The fermions $\psi(x)$ are not globally well-defined
geometrical objects on the Riemann surface $H(p,x)=0$.   First of all, they are
related to the chiral scalar through the usual bosonization formula
$$
\psi(x) = e^{\f(x)/g_s},
$$
which is not defined globally but has a natural interpretation as wave functions defined on
patches of the Riemann surface. We find that the fermions in various patches
transform to one another {\it not
geometrically but rather by Fourier-type transforms} dictated by
viewing $H(x,p)=0$ as a subspace of the quantum mechanical $x,p$ phase space. Even though we do
not have a deep explanation of this fact, we can motivate it and check
it in all the examples considered here. The fact that branes transform
as wavefunctions rather than as ordinary geometric objects on the
Riemann surface seems to be related to holomorphic anomaly for
topological strings \bcov\ and its interpretation
in terms of choice of polarization of quantum mechanical system,
as in the closed string context \wittenanomaly\ as we
will sketch at the end of this paper.
The fermionic formulation turns out to be a powerful viewpoint which
also leads to a unifying simple solution to the quantum amplitudes.
This encompasses not only the fact that fermions lead to the simplest
description of the $c=1$ non-critical bosonic string amplitudes \dmp ,
but also for the topological vertex.

{}From the viewpoint of Calabi-Yau geometry the complexity of the
model will be related to the number of asymptotic infinities in
the geometry of the Riemann surface $H(p,x)=0$.  For example in
the case of the $(1,r)$ Virasoro minimal model, $H=x-p^r$ coupled
to gravity, we have only one asymptotic infinity (represented by
$x\to\infty$).  For the $c=1$ string at self-dual radius and the
mirror of topological string on ${\bf P}^1$ (which we relate to
specific Calabi-Yau threefold geometry) we have two asymptotic
infinities given by $(p\pm x)\to\infty$.  For the topological
vertex we have three asymptotic regions $x\to\infty,p\to\infty,
(x\sim p+i\pi)\to\infty$. In general when we have $k$ boundaries
the amplitudes of the theory are captured by a state in the
$k$-fold tensor product of a chiral boson, whose positive
frequency modes represent the deformations of the complex
structure at infinity of the corresponding patch.  These states
will be the tau-function of some hierarchy captured by the
corresponding $\cW$-constraints.  The brane correlations are given
by suitable (``quantum'') Wick contractions of the free fermions.

The organization of this paper is as follows: In section 2 we discuss
the general setup of non-compact Calabi-Yau geometries and branes of
interest.  In section 3 we discuss deformations of complex structure
on this class of Calabi-Yau and the quantum Kodaira-Spencer theory in
this context.  In section 4 we discuss the back reaction of B-branes
on complex geometry, and identification of non-compact B-branes as
fermions.  We also discuss the action on the branes and how this can
be used to compute closed string amplitudes.  We consider two distinct
ways this can be done depending on whether one is using compact (leading
to ordinary matrix models) or
non-compact B-branes (leading to Kontsevich-like matrix models).
  In section 5 we present a number of examples and solve each
one viewed from closed string (i.e. KS) viewpoint, large $N$ duality
viewpoint and Kontsevich-like matrix model view point.  We will focus
on four classes of examples: $(r,s)$ minimal models
coupled to bosonic strings, $c=1$ bosonic string, topological string on
${\bf P}^1$, and finally the topological vertex:

$$
\vbox{\offinterlineskip
\halign{ & $#$\qquad \hfil  &  $#$ \hfill & $#$ & \hfil $#$ \cr
& (i)\qquad  & H(p,x)=p^r+x^s+... \ \ \   & \quad  \leftrightarrow \quad   & (r,s)\ {\rm minimal\ models\ coupled\ to\ gravity} \cr
&\phantom{X}&&&\cr
& (ii)\qquad & H(p,x)=p^2-x^2 \qquad \ \ & \quad  \leftrightarrow \quad
  & c=1\ {\rm self\ dual\ radius\ coupled\ to\ gravity}\cr
&\phantom{X}&&&\cr
&(iii)\qquad & H(p,x)=e^p+q e^{-p} +x  &\quad \leftrightarrow \quad & {\rm Mirror\ of\ topological\ string \ on }\ {\bf P}^1 \cr
&\phantom{X}&&&\cr
&(iv) \qquad & H(p,x)=e^p+e^x+1 \ \     &\quad \leftrightarrow \quad      &{\rm Mirror\ of\ }{\bf C}^3 \ ({\rm topological\ vertex})\cr
}}
$$
  In section 6 we comment on connections to non-critical
(super)strings.  In section 7 we end with some open questions and
concluding remarks.  In appendix A we collect some facts about the
$c=1$ scattering amplitudes and in appendix B we discuss some
amplitudes of the topological vertex as solutions to $\cW$-algebra
constraints.

\newsec{Topological strings on non-compact Calabi-Yau geometries}

In this paper we consider B-model topological strings on non-compact
Calabi-Yau manifold $X$ given as a hypersurface in ${\bf C}^4$
$$
zw - H(p,x)=0.
$$
These geometries allow a Ricci-flat metric that is conical at
infinity \ty . The holomorphic $(3,0)$ form $\Omega$ can
in this case be chosen to be
$$
\Omega = {dz \wedge dp \wedge  dx \over z}.
$$

One should consider B-model topological string theory as a
quantization of the variation of complex structures on $X$
\bcov .  If one only considers perturbation of the function
$H(p,x)$, keeping the dependence on $w, z$ fixed, the problem reduces
essentially to one (complex) dimension. In that case it helps to
consider the Calabi-Yau $X$ as a fibration over the $(p,x)$-plane,
with fiber the rational curve $zw-H=0$. This fiber clearly degenerates
--- it develops a node --- on the locus
$$
H(p,x)=0.
$$
This degeneration locus is therefore an affine non-compact curve.

Moreover the periods of $\Omega$ over three-cycles on $X$ reduce, by
Cauchy's theorem, to integrals of the two-form
$$\int_D dx\wedge dp$$
over (real two-dimensional) domains $D$ in the complex two-dimensional
$(p,x)$-plane, such that $\partial D \subset \Sigma$, where $\Sigma$
is the analytic curve $H(p,x)=0$. These integrals in turn reduce by
Stokes' theorem (coordinate patch by coordinate patch) to
$$\int_\gamma p\, dx$$
where $\gamma=\partial D$ denotes a one-cycle on the Riemann surface
$\Sigma$. Thus the complex structure deformations of the function
$H(p,x)$ are controlled by the one-form
$$\lambda =p\,dx.$$
This one-form is holomorphic in the interior of $\Sigma$, but will
have in general singularities at infinity.

There are a natural set of 2-branes (with a world-volume that has
complex dimension one) in this geometry \AV .  They are
parameterized by a fixed point $(p_0,x_0)$ in the $(p,x)$ plane, and
are given by the subspace
$$(z,w,p,x)=(z,w ,p_0,x_0)$$
where $(z,w)$ are restricted by
$$zw=H(p_0,x_0).$$
That is, the brane wraps the fiber of the fibration of $X$ over the
$(p,x)$ plane.  This thus gives us a one complex dimensional subspace
which generically we can identify, say, with the coordinate $z$. For
simplicity we drop the $0$ subscript from $p_0,x_0$ and denote them
simply by $(p,x)$.

The worldvolume theory on these branes is given by a reduction of
holomorphic Chern-Simons down to two (real) dimensions. Its kinetic
action will involve the term \AV :
$${1\over g_s}\int dzd{\overline z} \ p {\overline \partial}x$$
{}From this we see that in the brane probe the holomorphic symplectic
form $dp\wedge dx$ gives rise to field variables $p(z)$ and $x(z)$
which should be considered as canonically conjugated.  In fact, from
the action we read off that their zero modes, which can be identified
with the coordinates of the moduli space of these branes, will have
canonical commutation relation
$$
[x,p]=g_s.
$$
This Lagrangian structure and its quantum appearance will play a key
role in this paper.

There is a more restricted class of branes, which will be very important
for us. These correspond to fixing $(p_0,x_0)$ to lie on the Riemann
surface $C$, that is on the degeneration locus
$$H(p_0,x_0)=0.$$
In this case the equation satisfied for $(z,w)$ becomes
$$zw=0,$$
and we can choose either $z=0$ or $w=0$ to satisfy it.  In particular
the brane in the `bulk,' {\it i.e.}\ at arbitrary position
$(p_0,x_0)$, now splits to two intersecting branes on the boundary
given by $H=0$.  Taking any of these two branes, we can move it along
the points of the Riemann surface.  In other words, the moduli of
either of these branes is now given by a one complex dimensional space
that can be identified with the points on the Riemann surface $H=0$.
There is a sense in which these two types of branes `annihilate' one
another. What we mean by this is that, as discussed in \AV , the branes
in the bulk give rise to zero amplitudes in the topological B-model.
It is only after they split up on the Riemann surface that they
contribute to B-model amplitudes.  In this sense we can think of these
as being `brane/anti-brane' pairs.\foot{This is not strictly true as
we can define the notion of anti-brane for either of these two branes
to be the same brane geometry but counting the brane number by a
$(-1)$ factor, as in \van .}

\newsec{Classical and Quantum Complex Structure Deformation}

\def\tt{{\widetilde{t}}}
\def\tp{{\widetilde{p}}}
\def\tx{{\widetilde{x}}}
\def\tf{{\widetilde{\phi}}}

We will be considering the B-model topological strings on non-compact
Calabi-Yau geometries which quantize the complex structure. Including
the full $g_s$ corrections this corresponds to the quantum
Kodaira-Spencer theory of gravity \bcov.

At tree-level the B-model computes the variation of complex structure
through period integrals. In the case of a compact Calabi-Yau $X$ one
picks a canonical basis of three-cycles $(A_i,B_i)$ of $H_3(X)$ and
compute the periods
$$
s_i = \oint _{A_i} \Omega,\qquad
\cF_i = \oint_{B_i} \Omega.
$$
These periods are not independent, but satisfy the special geometry relation
$$
\cF_i(s) = {\p \cF_0 \over \p s_i}
$$
in terms of the prepotential $\CF(s)$.  Mathematically these
identities express the fact that the image under the period map of the
(extended) moduli space of $X$ forms a Lagrangian submanifold of
$H^3(X,\C)$.

Quantum Kodaira-Spencer theory associates to these data, using the
concepts of geometric quantization, a quantum wave function
$|X\rangle$ \refs{\wittenanomaly,\fivebrane}. In the full quantum
theory the $A$ and $B$ periods become canonically conjugate operators,
with commutation relations
$$
\left[ \oint_{A_i} \Omega, \oint_{B_j} \Omega \right] = g_s^2 \delta_{ij}.
$$
The closed string coupling $g_s^2$ plays here the role of $\hbar$. The
topological string partition function should be considered as the wave
function of the quantum state in the coordinate basis of the $A$-cycle
periods $s_i$
$$
Z(s) = \langle s | X \rangle.
$$
In this way the tree-level prepotential $\cF_0(s)$ gets replaced by
the full string partition function, including all the higher genus
quantum corrections
$$
Z(s) = \exp \cF(s),\qquad \cF(s) = \sum_g g_s^{2g-2} \cF_g(s).
$$
In the WKB approximation the $B$-cycle periods, that play the role of
dual momenta to the $A$-cycle periods $s_i$, are given by
$$
\oint_{B_i} \Omega = g_s^2 {\p \cF \over \p s_i} \sim {\p \cF_0 \over \p s_i}.
$$
All of this is highly reminiscent of the way chiral blocks in
two-dimensional CFT are described.

The CY geometries we consider in this paper are non-compact and are
given by the hypersurface
$$zw-H(p,x)=0.$$
In particular we will be considering complex deformations of the
Calabi-Yau involving varying $H(p,x)$ only.  We will consider
situations where the complex structure of the curve $H=0$ can be
changed at `infinity' only.  This is of course not generally the case,
for example when $H$ defines a higher genus Riemann surface, which has
normalizable moduli.  In such cases we decompose the Riemann surface
to pants and apply our constructions below to each of the components,
and deal with the more general situation by the gluing constructions.

Taking this into account, we thus consider situations where $H=0$
corresponds to a genus 0 surface with a number of boundaries.
Near each of the boundaries we choose a local coordinate $x$ such that
$x\rightarrow \infty$ at the boundary.  We also choose a symplectic
completion $p(x)$ such that the canonical one-form can be written as
$$\lambda =p\,dx.$$
We now want to consider variations of the complex structure at this
infinity $x=\infty$.  To this end we introduce a scalar field
$\phi(x)$ such that
$$\lambda =\partial \phi.$$
This we do at each boundary component.  In other words, we have the
relation $p(x)=\partial_x\phi$.  Classically, we choose the
expectation value
$$
\langle\phi (x)\rangle = \phi_{cl}(x)
$$
such that
$$\partial_x \phi_{cl} =p_{cl}(x)$$
where $p_{cl}(x)$ is obtained by solving the relation $H(p_{cl}(x),x)=0$
near the point $x=\infty$.

If we consider $H(p,x)$ as a Hamiltonian function on a phase space
$(p,x)$, the complex curve $\Sigma$ given
by $H=0$ is the level set of fixed energy and gets an interpretation
as a Fermi surface in the corresponding fermion theory. The one-form
$\lambda$ is nothing but the Liouville form $pdx$.  From this point of
view the equation $p_{cl}=\p_x\f$ is just the standard Hamilton-Jacobi
relation that gives the classical action $\f(x)$ for a solution with
$H=0$.

The field $\phi(x)$ should be considered as a chiral bosonic
scalar field, and arbitrary chiral deformations of it correspond
to complex deformations of the surface near $x=\infty$. When we
consider the most general deformation of the complex structure of
the surface near this point, according to the Kodaira-Spencer
picture, this is equivalent to a reidentification of $p$ and $x$.
Namely we can consider a general Laurent expansion around
$x=\infty$ of the form
$$p(x)=\partial \phi(x)= p_{cl}(x) + t_0 x^{-1} + \sum_{n>0} n t_n
x^{n-1} +
\sum_{n>0} {\cal F}_n x^{-n-1}.$$
(Note that in some of our examples $x$ is a periodic variable, in
which case we consider $e^{nx}$ as the expansion series instead of the
powers $x^n$, {\it i.e.}\ we make a Fourier expansion instead of a
Laurent expansion.)  Since the space is non-compact near $x= \infty$,
there are no restrictions on the coefficients $t_n$. Since they
multiply the non-normalizable modes of $\p\f$ they should be
considered as boundary values, or in the language of the AdS/CFT
correspondence, as coupling constants. (The zero-mode $t_0$, which is
log-normalizable and has no conjugated partner, should be discussed
separately, as we will.)

However the coefficients ${\cal F}_n$ that multiply the normalizable
modes, that are irrelevant at $x\to\infty$, are expected to be fixed
by the rest of the Riemann surface data. In fact, viewing $\phi(x)$ as
a quantum field, the positive and negative frequency modes are {\it
not} independent.  In particular they are conjugate variables, with
commutator proportional to $g_s^2$ coming from Kodaira-Spencer action,
just as was the case for the compact periods that we discussed before.
We can therefore view (in leading order in $g_s$) $\cF_n$ as
$$\cF_n=g_s^2 {\p  {\cal F} \over \p t_n},$$
where $\cF(t_n^i)$ is the free energy of the theory (which at
tree-level is given by $\F \sim \cF_0/g_s^2$).  Here we consider the
free energy as a function of the infinite set of couplings $t_{n}^i$,
where $i=1,\ldots,k$ runs over the number of boundary components and
$n\geq 0$. More precisely, in the full quantum theory the coefficients
$\cF_n$ are realized as the dual operators
$$\cF_n = g_s^2 {\p \over \p t_n}.$$

Equivalently, we can think of the free energy $\cF$ as defining a
state $|V\rangle$ in the Hilbert space ${\cal H}^{\otimes k}$,
where ${\cal H}$ denotes the Hilbert space of a single free boson
and $k$ is the number of asymptotic infinities.  The consistency
of the existence of a unique function ${\cal F}$ and the
consistency of these expansion in each patch completely fixes
${\cal F}$ as we will discuss later in this section.

This is best understood in terms of coherent states. If we introduce
the standard mode expansion of a chiral boson
\eqn\chiralboson{
\p\f(x) = \sum_{n \in \Z} \a_n x^{-n-1}, \qquad [\a_n,\a_m]=n g_s^2
\delta_{n+m,0},
}
then the coherent state $|t\rangle$ is defined as
\eqn\coherentstate{
|t\rangle = \exp \left(\sum_{n>0} t_n \a_{-n} \right) |0\rangle.
}
In this representation we have, with $k$ boundary components, the
relation
\eqn\freeenenergy{
\exp \cF(t^1,\ldots,t^k) =
\langle t^1| \otimes \cdots \otimes \langle t^k | V \rangle.
}
Note that in this representation of the state $|V\rangle \in
\cH^{\otimes k}$ as a coherent state wave function, the annihilation
and creation operators $\a_{\pm n}$ are represented as
\eqn\chiralbosonmodes{
\a_{-n} = n t_n,\qquad \a_n = g_s^2  {\p\over \p t_n}.
}

\subsec{Framing and the $\cW$ algebra}

Before discussing how ${\cal F}$ can be determined by this consistency
condition, we will discuss the notion of choice of coordinate, or what
we will call {\it framing}.

Given an asymptotic point $x\rightarrow \infty$, and the choice of the
one-form $\lambda =p dx$, we can ask how unique this choice of
$\lambda$ is?  We can in particular consider coordinate
transformations of the form
$$x\to x+f(p)$$
keeping $p$ fixed. These are canonical transformations in the sense
that the pair $(p, x+f(p))$ is still symplectic conjugated.  This type
of change of variables we call {\it framing}.

Consider a general analytic expansion of $f(p)$,
$$f(p)=\sum_{n\geq 0} a_n p^n.$$
For a quantum chiral scalar any change of the local coordinate $\delta
x = \epsilon(x)$ is implemented by the operator
$$\oint \epsilon (x) T(x) dx$$
acting on the Hilbert space $\cal H$, where
$$T(x)={1\over 2} (\partial \phi)^2 $$
is the energy momentum tensor.  In the case at hand we have
$$\epsilon (x)= f(p(x))=\sum_{n\geq 0} a_n p^n(x)=\sum_{n\geq 0}
a_n (\partial \phi)^n$$
thus the corresponding quantum operator that implements this framing
is given by
\eqn\framingoperator{
{1\over 2}\sum_{n\geq 0} a_n \oint dx \, (\partial
\phi)^{n+2}=\sum_{n\geq 0} a_n W_0^{n+2},
}
which is given by the linear combination of zero modes of a
$\cW_{1+\infty}$ algebra.  In one of the examples relevant for this paper
(the topological vertex) the $x$ and $p$ are periodic variables and
the only non-trivial framing deformation is of the form
$$
f(p)=r p
$$
where $r$ is an integer.  This leads to the action of the zero mode of
the $W^3$ generator on the Hilbert space,
\eqn\vertexframingoperator{
W^3_0 = \oint dx (\p\f)^3,
}
as we will note later.\foot{The relevance of this $\cW_\infty$ algebra
in the context of ${\cal N}=2$ gauge theory has been noted in
\no .  Its modification on the topological
vertex as turning on arbitrary Casimirs on the edges as propagators
has been noted by \ref\iqok{A. Iqbal and A. Okounkov, private
communication.}.} Notice that this operator, when written as a differential
operator on coherent wave state functions, is the ``cut and join'' operator
considered for example in \llz.


\subsec{Broken $\cW$-symmetry and Ward identities}

We will now turn to the general philosophy that we will use in this
paper to determine the full partition function in the relevant
examples. The starting point here is the underlying symmetry of the
problem. From the point of view of the three-dimensional Calabi-Yau
$X$ these symmetries are obviously given by the global
diffeomeorphisms that preserve the choice of the holomorphic
volume-form $\Omega$ (or equivalently the choice of complex
structure). When we consider the reduction the two dimensions, these
symmetries are implemented as diffeomorphisms of the $(p,x)$ plane
that preserve the symplectic form or holomorphic area $dp \wedge
dx$. Equivalently they are therefore given by general holomorphic
canonical transformations of the phase space variables $(p,x)$.

The corresponding Lie algebra of infinitesimal transformations is
given by the infinite-dimensional algebra $\cW_{1+\infty}$. The
Hamiltonian vector fields that generate these transformations can be
locally identified with general polynomials $f(p,x)$. The
infinitesimal action of these Hamiltonians is just
\eqn\areadiff{
\delta x  = {\p f(p,x) \over \p p},\qquad
\delta p  = - {\p f(p,x) \over \p x}.}
(Subsequently, we will often assume these canonical transformation to
be linear transformations, {\it i.e.}, to be $Sp(2,\R) \cong SL(2,\R)$
transformations, but that is not necessary at this point in the
discussion.)

However these symmetry considerations do not yet take into account the
presence of the curve
$$
H(p,x)=0,
$$
which is the locus where the fibration over the $(p,x)$ plane
degenerates. Only if $H$ vanishes identically (or is a constant), the full symplectic
diffeomorphism group will act as unbroken symmetries. If $H$ is not
identically zero, most of this symmetry group will be broken.  A
typical transformation will deform the level set $H(p,x)=0$ and
therefore will generate a deformation of the complex structure.  In
fact, the Kodaira-Spencer field $\phi(x)$ can be viewed as the
``Goldstone boson'' for these broken symmetries.

Typically we have one Goldstone boson for each broken symmetry.  Here,
the story is slightly different:  We consider the Cartan subalgebra
of $\cW_{1+\infty}$ and realize it as the analog of ``Goldstone bosons''.
In the $x$-patch these correspond to generators $x^n$, giving the
$t_n$ generators.  The rest
of the generators of $\cW_{1+\infty}$ are not independent and can be
written in terms of the corresponding $t_n$ modes.
For example,
another natural Cartan subalgebra given by generators $p^n$, which
as we discussed in the last section are identified with the framing, can
be written in terms of the zero modes of $(\partial \phi)^n$.
This is similar to the situation considered in \lowm\ where some
spacetime symmetries are broken and only a subset of broken symmetries
are realized as Goldstone bosons (the simplest example being
a flat D-brane which breaks both translational and rotational
symmetries but only translational symmetries are realized
as massless Goldstone bosons on it).

Precisely because of this realization of the symmetry algebra we
have a huge left-over constraint:  One for each generator of
the $\cW_{1+\infty}$,
which are far more than the Cartan generators.  Apart from this fact, the
situation is just as in other physical applications: even though the
symmetry group is broken, the broken symmetries are still important --
they give rise to Ward identities that can be used to constrain
(and sometimes
even solve) the scattering amplitudes of the Goldstone bosons. This
will be the approach that we take in this paper. We will implement the
Ward identities of the broken $\cW$-symmetries and thus solve for the
full string free energy $\cF(t)$ as a function of the coupling
constants or deformation parameters $t_n$ at the boundaries.  We will
now sketch the general features of this approach, leaving details to
later sections when we discuss explicit examples and when can use the
full power of the reformulation in terms of D-branes.

Note that the subgroup of $\cW_{1+\infty}$ of the form $H(p,x)f$ for any
$f$ will still be a symmetry even after we shift the vacuum.  This is
responsible, as we will discuss later for the fact that the brane
amplitudes are annihilated by $H(p,x)$.

Let us first consider the classical (tree-level) situation. So we
ignore all $g_s$ effects. We consider deformations of the curve
$H(p,x)=0$, where we allow singularities at the various points at
infinity. We parameterize these points at infinity with local
coordinates $\{x_i=\infty\}$, with $i=1,\ldots,k$ and $k$ the
number of boundary components. At each of these points we have an
expansion of the KS field of the form
\eqn\deform{
p_i =\partial \phi_i = p_i^{cl}(x_i) + t_0^i x_i^{-1} + \sum_{n>0} n
t^i_n x_i^{n-1} + g_s^2 \sum_{n>0} {\p \cF \over \p t^i_n} x_i^{-n-1},
}
for a single, global function $\cF(t^1_n,\ldots,t^k_n)$.

The local symplectic coordinates $(p_i,x_i)$ around the points
$x_i=\infty$ are related by canonical transformations, as we will see
in more detail in a moment. For a two-dimensional phase space general
canonical transformations are just area-preserving
diffeomorphisms. Of course there is at each base point $x=\infty$ a local
framing ambiguity, given by the Cartan subalgebra of $W_{1+\infty}$
that fixes the conjugate variable $p$. This is captured by the action
of the generators on the local coherent state $|t^i\rangle$, as we
discussed in the previous section.

Now consider the action of a $\cW_{1+\infty}$ element generated by a
Hamiltonian $f$. In local coordinates $(p_i,x_i)$ this Hamiltonian is
given by the function $f(p_i,x_i)$. Let us assume that it is of the
form
$$
f(p_i,x_i) = p_i^n x_i^m.
$$
This element implements the transformation
$$
\delta x_i = n p_i^{n-1} x_i^m,
\qquad
\delta p_i = -  m p_i^n x_i^{m-1}.
$$
Ignoring all quantum normal ordering ambiguities, which
we will return to after we discuss branes and fermions in the next
section, this generator is
implemented in the KS theory as the mode $W^{n+1}_m$ of the
$\cW$-current
$$
W^{n+1}(x) \sim {1\over n+1}(\p\phi)^{n+1},
$$
namely
$$
W^{n+1}_m = \oint_{{\cal C}_i} x_i^m W^{n+1}(x) \sim \oint_{{\cal C}_i}
 x_i^m {1\over
n+1}(\p\phi)^{n+1}
$$
Here the contour ${\cal C}_i$ encloses the given puncture. When
written in terms of the local parameters $t_n^i$ these
$\cW$-generators act as order $n+1$ differential operators.
Particularly relevant is the case of the spin one current
$W^1=\p\phi$. Its modes correspond to the Hamiltonians $f(p,x)=x^m$.
They generate the pure deformations of the field $\p\f(x)=p$
$$
\delta x =0, \qquad \delta p = m x^{m-1}.
$$
That is, they generate (locally) the linear flows $\p/\p t_n^i$.

In this way we get a local action of the $\cW$-algebra at each
puncture.  But there is one very important and non-trivial
relation between all these actions: they should all parameterize
the deformations of the same analytic curve. That is, if we
introduce a deformation \deform\ at the puncture
$P_i=\{x_i=\infty\}$, this will induce deformations at all other
punctures $P_j=\{x_j=\infty\}$. The deformed functions $p_j(x_j)$
which are given in terms of the local expansions of the
Kodaira-Spencer field $\f$ should all describe, in different
coordinate patches of course, one and the same analytic curve.

This condition relates the various local actions in terms of a global
Ward identity. This identity takes the symbolic form
\eqn\symbolicward{
\left \langle\sum_i \oint_{C_i} W \right\rangle =0.}
Equivalently, one can think of deforming the contour $\cC_i$ around
$P_i$ over the Riemann surface into contours $\cC_j$ encircling the
other punctures $P_j$.  We claim that this Ward identity is sufficient
to solve for the partition function. However, in order to do so, we
first have to relate the mode expansion of the $\cW$ currents in the
local patches. For this we have to know how the Kodaira-Spencer field
$\f(x)$ transforms from patch to patch.

\subsec{Canonical transformations and Kodaira-Spencer field}

For simplicity let us assume there are two punctures with local
coordinates $x$ and $\tx$. At each puncture we have the
canonical one-form that can be written as $p\,dx$ respectively $\tp
\,d\tx$.  These two one-forms should coincide up to a gauge
transformation, so we have a relation of the form
$$
pdx-\tp d\tx= dS.
$$
This expresses the fact that the pairs $(p,x)$ and $(\tp,\tx)$ are
related by a canonical transformation, if we extend the coordinate
transformation to the full $(p,x)$ plane. That is, we have
$$
dp \wedge dx = d\tp \wedge d\tx.
$$
That this symplectic two-form is preserved is of course also obvious
from the fact that is given by the reduction of the holomorphic
three-form of the Calabi-Yau space.

In general such a canonical transformation is given in terms of a
generating function $S(x,\tx)$ that satisfies by definition
\eqn\genfunc{
p = {\p S(x,\tx) \over \p x},\qquad
\tp  = - {\p S(x,\tx) \over \p \tx}.
}
At each puncture we can introduce the local Kodaira-Spencer scalar
fields
\eqn\fields{
p(x)  ={\p\f \over \p x},\qquad
\tp(\tx)  ={\p\tf \over \p \tx}.
}
If we plug-in this form for $p$ and $\tp$ into \genfunc\ we get
\eqn\gener{
\eqalign{
{\p \f(x) \over \p x} & = {\p S(x,\tx) \over \p x}, \cr
{\p \tf(\tx) \over \p \tx} & = - {\p S(x,\tx) \over \p \tx}.}
}
Note that in this formulation the variable $\tx$ starts out as a
function of both $x$ and $p$. Once we give $p=\p\f(x)$ the relation
between $\tx$ and $x$ is implicitly determined by \gener.  In the
final equations the function $S$ should be regarded as given once and
for all (it just determines the coordinate transformation) and the two
functions $\f$ and $\tf$ should be considered as the variables that are
expressed into each other.

Each of these fields has a mode expansion of the type \deform\ given
(at tree-level) in terms of the genus-zero free energy $\cF_0$ and the
couplings
\eqn\px{
\eqalign{
\p \f(x) & = \sum_{n>0} n \, t_{n} x^{n-1} + \sum_{n>0} {\p \cF_0 \over  \p t_{n}}
x^{-n+1} \cr
\p\tf(\tx) & = \sum_{n>0}n \, \tt_{n} \tx^{n-1} +
\sum_{n>0}
{\p \cF_0 \over \p \tt_n} \tx^{-n+1}.\cr} }
Given the relation between the fields $\f(x)$ and $\tf(\tx)$ we can
clearly obtain in this way constraints on the free energy $\cF_0(t,\tt)$.

It is interesting to consider a small fluctuation $\f_{qu}$ around the
classical value $\f_{cl}$. If we write
$$
\p\f=\p\f_{cl} + \p\f_{qu},
$$
one easily verifies\foot{To see this, note that expanding \gener\
about the classical solution we have
$$
\p\f_{qu}(x)={\p^2 S\over \p x \p \tx}(x,\tx_{cl})\;\delta \tx,\quad
\p\tf_{qu}(\tx)=-{\p^2 S\over \p x \p \tx}(x_{cl},\tx)\;\delta x
$$
where $\delta x$, $\delta \tx$ are the corresponding small
fluctuations in $x$, $\tx$.  It follows that
$\p\f_{qu}(x)=-\p\tf_{qu}(\tx_{cl}){\delta \tx \over \delta x}$, and
since $\tx=\tx_{cl}(x) + \delta \tx = \tx_{cl}(x_{cl}(\tx)+\delta x)
+\delta \tx$ implies that $\delta\tx = -{\p \tx_{cl} \over \p x}\delta
x$, the claim follows.}  that the field $\p\f_{qu}$ transforms as
$$
\p\tf_{qu}(\tx) d\tx = \p\f_{qu}(x) dx.
$$
That is, the field $\f_{qu}$ is a globally defined scalar field on the
Riemann surface. Of course for finite values the transformation rules
for $\f_{qu}$ are highly non-linear.

We will see in the next section that the correct physical interpretation
of equations \gener\ is as
the semi-classical ($g_s \to 0$) saddle-point approximation of the
relation
\eqn\trans{
e^{\tf(\tx)/g_s} = \int dx \ e^{-S(x,\tx)/g_s} e^{\f(x)/g_s}.
}

We now know, at least at tree-level, how both the coordinate $x$ and
the field $\p\f=p$ transform from patch to patch. Therefore these
formulas can now be used to determine the transformation of the
$\cW$-currents from patch to patch. From these transformation laws we
can then read off the Ward identities and in the end solve for the
free energy, as we will demonstrate in many concrete examples later.

\subsec{Linear transformations and quantization}

As a small aside we will make the above formalism a bit more
transparent by considering the case of a linear canonical
transformation. So let us consider a $SL(2, {\bf R})\cong Sp(2, {\bf R})$
transformation
\eqn\lin{
\eqalign{ \tp & = a p + b x \cr
          \tx & = c p + dx  \cr  }
}
with
$$
g = \pmatrix{a & b \cr c & d \cr} \in Sp(2, {\bf R}), \qquad ad - bc=1.
$$
Now these relations can be written (for $c\not=0$) as
$$
\eqalign{  p   & = {1\over c}\left(-dx + \tx\right) \cr
           \tp & = {1\over c}\left(x + a \tx\right) \cr}
$$
so that the generating function is given by
$$
S(x,\tx) = {1\over 2 c} \left(-d x^2 + 2 x\tx - a \tx^2\right)
$$

In this case one can straightforwardly go to the quantization, where
$x$ and $p$ become conjugate variables with $g_s$ playing the role of
$\hbar$
\eqn\heisenberg{
[x,p]=g_s.
}
We now want to consider these variables as operators acting on
wave-functions $\Psi(x)$ with
$$
p=- g_s {\p \over \p x}.
$$

As is well-known, linear canonical transformations can be
unambiguously carried over to the quantum case. There is an (almost)
unique lift of the $Sp(2,\R)$ element $g$ to a unitary operator $U(g)$
on the quantum mechanical Hilbert space $V$ of $L^2$
wave-functions. This representation is known as the {\it metaplectic}
representation.  The metaplectic group $Mp(2)$ is a two-fold cover
of the symplectic group $Sp(2, {\bf R})$.

It might be helpful to remind the reader that the metaplectic
representation can be considered as the bosonic analogue of the spinor
representation of the group $Spin(n)$, defined as a double cover of
$SO(n)$. Indeed, for a given a {\it symmetric} form $h_{ij}$ spinors
are obtained as representations of the Clifford algebra
$$
\g_i \g_j + \g_j \g_i = h_{ij}.
$$
One then writes half of the $\g_i$ as fermionic creation operators
$\theta_a$ and the other half as fermionic annihilation operators
$\pi_a=\p/\p\theta_a$. A spinor is then simply a function
$\Psi(\theta)$ of these anti-commuting variables. The generators of
$Spin(n)$ are written as anti-symmetric quadratic expressions in the
$\theta_a$ and $\pi_a$.

Similarly, in case of a (non-degenerate) {\it anti-symmetric} form
$\omega_{ij}$ one now starts with a representation of the Heisenberg
algebra
$$
\xi_i \xi_j - \xi_i \xi_j = \omega_{ij}.
$$
After one picks a polarization in terms of coordinates $x_a$ and
momenta $p_a=\p/\p x_a$, the representations of this algebra are given
by square-integrable wavefunctions $\Psi(x)$.  Because the $x_a$ are
bosonic, this representation is infinite-dimensional, in contrast with
the fermionic spinor representations. In complete analogy, the
metaplectic representation of the group $Mp(n)$, now defined as the
double cover of $Sp(n)$, is generated by the symmetric quadratic
functions in the variables $x_a$ and $p_a$.

In the two-dimensional case the infinitesimal generators of $Sp(2)$
are given by
$$
\eqalign{
J_+ & = x^2,\cr
J_0 & = xp+px=-g_s\{x,\p_x\},\cr
J_- & = p^2=g_s^2 \p_x^2.\cr}
$$

In the metaplectic representation the kernel of the matrix $U(g)$ is
essentially given by the exponential of the generating function
$S(x,\tx)$ that we have just determined. More precisely, including the
full $g_s$ dependence, we have the transformation law
$$
U(g) \Psi(\tx) =
\int {dx \over \sqrt{2\pi g_s c}} \;\exp\left[{1 \over 2 g_s c}
\left(d x^2 - 2 x \tx + a \tx^2\right)\right] \Psi(x).
$$
Note that all these integrals should be considered as contour
integrals in the complex $x$-plane, where the choice of contour is
determined such that the integral makes sense.  There is a square-root
ambiguity in this action that in the end requires the double-cover.

{}The case we will use often is when the linear canonical
transformation corresponds to the element of $SL(2,\Z)$,
$$
S = \pmatrix{ 0 & -1 \cr
              1 &  0 \cr} \in SL(2,{\bf Z})
$$
that interchanges the coordinate and the momentum
$$
\eqalign{\tp & = - x, \cr
         \tx & =   p. \cr}
$$
This is of course quantum mechanically implemented by the Fourier
transform
$$
U(S) \Psi(\tx) = \int {dx \over \sqrt{2\pi g_s}} e^{x\tx/g_s}
\Psi(x).
$$
%

\newsec{B-Branes}

Branes have played a key role in a deeper understanding of
superstrings. It is thus not surprising that also for topological
strings they play a key role.  As we will find, in terms of brane
degrees of freedom, the topological string amplitudes become very
simple.

In the context of superstring target space, branes are defined by
their impact on gravitational modes, as sources for certain fields.  A
similar story is also true for topological string for branes.  In
particular consider a one complex dimensional subspace inside the
Calabi-Yau with $N$ B-branes wrapped over it.  Then as discussed in
\akmvII\ this affects the closed string modes by
changing the periods of the holomorphic threeform $\Omega$. Namely,
let $C$ be a 3-cycle linking the B-brane world-volume.  Then we have
the following change in $\Omega$
\eqn\bacr{
\Delta \int_C \Omega = Ng_s.}
So the operator that creates a brane shifts the value of the period
(and therefore the complex structure) in the dual cycle.  Note that
this back reaction is invisible at tree-level.

 The importance of the branes for us is to use the action on the
world-volume of the branes to find the closed string amplitudes.  This
is the familiar story that integrating out the open string sector
gives the closed string results for the deformed geometry. Symbolically
we can write this effect as
\eqn\closedopenschematicI{
Z_{closed}(m)Z_{open}=Z_{closed}(m')
}
where $m$ denote the closed string moduli of the Calabi-Yau where the
branes live and $m'$ is the deformed string moduli including the back
reaction of the branes.  This will depend on the location of the
brane, through \bacr .
The back reaction can be obtained by integrating
out the open string sector.  In other words we have
\eqn\closedopenschematicII{
Z_{open}=Z_{closed}(m')/Z_{closed}(m).
}
In many cases of interest the full relevant moduli
dependence is in $m'$ and this becomes an efficient
method to compute closed string amplitudes.

We will be considering two kinds of B-branes: B-branes wrapped over
(i) compact cycles or (ii) non-compact cycles.  In the first case of
compact cycles, where we consider
branes wrapped on ${\bf P}^1$ cycles, we get geometric
transitions where a ${\bf P}^1$ shrinks and the Calabi-Yau undergoes
the conifold transition with an $S^3$ emerging. In that case we obtain
a new homology class and a corresponding new period integral. In the
non-compact case the geometry of the brane is ${\bf C}$ and the
geometry of Calabi-Yau is modified at infinity. It this latter case
that we will predominantly focus on in this paper. We can also have
both types of branes present and we will also briefly comment on that.

\subsec{Compact Branes}

The geometry will contain ${\bf P}^1$'s near each of
which the CY looks locally like
${\cal O}(-1)\oplus {\cal O}(-1)\rightarrow
{\bf P}^1$, where the B-brane is wrapped over ${\bf P}^1$. The three cycle
surrounding ${\bf P}^1$ in this case is an $S^3$.  The gravity back reacts,
as discussed above, by
$$
\int_{S^3} \Omega =Ng_s.
$$
This suggests, as was first conjectured in the mirror context
in \gvI\ that
the Calabi-Yau undergoes a transition where ${\bf P}^1$ shrinks
and $S^3$ grows and the size of $S^3$ is given by
$$S=Ng_s.$$
A class of examples which exhibits this geometry is given
by the Calabi-Yau defined as a hypersurface
\eqn\btra{zw-p^2+W'(x)^2=0,}
where $W'(x)$ is a polynomial of degree $n$ in $x$.
Near each critical point $W'(x)=0$ we have a conifold singularity
which we can blow up to a ${\bf P}^1$.
Around each of the ${\bf P}^1$'s we can wrap $N_i$ B-branes, as $1=1,...,n$.
Then as conjectured in \civ\ this undergoes
a transition to a geometry with $n$ $S^3$'s replacing the $n$ ${\bf P}^1$'s
$$zw-p^2+W'(x)^2+f(x)=0,$$
where $f(x)$ is a polynomial of degree $n-1$ in $x$, whose coefficients
are fixed by the condition that the size of each $S^3$ is
$$S_i=N_i g_s.$$
In the context of topological string this duality was explored
in \dvI .  In particular the open string field theory
was identified with a 2d holomorphic Chern-Simons theory which
was shown to reduce to a matrix model with action
$$S={\rm Tr}\, W(\Phi)/g_s,$$
and thus it leads to the conjecture \dvI :
\eqn\dvconjecture{
Z_{closed}(S_i)=\int D\Phi \;\exp(-{\rm Tr}\,W(\Phi)/g_s)
}
We will explore aspects of this correspondence later
in this paper.

\subsec{Non-compact Branes}

Consider non-compact branes that we discussed before, whose moduli are
parameterized by a point on the Fermi surface $H(p,x)=0$.  In the
presence of these branes, \bacr\  implies
a change in the integral of the 1-form $\lambda=pdx$, which is the
reduction of $\Omega$ (integrated along two of the normal directions).
In this reduction the period of $\lambda$, integrated around the point
$P$ on the Riemann surface where the brane intersects, receives a
non-trivial contribution:
$$\oint_P\lambda =\oint_P \partial \phi =g_s.$$
Let $\psi(P)$ denote the operator creating a brane at the point $P$ on
the surface.  Then we have the following identity inside correlation
functions
$$\langle\cdots\oint_P \partial \phi \ \psi(P) \cdots\rangle
=g_s\langle\cdots\psi(P)\cdots\rangle $$
which implies that the brane creation operator affects
the closed string sectors by
\eqn\ferm{\psi (z)=  \exp(\phi (z) /g_s).}
This means that the brane is the fermion associated to $\phi$ by the
standard bosonization rules. This fact was pointed out in \akmvII.  Note
that \ferm\ is also consistent with the fact that the classical action
for the fermionic brane at position $z$ is given by
\AV
\eqn\actionaschiralboson{
S(z) ={1\over g_s}\int^z \lambda=\phi(z)/g_s.}
Similarly the anti-brane is defined by the condition that it gives the
opposite change in the period integral which means that it is given by
the conjugate fermion:
\eqn\conjugatedferm{\psi^*(z)= \exp(-\phi(z)/g_s )}
The action for the anti-brane is negative that of the brane and we have
$Z_{brane}=1/Z_{anti-brane}$.

Now consider the Kodaira-Spencer theory in the context of the
discussion of the previous section where the relevant geometry is
described by the deformation of a Riemann surface $H(p,x)=0$.  Let
$x\rightarrow \infty$ denote a coordinate for asymptotic infinity of
the Riemann surface.  Suppose we put branes at positions $x^i$ near
this asymptotic patch.  Then the gravitational backreaction is given
by
$$
\prod_i \psi(x^i)=\prod _i \exp(\phi (x^i)/g_s)
=\prod_{i\not= j} (x^i-x^j) \exp\left(\sum_i \phi (x^i)/g_s\right).
$$
If we compute the expectation value $\langle \partial \phi(x) \rangle$
in this background we find
\eqn\eva{\langle \partial \phi(x) \rangle=g_s\sum_i {1\over x-x_i}=
g_s\sum_{n>0}{ x_i^{-n} x^{n-1}}}
which means that, apart from the prefactor which measures
the interaction between the branes, we have turned on a background
given by the couplings
$$t_n={g_s\over n} \sum_i (x^i)^{-n}$$
for $n>0$. The momentum of $\phi$ (related to the $t_0$ mode) is also
shifted by the number of fermions we put in.  We could also consider
putting anti-branes, which would give a similar formula as above,
except that we have
$$t_n={g_s\over n}\sum \pm (x^i)^{-n}$$
where $\pm$ is correlated with whether we have put a brane or
anti-brane at $x^i$ and also we get a prefactor of $(x^i-x^j)^{\pm 1}$
depending on whether we put branes of the same type or opposite type
at $x^i$ and $x^j$.

We can now recast the amplitudes in terms of branes.  Namely, if we
put $N$ branes at positions $x^1,\ldots,x^N$ in the local $x$-patch,
the corresponding correlation function is given by
\eqn\onep{\langle N | \prod_{i=1}^{N} \psi(x^i) |V\rangle =
e^{\cF(t)}\;\prod_{i<j} (x^i-x^j) .}
Here we have considered the case with one asymptotic infinity. If we
have $s$ asymptotic infinities, we can consider putting one stack of
branes for each asymptotic region and recover ${\cal F}$.
We can also use this relation to write the state $|V\rangle$ directly
in the fermionic basis. We will return to this below.

\subsec{Branes and wave functions}

As we discussed above, inserting a fermion $\psi(x)$ at a point on the
Riemann surface corresponds in the Calabi-Yau manifold, to
inserting a B-brane there. Since the B-branes are globally well
defined objects this would seem to imply that the fermions are free.
As we will explain below, this is basically true, however the relation
to B-branes implies that the fermions have rather unusual property in
going from patch to patch. This was in fact discussed in the previous
section and we will restate it here in terms
of the wave functions for fermions.

Let us consider the world-volume theory on the
brane.  In fact, let us insert a B-brane in an asymptotic patch $x_i
\rightarrow \infty$.  Since the D-brane is non-compact the partition function of
the D-brane is a wave function $Z_{open}(x_i)$ depending on $x_i$,
since this is what is fixed at infinity on the world-volume of the
D-brane. Let us denote this by
$$
Z_{open}(x_i) = \Psi(x_i).
$$
On the other hand, as we have argued above, in the Kodaira-Spencer
theory this corresponds to inserting a fermion $\psi=e^{\f/g_s}$ at
$x=x_i$, {\it i.e.}\ we have
\eqn\ferm{
\Psi(x_i)=\langle \psi(x_i)\rangle.
}
Since $\Psi(x_i)$ is a wave function, it transforms like one in going
from patch to patch.

More precisely, in this patch, we really have a symplectic pair of
variables $(x_i,p_i)$. In terms of the theory on the B-branes the
variables $x_i$ and $p_i$ correspond to the zero modes of fields that
are canonically conjugate, so $[p_i,x_i]=g_s$. The wave function
$\Psi(x_i)$ forms a metaplectic representation of this algebra, and by
\ferm , so does $\psi(x_i)$.

Consider another patch with symplectic pair of coordinates $(p_j,x_j)$
and the corresponding fermion.  Than we have that $(p_i,x_i)$ and
$(p_j,x_j)$ are related by a canonical transformation preserving the
symplectic form $dx_i\wedge dp_i =dx_j\wedge dp_j$, with a generating
function $S(x_i,x_j)$.  This acts in the usual way on the wave
functions of $\Psi_i$, so $\psi_i(x_i)$ must transform in the
same way as well,
\eqn\wf{
\psi_j(x_j)=\int dx_i \;e^{-S(x_i,x_j)/g_s}\; \psi_i(x_i).
}
In terms of the bosons $\psi(x)=e^{\phi(x)/g_s}$, this is what we
anticipated in \trans.

In particular, if we restrict to linear transformations $g \in Sp(2, {\bf R})$,
we find that the fermion field $\psi(x)$ transforms in the metaplectic
representation
$$
\psi_j(x_j) = U(g) \psi_i(x_i).
$$

This transformation property is immediately clear at tree-level if we
make use of the relation to the Kodaira-Spencer field through the back
reaction.  Using the bosonization/fermionization formulas we see that
the one-point function of the brane creation operator
$$
\Psi(x)=\langle \psi(x)\rangle
$$
can be expressed as
$$\Psi(x) = \exp {1\over g_s} \int^x p(x) dx = \exp {1\over g_s} \f(x).$$
But this is just the WKB approximation for a wave function! The
transformation rules for the boson $\f(x)$ that we found in section 3
therefore immediately imply the transformation rules for the
fermion/brane. In fact, these rules become more transparent from the
brane perspective. Starting from the {\it linear} action of the
coordinate transformation in the Hilbert space of brane wave function
$\psi(x)$, one derives the {\it non-linear} action on the space of
Kodaira-Spencer fields $\f(x)$.

Recall that the boson $\phi(x)$ has a classical piece, corresponding
to the background geometry, and a fluctuating quantum piece:
$$
\phi(x) = \phi_{cl}(x) + \phi_{qu}(x),
$$
The classical piece is given by the integral of the canonical one-form
on the Riemann surface, $\phi_{cl}(x) = \int^{x} p\,dx$. It is the
field $\phi_{qu}(x)$ that creates the quanta of the Kodaira-Spencer
field. Similarly we have to distinguish the classical contribution
to the fermions that create branes/anti-branes
$$
\psi(x) = e^{\phi_{cl}(x)/g_s}\psi_{qu}(x), \qquad
\psi^*(x) = e^{-\phi_{cl}(x)/g_s}\psi^*_{qu}(x).
$$
Note that only the full expression $\psi(x)$ transforms in the
metaplectic representation. So, in all the above formulas one should
always subtract the classical contribution to find the transformation
rules for the quantum field $\psi_{qu}(x)$.

In particular, in the case of small quantum fluctuations we see that
we can approximate \wf\ by a Gaussian integral. This implies that
in this approximation $\psi_{qu}$ transforms as
$$
\psi^{qu}_j(x_j) (dx_j)^\hf = \psi^{qu}_i(x_i) (dx_i)^\hf.
$$
So, in this limit the fermions {\it do} transform as actual global
spin $1/2$ field on the Riemann surface. This fact is well-known in
quantum mechanics --- wavefunctions transform as half differentials.
This is directly related to another fact. The loop momenta
$$
\oint \p\f = \oint p\,dx
$$
of the boson $\f(x)$ are globally well-defined. The invariance of
these periods of the Liouville form under canonical transformations is
well-known in classical and quantum mechanics. In our present context
they are just the projections of the period of the holomorphic 3-form
$\Omega$ on the Calabi-Yau threefold.
In terms of the branes/fermion we have
$$
\p\f = \psi^* \psi
$$
and therefore the periods are given by the ``norm'' of the wave
function
$$
\oint \p\f= \int dx \,\psi^*(x) \psi(x).
$$
In order to make sense of this expression in different coordinate
patches the fields $\psi,\psi^*$ should transform as a
half-differential -- a fact well-known from quantum mechanics.

Coming back to the one-point function $\Psi(x)=\langle \psi(x)
\rangle$, we note that it is not given by an arbitrary wave
function. Semi-classically it obviously satisfies the Schr\"odinger
equation
$$
H(p,x)\Psi(x) =0
$$
where the Hamiltonian is given by the equation of the Riemann surface!
As discussed before, this is a reflection of the unbroken part of the
$\cW$-algebra symmetry when we shift to the $H(p,x)$ background.
We have found that this fact generalizes in many settings to the full
quantum theory. For example, for the topological vertex this fact
immediately leads to the quantum dilogarithm as giving the full
quantum amplitudes with one set of Lagrangian branes in ${\bf C}^3$,
as we will discuss later. In general however, there are normal
ordering ambiguities in writing $H(p,x)$, which make this difficult to
use. We will return to this below, where we will provide a way to
resolve these normal ordering ambiguities.

\subsec{The quantum free energy ${\cal F}$ and B-branes}

In this subsection we explain how one can use the formulation in terms
of branes and the $\cW_{1+\infty}$ symmetries of the theory to compute
the quantum free energy, in terms of an infinite sequence of Ward
identities that the amplitude $|V\rangle$ satisfies.  This in particular
provides the correct normal ordering prescription for the
$\cW_{1+\infty}$ symmetry generators.

Suppose we pick a patch, say $x_i\rightarrow \infty$ corresponding to
a point $P_i$ on the Riemann surface.  In this patch, consider an
action of the $\cW_{1+\infty}$ generator given by the Hamiltonian
$$
f(x_i,p_i) = x_i^m p_i^n.
$$
Quite generally, such a Hamiltonian is represented in the fermionic
representation as
$$
W_m^{n+1} = \oint_{P_i} \psi^{*}(x_i)\;x_i^m p_i^n\;\psi(x_i).
$$
As we discussed in section 3, the $\cW$ actions preserve the
symplectic form and they generate broken symmetries of the theory
corresponding to repameterizations of the Riemann surface, provided
their action is regular.  This action generalizes the framing
ambiguity which also acts as $\cW$-algebra in the fermionic
basis. Namely \wal\ the generators are given by fermion bilinears
$W_0^{n+1}$.

In the case of one puncture, the $\cW$ symmetry implies that its
generators annihilate the state:
\eqn\wards{
\oint_{P} \psi^{*}(x)\;x^mp^n\;\psi(x)|V\rangle =0.}
When there is more than one puncture, the symmetry generators do not
annihilate the state $|V\rangle$ but get related to shifting of coordinates on
other punctures:
\eqn\ward{
\oint_{P_i} \psi^{*}(x_i)\;x_i^mp_i^n\;\psi(x_i)|V\rangle
= -\sum_{j\neq i}^s\oint_{P_j}
\psi^{*}(x_j)\;x_i(x_j,p_j)^mp_i(x_j,p_j)^n \;
\psi(x_j)|V\rangle.
}
In the above equation, $x_i=x_i(x_j,p_j)$ and $p_i=p_i(x_j,p_j)$
correspond to the canonical transformations of coordinates between the
different patches, so \ward\ is simply a consequence of how the
fermions transform. In all cases we will restrict to $SL(2,{\bf Z})$
transformations.  The transformations relating different patches could
have been relaxed to arbitrary symplectic transformations, however the
normal-ordering ambiguities in terms of defining these in general give
rise to many different solutions related by subtle quantum
ambiguities, so we will not consider this.

The Ward identities that the symplectic reparameterization imply are
sufficient to fix the state $|V\rangle$ (Strictly speaking this is
true for Riemann surfaces of genus zero. There is a subtlety here for
higher genus Riemann surfaces, to which we will return later).  To
solve the Ward identities for $|V\rangle$ it is useful to bosonise the
fermions, and evaluate \ward\ in an arbitrary coherent state $\langle
t|= \langle t_1| \otimes \ldots \otimes \langle t_s|$.  This gives
rise to a set of differential equations for the free energy, as we
have for example,
$$
\eqalign{ &
\langle t|\oint_{P_i} e^{-\phi(x_i)/g_s}
{x_i}^m\;\;p_i^n\;
e^{\phi(x_i)/g_s}|V\rangle \cr
& \qquad =
\oint_{P_i} e^{-\phi(x_i;\, t_k, {\del_k })/g_s}\;
({g_s \del_{x_i}})^n\;x_i^m\;e^{\phi(x_i;\,t_k, {\del_k)/g_s}} e^{{\cal F}(t)}.\cr }
$$
These can be solved recursively, genus by genus: expanding \ward\ in
power series in $g_s$ as well as the free energy, ${\cal F}(t,g_s)
=\sum_g F_{g}(t)\; g_s^{2g-2}$, one first gets a set of differential
equations satisfied by $\cF_{0}(t)$. Solving this, at next order
$g_s^2$ order we get a Ward identity for $\cF_{1}(t)$ and so on.

We discussed above that the generators of $\cW_{1+\infty}$ algebra
that give rise to symmetries of the theory correspond to those
symplectic coordinate transformations which are regular on the Riemann
surface. Note that, in the classical limit -- more precisely, in the
limit where the Kodaira-Spencer theory becomes an ordinary theory of a
globally defined free chiral boson on a Riemann surface -- the
generators must correspond to symmetries of this classical
theory, which are well known.  The Ward identities \ward\ reduce in
this limit to
$$
\int_{P_i} x_i^m p_i(x_i)^n \del \phi(x_i)=
-\sum_{j\neq i}^s\int_{P_j} x_i(x_j)^m p_i(x_j)^n\;\del \phi(x_j).
$$
where $p_i=p_i(x_i)$ is the classical equation that $p_i$ satisfies on
the Riemann surface. This generates a symmetry provided $x_i^m
p_i(x_i)^n$ are holomorphic on the punctured Riemann surface.  In the
following sections we will see many examples of how $\cW_{1+\infty}$
symmetries can be used to find $|V\rangle$.

In the cases that we consider in this paper, the state corresponding
to Kodaira-Spencer theory on the Riemann surface with punctures is a
Bogoliubov transformation of the fermionic vacuum, corresponding to
fermions being free.  We conjecture that this is generally
the case; it should be possible to derive this from the
$\cW_{1+\infty}$ constraints.

Introduce the familiar mode expansion of the fermions
$$
\psi(x_i) = \sum_{n\in {\bf Z}} \psi^{i}_{n+1/2} x_i^{-n-1},\qquad
\psi^{*}(x_i) = \sum_{n\in {\bf Z}} \psi^{*i}_{n+1/2} x_i^{-n-1}
$$
for $i=1,\ldots s$ where $s$ is the number of punctures,
with anticommutation relations
$$
\{ \psi^i_n, \psi_m^{j*} \}=\delta^{ij}\delta_{n+m,0},
$$
in such a way that the positive modes annihilate the vacuum:
$$
\psi_n^i |0 \rangle = \psi_n^{i*} |0 \rangle =0, \,\,\,\,\, n>0.
$$
Then the state $|V\rangle$ can be written as
\eqn\bilinearform{
|V\rangle = \exp \left[ \sum_{i,j=1}^s \sum_{m,n=0}^\infty
a_{mn}^{ij} \psi^i_{-m-1/2} \psi^{j*}_{-n-1/2}\right]|0\rangle.}
Having computed the state $|V\rangle$ we can explicitly check
whether this holds.

The knowledge of amplitudes corresponding to inserting B-branes in
only one patch, can be used to compute the whole of $|V\rangle$.
Consider the two-point functions $\langle
0|\psi(x_i)\psi^{*}(x_j)|V\rangle.$ For fermions in the same patch, we
can write this as
$$
\langle 0|\psi(x_i)\psi^{*}({\tx}_i)|V\rangle =
G(x_i, {\tx}_i)\Psi(x_i)\Psi^{*}(\tx_i)
$$
where $G(x_i,\tx_i)$ is the Green's function of free-fermions on a
quantum Riemann surface.
%
%
Moreover, the knowledge of $\Psi(x_i)$ and $G(x_i,{\tx}_i)$ at any one
of the punctures allows us to compute, by ``parallel transport'', all
other two-point functions $\langle 0|\psi(x_j)\psi^{*}(x_k)|V\rangle$
for $j,k\neq i$.
Using \wf\
$$
\psi(x_j)=\int d\tx_i \;e^{-S(\tx_i,x_j)/g_s}\; \psi(\tx_i).
$$
we have
$$
\langle 0|\psi(x_j)\psi^{*}(x_k)|V\rangle=
\int dx''_i \int dx'_i\;e^{S(x''_i,x_k)/g_s-S(x'_i,x_j)/g_s}\;
{\Psi(x'_i)\Psi^{*}(x''_i)G(x'_i,x''_i)}.
$$
Moreover, it follows that it suffices to check the bilinearity in one
patch only \foot{This is because the defining property of $|V\rangle$
is that the correlation functions of fermions are given by
determinants of $G(x_i,x_j)$ and this is preserved by canonical
transformations.}.

\subsec{Non-compact brane probes and Kontsevich-like matrix models}

As discussed in the previous section we can view the closed string
deformation of geometry, in the patch with local coordinate $x_j$,
parameterized by the couplings $t_n^i$ as being induced by branes at
$x^i=x^i_j$, $j=1,\ldots N$.  In terms of string theory description
what this means is that if we denote the partition function of closed
strings by $Z_{closed} (t_n^i)$ and that of the open string sector by
$Z_{open}$ then
$$
Z_{closed}(0)Z_{open}(x^i_j)=Z_{closed}(t_n^i),
$$
where the $t_n^i$ are determined from the geometry of the branes
located at $x_i^j$ in each asymptotic geometry given by
$x^i\rightarrow \infty$.  These parameters are related to each other
by
$$
t_n^i={g_s\over n}\sum_j (x^i_j)^{-n}.
$$
This in particular means that the $t_n^i$ dependence of $Z_{closed}(t_n^i)$
can be computed by integrating the field theory living on the brane,
which is given by $Z_{open}(x_j^i)$.  This suggests that another way
to compute the closed string amplitudes is to use the open string sector
which deforms it.

The action for the open string sector has already been studied \AV.
Let us, for simplicity first write it for one brane probe, located at
$x=x_1$ which is defined near the asymptotic region $x\rightarrow
\infty$.  Then the tree level action, as a function of varying the
position of the brane to a generic point $x$ is computed by
considering the disc diagram in the B-model and is given by
integrating the 1-form $\lambda=xdp$
\eqn\kmm{
S(p)={1\over g_s} \left(\int^p x(p') dp' -x_1 p\right)=
{1\over g_s}\Bigl[W(p)-x_1 p\Bigr],
}
where $x(p)$ is found by solving $H(p,x)$ and
$$
W(p)=\int^p x(p') dp'.
$$
Note that the extremum of the action gives the classical solution
$$
dS/dp=0\rightarrow x(p)=x_1.
$$
which is consistent with the classical position of the brane. If we have
instead of one brane, $N$ branes at positions
$$\Lambda =(x_1,..., x_N),$$
we can write a matrix model version of the above action where we treat
$p$ as an $N\times N$ matrix, which we denote by $P$ and $\Lambda$ as
a diagonal $N\times N$ matrix.  We thus have a matrix action
$$
S(P)={1\over g_s}\Tr \Bigl[W(P) - \Lambda P\Bigr],
$$
In general there can be quantum corrections to this classical
action, which can include multi-trace contributions.
In some cases there is no quantum correction in which
case we get a very simple matrix model description of the
closed string amplitude:
$$
Z_{closed}(t_n)=\int DP\  \exp{\Bigl[{1\over g_s}{\rm Tr}( W(P) - \Lambda
P)\Bigr]}$$
This is a Kontsevich-like matrix model, where the source
term has the data about the classical position of the branes
and satisfies
$$
t_n={g_s\over n}\Tr\, \Lambda^{-n}$$
This is assuming that $\partial \phi$ has integral expansion powers in
$x$, as in \eva .  However, in some examples (such as the KdV
hierarchy) we may have for geometric reasons a monodromy in $\partial
\phi$ as $x\rightarrow e^{2\pi i }x$, in which case the above expansion
modes $n$ will be fractional.
Later in this paper we will show that the original example of
Kontsevich arises from exactly such a picture.  Note that we do not
need $N$ to have a large value for this relation to be true.  It is
true for any $N$.  Of course for finite $N$ we only get an
$N$-dimensional subspace of all the allowed values of $t_n$'s.

We can in fact state the above from a slightly different
perspective, but in a way which includes the quantum corrections.
Namely, consider $N$ B-branes in some patch where we
have, say $(x_i, p_i)$ as coordinates, and relate this to B-branes in
different coordinates $(x_j,p_j)$, related to the
original ones by a symplectic transformation generated by $S(x_i,x_j)$
which is quadratic in the variables.
$$
S(x_i,x_j) = (-{d\over 2} x_i^2 + x_i x_j - {a \over 2} x_j^2)/c.
$$
We can then write the amplitude
$$
\langle N| \psi(x_{j,1})\ldots \psi(x_{j,N})|V\rangle =
\prod_{ n<m}(x_{j,n}-x_{j,m}) e^{{\cal F}(x_j)}
$$
in terms of
$$
\langle N| \psi(x_{i,1})\ldots \psi(x_{i,N})|V\rangle
$$
as
$$
\langle N| \psi(x_{j,1})\ldots \psi(x_{j,N})|V\rangle =
\int \prod_{n=1}^{N} dx_{i,n} e^{-\sum_{n=1}^N S(x_{i,n},x_{j,n})}
\langle N| \psi(x_{i,1})\ldots \psi(x_{i,N})|V\rangle.
$$
But this is just an eigenvalue representation of the matrix integral
$$
Z(X_j) =
\int DX_i e^{-{\rm Tr} \, S(X_{i},X_{j})}Z(X_i)
$$
where $Z(X_i)=e^{{\cal F}(X_i)}$ expressed in terms of
${\rm Tr} \, X_i^n =\sum_m x_{i,m}^{n}$. We have used here the well known expression for
the Itzykson-Zuber integral \iz.
As we discussed above, in some cases, such as the $(m,1)$ minimal models,
one can find variables $(p_i,x_i)$ such that $Z(X_i)$ is simple and the
above is an effective way to evaluate the amplitudes corresponding to
other patches.

\subsec{Compact and non-compact branes}

We can also consider both non-compact branes and
compact branes at the same time.   This will give rise
not only to a transition
in the compact parts of the geometry but influence the
asymptotic geometry of the Calabi-Yau as well.  As for
the action for the branes this means that in addition
to the matrix model action $W(\Phi)/g_s$ describing the compact branes, and the
corresponding contribution of the non-compact brane
we also have to integrate
out the open strings stretched between the compact and the
non-compact branes.  This is the mirror situation to
that considered in \ovknot\ which leads to
determinant of the mass of the open strings stretched between them.
In the above geometry \btra , if the non-compact B-brane is at $x$
this leads to the mass
$$
M=x-\Phi,
$$
which thus leads to modification of $Z$ by the inclusion of
$\det(x-\Phi)$, or by the changing of the action by
$$
\delta S={\rm Tr}\, \log(x-\Phi).
$$
Note that this is up to a factor of $2$ the same as adding one
extra eigenvalue to $\Phi$ and freezing
it at $x$.

\subsec{Fermion number flux and higher genera}
Let us consider the Riemann surface $\Sigma:\ H(p,x)=0$ of genus $g$.  If
$g>0$, we can obtain the amplitudes
 by sewing of $g=0$ case, as is familiar in the operator formulation
 of Riemann surfaces.  However, there is one subtlety with this
 sewing prescription:
we can ask what is the fermion number (the momentum
of the $\phi$ field) going around the loops (say along the $A$-cycles)
of the Riemann
surface.  In fact we have to
have a particular treatment of the momenta around loops to get
a unique answer for our amplitudes in such cases.  This issue
does not arise when $g=0$, and in such cases any fermion
operators we insert at the boundaries can be absorbed by shifting
the vacuum fermion number at the corresponding Hilbert space,
as discussed before.

There are two natural choices one can make:  One is to set
the momenta in the loops to be zero.  This case turns
out to be the natural choice for computation of the B-model amplitudes.
However, it is not so natural from the viewpoint of free fermions:
If we wish to retain a free fermionic description for higher genus
as well, it is natural {\it not to fix the fermion number} around the loops.
We introduce $g$ new variables $\theta_1,...,\theta_g$ and
consider
$$Z(\theta_i)=\sum_{N_i}Z_{N_1,...,N_g} {\rm exp}(i N_i \theta_i)$$
This has the advantage that the amplitudes can still be constructed
using free fermions (i.e. the vertex is still exponential of
bilinear fermion terms instead of sum of such exponentials).
Of course we can obtain the answer for the case where we have fixed
the momenta to be zero from this expression by considering
$$Z_{N_i=0}=\int \prod_{i=1}^g d\theta_i \ Z(\theta_i)$$

It may appear that there is more information in $Z(\theta_i)$
than in $Z_{N_i=0}$.  However this is not the case, because there
 is some redundancy in $Z_{N_i}$.  This can be seen as
follows:  Consider adding a brane/antibrane at a point
on the Riemann surface $\Sigma$ and move the brane around
a non-trivial cycle, say $B_i$ and bring it back to annihilate
the anti-brane.  In doing so we have changed the fermion number
flux through the $A_i$ cycle by one unit.  This can be equivalently
viewed as changing the moduli of the Calabi-Yau
(this was also observed in a related context in \ref\nei{
A. Neitzke and C. Vafa, work in progress.}),
and in
particular the Riemann surface $\Sigma$.  Note that as the brane
traverses along the $B_i$ cycle on the Riemann surface,
it spans a 3-cycle $C$ in
the Calabi-Yau.
According to \bacr\ this changes the integral of $\Omega$
along three-cycle $C'$ by
$$\Delta \int_{C'} \Omega =\#(C\cap C')\;g_s$$
Let us denote the moduli of the Calabi-Yau by $\int_{C_i}\Omega =S_i$
and let the intersection number of $C_i$ and $C$ be $\#(C_i\cap
C)=m_i$.
Then we have
\eqn\changeflux{
Z_{N_i}(S_i)=Z_{N_i=0}(S_i+N_im_i g_s)}
and
$$Z(\theta_i)=\sum_{N_i}{\rm exp}(i N_i \theta_i)\ Z_{N_i=0}(S_i+N_im_i g_s).$$
%

\newsec{Examples}

In this section we consider four classes of examples
to illustrate the ideas developed in the previous sections.

In the first class of examples we consider CY geometries of the form
$$H(p,x)=\sum_{i=0}^r p^i f_i(x)$$
where $f_i(x)$ is a power series in $x$ (possibly a finite
polynomial).  These geometries can also be viewed as being
obtained from matrix models associated with $A_{r-1}$ quiver
diagrams, with $r-1$ adjoint matrices with potentials $W(\Phi_i)$
interacting with bifundamental fields.  The case $r=1$ corresponds
to the KP hierarchy. We also discuss the relation of these
theories to $(r,*)$ minimal models.  The case $r=1$ can be viewed
as taking a particular limit of these geometries.  We show in the
case of $(1,p)$ models how the brane probe picture gives rise to
the Kontsevich-like models (the case $p=2$ is the original
Kontsevich model).  These are also equivalent to the $p$-th,
${\cal N}=2$ minimal model coupled to topological gravity
\refs{\dvv ,\wal}.  We also explain
 the meaning of Virasoro (or more generally
$\cW$-constraints) for these models, as simply encoding the fact that
Calabi-Yau is non-singular.

In the second example we consider CY geometry corresponding to the
conifold $H(p,x)=xp-\mu$ and its deformations.  As has been noted in
\goshv\ this is equivalent to topological string for $c=1$ bosonic
string at self-dual radius.  We show how one can interpret the results
of \dmp\ from the perspective of the present paper, as well as how the
Toda hierarchy arises from the Calabi-Yau geometry.

In the third example we consider the mirror of A-model topological
string on ${\bf P}^1$ coupled to topological gravity.  This is a model
in which $H(p,x)=x+e^p+ q e^{-p}$ where $x\in {\bf C}$ and $p\in {\bf
C}^*$.

In the fourth example we consider the mirror of A-model topological
string on ${\bf C}^3$.  This is the case where $H(p,x)=e^x+e^p-1$.  We
recover the topological vertex
\akmvII\ from the perspective of this paper from various approaches.
We also give the generalization of this to B-model mirrors of
arbitrary toric Calabi-Yau manifolds.

\subsec{Example 1: Matrix models and the KP hierarchy}

It might be useful to first consider the case of geometries that
naturally appear in matrix models and their double scaling limit. Let
$W(x)$ be a polynomial of degree $n+1$.
Consider, as in \dvI , $N$ compact B-branes in
$$
zw + p^2-W'(x)^2=0
$$
where the B-branes are wrapping $n$ ${\bf P^1}'s$ at $W(x)'=0$, as we
discussed in section 4.1.
The open topological string is described by the
saddle-point expansion of the $N\times N$ matrix integral
\eqn\onematrix{
Z = {1\over {\rm vol}\,U(N)} \int d\F_{N\times N} \cdot \exp\left[{{1\over g_s}\Tr\,W(\F)}
\right].
}
In the 't Hooft limit $N\to\infty,g_s\to 0$ with $g_s N$
fixed, this describes closed topological strings in the Calabi-Yau geometry
$$
zw + p^2 -W'(x)^2 -f(x) =0.
$$
Here
$$
W(x) = \sum_{i=0}^{n+1} t_i x^i
$$
is a given polynomial, and $f(x)$ is a polynomial of degree $n$ --- the
quantum correction that is determined by the saddle-point around which
the expansion is performed. More precisely, $f(x)$ is encoded by the
distributions $N_i$ of the eigenvalues of $\f$ among the $d$ critical
points of $W$ through the 't Hooft couplings $s_i= g_sN_i$.

The conjugate variable $p(x)$ appears naturally as (the expectation
value of) the resolvent of the matrix model
$$
p(x) = W'(x) + 2g_s \Tr{1\over x-\F}.
$$
The corresponding meromorphic one-form  can be written as
$$
\lambda = p(x) dx =d\f(x)
$$
with
$$
\f(x) = W(x) + 2 g_s \Tr\log(x-\F).
$$
This collective fields has an interpretation as the effective action
of a single extra eigenvalue in the background of the $N$ dynamical
eigenvalues of the matrix $\F$. Identifying the one-point functions in
the matrix model with derivatives of the partition function (in the
genus zero limit, with $\log Z = \sum_{g\geq 0} g_s^{2g-2} \cF_g$)
gives the relation
$$
\langle \Tr\,X^n\rangle = g_s {\p \cF_0 \over \p t_n}
$$
Using this fact we can write the mode expansion of $\p\f$ in the, by
now familiar, form
\eqn\modes{
\p\f(x)= \sum_{n>0} n\, t_n x^{n-1} + 2
g_s^2 \sum_{n>0} {\p\cF_0 \over \p t_n} x^{-n-1}
}
The hyperelliptic curve
$$
H(p,x)=p^2 - W'(x)^2 -f(x)=0
$$
that gives the relation between $x$ and $p$ emerges as the large $N$
limit of the loop equations of the matrix model.

Note that this curve has two set of moduli. The unnormalizable ones
correspond to deformations of the couplings $t_k$ in the matrix model
potential. The corresponding deformations of the meromorphic one-form
$\delta \lambda$ are given by Abelian differentials of the second
kind, that behave as $\sim x^{n+k} dx$ in the limit $x\to\infty$. The
couplings $s_k$ induce normalizable deformations $\delta \lambda \sim
x^k dx$ that are given by Abelian differentials of the first kind and
correspond to variations of the quantum correction $f(x)$.  We will
focus our discussion here on the unnormalizable deformation $t_k$ that
are visible at $x\to\infty$.

In the standard interpretation of the (genus zero) loop equations one
considers the general deformation
$$
p(x) = p_{cl}(x) + \p\f(x)
$$
where $\p\f$ has the mode expansion \modes. This includes terms
$x^{-n}$ that have poles in the interior $x\to 0$. However, the
deformation should be such that the corresponding hyperelliptic curve
stays smooth away from infinity. Therefore the stress tensor
$$
T(x) = \p\f^2 = p^2 = \sum_{n\in Z} L_n x^{-n-2}
$$
should be regular in the interior. In particular, the modes $L_n$ with
$n\geq -1$ should vanish. These are the famous Virasoro constraints.

In the full quantum theory these relations still hold, but now are
implemented as operator relations. That is, one writes the partition
function as a wave function in a coherent state basis
$$
Z = \langle t| V \rangle
$$
and requires
$$
L_n|V\rangle =0,\qquad n\geq -1.
$$
Now $L_n$ are the modes of the quantum stress tensor.

By generalizing to matrix chains, more precisely quiver matrix models
of type $A_{r-1}$ \refs{\ckv ,\cfikv ,\dvII }, one obtains curves that are
$r$-folded covers of the eigenvalue $x$-plane given by a schematic
equation of the form
$$
p^r+ \ldots + F(x)=0.
$$
In that case the corresponding matrix model partition function is not
determined by the quadratic Virasoro constraints, but by
$W$-generators of order $r$ in the collective field $p=\p\f(x)$.

\subsec{Branes in the Gaussian matrix model}

In the light of the discussion we have had in the
previous sections, we now revisit the matrix model and interpret it in
terms of branes.  Quite generally, putting a brane in the matrix model
is implemented by shifting the couplings in the superpotential
$$
t_k \to t_k + {g_s\over k} x^{-k}.
$$
The brane creation operator is given by
$$
\psi(x) = e^{\f(x)/g_s} =  e^{W(x)/g_s} \det(x-\F)
$$
where we recognize the classical and the quantum
contribution. Partition functions of branes are therefore represented
as correlation functions of characteristic polynomials in the matrix
model. For example, a $M$-point function is given by
$$
\langle M| \psi_{qu}(x_1) \cdots \psi_{qu}(x_M) |V\rangle
= \Bigl\langle \det(x_1-\F)\cdots \det(x_M-\F)\Bigr\rangle,
$$
where $\langle \cdots \rangle$ denotes a normalized expectation value
in the matrix model.

Using the methods of orthogonal polynomials one can easily evaluate
such correlators. Let $\{P_i(x)\}_{i \geq 0}$ be a basis of orthogonal
polynomials for the matrix model \onematrix, normalized such that
$P_i(x) =x^i+ \ldots$, and let
$$
\Psi_i(x) = P_i(x) e^{W(x)/2g_s}
$$
be the corresponding ``wave functions.'' Then one expresses the brane
$n$-point function as an $n \times n$ Slater determinant \minmat\
$$
\Bigl\langle \det(x_1-\F)\cdots \det(x_M-\F)\Bigr\rangle =
{\det P_{N+j-1}(x_i) \over \Delta(x)},
$$
with Vandermonde determinant
$$
\Delta(x) = \det x_i^{j-1}
$$
Adding the classical contribution we get
$$
\Bigl\langle \psi(x_1) \cdots \psi(x_M) \Bigr\rangle =
{\det \Psi_{N+j-1}(x_i) \over \Delta(x)}.
$$
In particular for the one-point function we have
$$
\langle\psi(x)\rangle = \Psi_N(x).
$$

This relation is particularly illuminating in the simple case of the
Gaussian model
\eqn\gaussian{
Z = {1\over {\rm vol}\,U(N)} \int d\F \cdot \exp\left[-{{1\over 2g_s}\Tr\,\F^2}
\right].
}
This corresponds to the deformed geometry
$$
H(p,x)=p^2+x^2-\m=0
$$
with $\m=g_sN$. In this case (and only in this case) the orthogonal
polynomials are actual eigenfunction of a Schr\"odinger operator. The
polynomials $P_i(x)$ are Hermite polynomials, and therefore the
corresponding wave functions satisfy
$$
\left(-g_s^2 {\p^2 \over \p x^2}+ x^2\right)\Psi_k(x) =
g_s(k+ \hf) \Psi_k(x).
$$
Note that in this case, where we recall that $\mu = g_s N$, we can
verify that the one-point function $\langle\psi(x)\rangle=\Psi_N(x)$
indeed satisfies (up to the quantum shift $N \to N + \hf$)
$$
H \Psi_N = 0.
$$

This relation is not as straightforward in the more general
non-Gaussian matrix model. In that case the corresponding Riemann
surface has genus $g>0$. Therefore there are non-trivial loop momenta
or fluxes
$$
{1\over 2\pi} \oint_{A_i} p\,dx = g_s N_i
$$
As we discussed before the existing of this ``bulk'' moduli makes the
free fermion formulation less straightforward, since we now have to
project on fixed loop charges going through the handles of the Riemann
surface. This is directly related to the fact that the saddle-point
approximation of the matrix model is not captured by the method of
orthogonal polynomials. It would be interesting to push this
connection further.

Returning to the Gaussian matrix model, let us now derive the
Kontsevich matrix model description of this.
This  corresponds to putting $M$ non-compact
B-branes in the geometry
\eqn\qg{
zw = p(p-x)-\mu.
}
near $x\rightarrow \infty$ (where we changed variables apropriately).
This, as explained in section 4.5.
is described by the Kontsevich-type matrix model with action
$$
S = \Tr\left[\Lambda X - \int X(P) dP\right],
$$
where $x({p})$ solves $p(p-x)=\mu$. So we get the matrix model
\eqn\km{
\int_{M\times M} DP \,det(P)^{N }e^{\Tr\left[\Lambda P - P^2/2\right]/g_s}
}
where we have used that $\mu/g_s = N$.

This also leads us to a physical interpretation of
the claim \minmat\ of the equivalence of a general 1-matrix model
\eqn\gm{
\int DX_{N\times N}\; e^{[\,\Tr\, X^2/2 \,+\, \sum_{n>2} t_n \,\Tr\,X^n\,]/g_s}
}
with this particular Kontsevich-type matrix model \km .

Namely, as we have explained
above, the matrix model in \gm ,
for small $t_n$ is describing $N$ compact B-branes on the ${\bf P}^1$
in the conifold
$$
zw = p(p-x)
$$
and deformations by $t_n$'s can alternatively be viewed in terms of
turning on non-normalizable deformations
$$
zw = p(p-W'(x))
$$
where
$$
W'(x, t_n)=x - \sum_{n>2} t_n x^{n-1}
$$
or placing non-compact B-branes at $x = x_i$, $\sum_i x_i^{-n} =\Tr\,
\Lambda^n$, as we discussed above in addition to the compact ones.
The large $N$ dual description of this can, on
the one hand, be viewed as closed string theory on
$$
zw = p(p- W'(x))- \mu W'(x)/x
$$
where $\mu=Ng_s$ and the deformation of the geometry corresponds to
growing a single $S^3$ at $x=0$, corresponding to $t_n$'s being small
deformations around the deformed conifold background $zw=y(y-x)-\mu$.
On the other hand, it is given by placing B-branes at fixed $x$
in \qg . Thus we have reproduced the
result  shown in \minmat\ using the methods of orthogonal
polynomials.

\subsec{Double scaling limits}

The above discussion is perhaps a bit confusing, since (for the one
matrix model) the corresponding hyperelliptic Riemann surface has two
asymptotic ends, distinguished by the sign of $p(x)\sim \pm x^d$ in the
limit $x\to\infty$, or equivalently by the choice of the sheet of the
two-fold cover. A priori one can have independent deformations at both
end, as we will see in a moment when we consider the $c=1$ string and
the Toda hierarchy. Above we treated both ends symmetrically.

This situation simplifies if we take a double scaling
limit \refs{\bk ,\ds ,\GrossAW}.
Geometrically this corresponds to considering the local
geometry where we send $2k+1$ of the original $2d$ branch points close
together and zoom in at that region. (Equivalently, one can send the
remaining branch points to infinity.) In that case the Riemann surface
takes the form
$$
p^2 + x^{2k+1} + \ldots =0,
$$
where the ellipses indicate terms of lower order in $x$. In such a
geometry there is only one asymptotic region $x\to \infty$, where the
variable $x$ is a good coordinate.

This particular double scaling limit of the matrix model is well-known
to correspond to a non-critical bosonic string theory based on the
$(2,2k+1)$ minimal model coupled to two-dimensional gravity
\refs{\dvv ,\DouglasDD }. In
particular, in the simplest case where we zoom in on a {\it single}
branch point, we have the geometry
$$
p^2+x=0,
$$
which corresponds to the $(2,1)$ model or pure topological gravity.

In this case we can introduce again the Kodaira-Spencer field $\f(x)$,
which now has a {\it twisted} mode expansion, because the global
geometry enforces that at infinity
$$
p(e^{2\pi i} x ) = - p(x).
$$
If we expand this twisted boson as
$$
p(x)= \p\f(x) = \sum_{n \geq 0} n\, t_n x^{n-{1\over2}} + 2
g_s^2 \sum_{n \geq 0} {\p \over \p t_n} x^{-n-{3\over 2}}
$$
we make contact with the usual description of topological gravity,
where the coupling $t_n$ couples to the descendant $\tau_n$ of the
identity operator.

Besides the twisting, we have precisely the same argument that fixes
the solution of this deformation problem. The function $p(x)^2$ should
again be smooth in the interior of the $x$-plane. Apart from
$x=\infty$ there should be no extra poles, {\it i.e.}\ no new
asymptotic ends. This then gives again the Virasoro constraints
$$
L_n |V\rangle=0.
$$
It is well-known that these constraints are equivalent to statement
that the partition function $Z$ is a tau function of the KdV hierarchy
that satisfies the so-called string equation
\refs{\GrossAW ,\DouglasDD ,\WittenHR }.

\subsec{The $(m,1)$ minimal models}

One can also consider the double-scaled quiver matrix model that gives
rise to the $(r,s)$ minimal model coupled to gravity. This leads to
the effective geometry
$$
p^r + x^s + \ldots =0.
$$

In particular, consider $(m,1)$ topological minimal models.
The corresponding Riemann surface is given by
$$
p^{m} + x = 0
$$
before turning on the deformations.  This is known to be governed by a
generalized KdV hierarchy with Lax operator, $L(t=0) =L_0$ at zero
times, given by
$$
L_0 = D^{m}+ x
$$
where $D = {\del \over \del x}$.  Notice that the Lax operator is the
equation of the Riemann surface with $D=p$.  Moreover, the one-point
function, corresponding to inserting a non-compact B-brane at fixed
value of $x$
$$
\Psi(x,t) = \langle t| \psi(x)|V \rangle,
$$
is known in the integrable systems literature as the Baker-Akhiezer
function, and $\Psi(x) = \Psi(x,0)$ is annihilated by $L_0$
$$
L_0 \Psi(x) =0
$$
(We can relax the eigenvalue to be arbitrary, but this can be
reabsorbed by shifting $x$).  Turning on the times of the KdV
hierarchy can thus be identified with placing non-compact B-branes at
fixed values of $x$.  In terms of the $(m,1)$ minimal model, this is
computing the topological string partition function deformed by the
gravitational descendants.

We now want to derive the Kontsevich matrix model which also describes this.
To do so, consider first a  B-brane in the $p$ patch. In this case,
we have
$$
x = \del \phi(p) = p^{m} + \sum_{n>0} n t^{n} p^{n-1} + {\del {\cal
F}\over \del t_n} p^{-n-1}
$$
in terms of the classical piece and the quantum one.  In this case,
there is no quantum correction, ${\cal F}=0$ and the classical result
is exact.  This is because the Ward identity
$$
\oint\psi^*(p) p^{n}\psi(p)|V\rangle =0
$$
for $n>0$ implies that $\del \cF / \del t_n=0$ for any $n$ which
together with the Riemann surface equation as an initial condition,
implies that ${\cal F}$ is zero all-together.  We see that
$$
\Psi(p) = \langle \psi(p) \rangle= \exp\left[ {1\over g_s} {p^{m+1}\over m+1}\right]
$$
where ${p^{m+1}\over m+1}$ agrees with the classical action $\int x(p)
dp$. Moreover, we have that
$$
H(p,x)\Psi(p) = (p^m + x)\Psi(p) =0
$$
with $x = -g_s {\del \over \del p}$.

To describe the topological $(p,1)$ minimal model deformed by various
observables, as we discussed above, we need to consider non-compact
B-branes at fixed values of $x$.  We can do this by relating them to
B-branes in the $p$-variable, which are simple, and considering
symplectic transformations.  The canonical transformation that
relates the two descriptions is just the $S$ transformation, {\it
i.e.}, the Fourier transform
$$
\Psi(x) = \int dp \;e^{- xp/g_s}\Psi(p)
=\int dp \;\exp\Bigl[\bigl(- xp +{p^{m+1}\over m+1}\bigr)/g_s\Bigr].
$$
We can now see how the Kontsevich matrix model arises, namely, by
putting $N$ B-branes at fixed value of $x$.  This leads, as discussed
before, to a Kontsevich-like matrix model given by \kmm\ with action
$$
S={1\over g_s}
\Tr[-\Lambda P + \int X(P)dP]={1\over g_s}\Tr\left[-\Lambda P +{1\over m+1} P^{m+1}
\right].
$$
The matrix $\Lambda$ fixes the classical $x$ positions of the branes.
The couplings
$$t_k={g_s\over k+{1\over m}}{\rm Tr} \Lambda^{-k-{1\over m}} =
{g_s\over k+{1\over m}}\sum_i x_i^{-k-{1\over m}}$$
encode the deformed geometry by
$$
p=x^{1\over m}+\sum_{k=0}^{\infty}{t_k}x^{k+{1\over m}}+O(1/x),
$$
where we have taken the asymptotic monodromy of $p$ into account.

This model is equivalent to bosonic $(m,1)$ minimal model coupled to
gravity.  On the other hand it is also known that this is equivalent
to the B-model topological string coupled to minimal ${\cal N}=2$ model
\refs{\kekeli ,\dvvts}, which is given by LG theory
with one chiral field $P$ with superpotential \mvw
$$
W(P)=P^m.
$$
{}From the above relation we conclude that if we consider coupling of
the B-model theory with superpotential $W(P)$ to topological strings
then this is equivalent to considering B-model topological strings on
the Calabi-Yau 3-fold
\eqn\cts{zw-x-W(P)=0.}
This may also be possible to explain more directly by showing that the
worldsheet theory of Calabi-Yau and that of the theory with
superpotential $W(P)$ coupled to topological gravity are the same.  In
fact there is some evidence that this is not too difficult.  It has
been shown in \OVblackhole\ that the conformal theory of strings
propagating on the $A_{m-1}$ singularity of $K3$
$$zw- P^m=\mu$$
is given (up to a ${\bf Z}_m$ orbifold) by the tensor product of
the $N=2$ SCFT with superpotential $P^m$ times a Liouville system.
The geometry we are considering here is a simple extension of this
where $\mu$ is varying over an extra dimension parameterized by
$x$, $\mu(x)=x$.  It would be interesting to complete this
identification. We will see another example of this when we
consider below topological strings on ${\bf P}^1$.

\subsec{Example 2: $c=1$ strings and the Toda hierarchy}

The $c=1$ non-critical string compactified on the self-dual radius
$R=1$ is well-known to be equivalent to B-model topological strings on
the conifold. In this case we are dealing with Hamiltonian
$$
H = p^2 - x^2 - \mu.
$$
We will change variables to
$$
x,y =p\pm x
$$
so that the level set is given by the curve
$$
xy = \mu.
$$
\ifig\cone{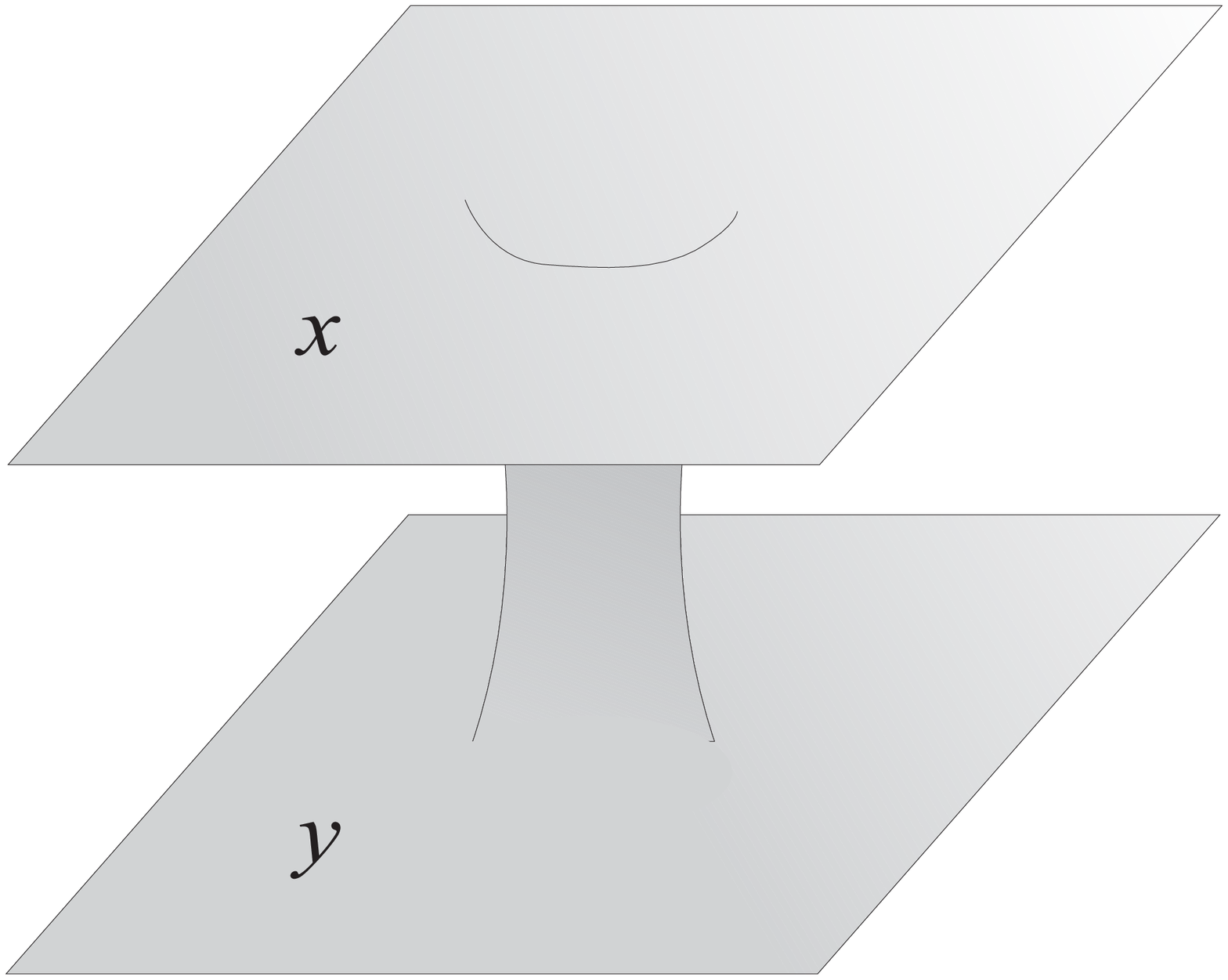}{55}{
The geometry relevant to the $c=1$ string at selfdual radius is the
complex curve $xy=\mu$.}
Note that this curve has two asymptotic regions (see \cone) described
by $x\to\infty$ and $y\to\infty$. These regions correspond to the
incoming and outgoing excitations in the usual $S$-matrix formulation
of the $c=1$ matrix model.

At tree-level this model describes the Kodaira-Spencer theory of
the curve $xy=\mu$. Here one parameterizes these deformations as
Laurent expansions of $y(x)$ in the $x$-patch and $x(y)$ in the
$y-$patch by
$$
y = -\del_x \phi(x), \qquad x =\del_y {\tf}(y).
$$
Here the $\phi$'s have a classical piece $\phi_{cl}(x),\; {\tilde
\phi}_{cl}(y)$, corresponding to the one-form on the Riemann surface,
and a quantum piece, corresponding to the Kodaira-Spencer field
$\phi_{qu}(x), \;{\tf}_{qu}(y)$ in the two patches.  The classical
contributions read
$$
\del_x \phi_{cl}(x)= -{\mu \over x}, \qquad \del_y {\tf}_{cl}(y)={\mu \over y},
$$
whereas the quantum parts have mode expansions
$$
\eqalign{
\del_x \phi_{qu}(x) & = \sum_{n>0} n \, t_{n} x^{n-1} +
\sum_{n>0} {\del\over \del t_n} \; x^{-n+1}, \cr
\del_y {\tf}_{qu}(y) & =\sum_{n>0}n \, \tt_{n} y^{n-1} +
\sum_{n>0}
{{\del\over \del {\tt}_n}}\; y^{-n+1}.\cr}
$$
The coefficients $t_n,\tt_n$ of the unnormalizable modes (in the limit
of $x,y \to \infty$) should be seen as coupling constants, while the
coefficients of normalizable modes have an interpretation as one-point
functions of the dual operators $\del_{n}$.
Note that the couplings and one-point functions can be written, in the
classical limit, as
periods ($\oint \equiv {1\over 2\pi i} \int$), e.g.
\eqn\periods{
\eqalign{
t_{n} & =  {1\over n} \oint_{x\rightarrow \infty} y x^{-n} dx \cr
{\p F_{0} \over \p t_{n}} & = \oint_{x\rightarrow \infty}  y x^{n} dx
}}
With this substitution the consistency of the equations
$$
y=\del\phi(x)=f(x), \qquad x=\del {\tf}(y)=f^{-1}(y),
$$
is equivalent to the fact that the function $\cF_0(t_n,\tt_n)$ solves
the dispersionless limit of the two-Toda hierarchy.

One can also easily solve this model in terms of Ward identities, at
least at tree-level. Note that in this case the curve has topology
$\C^*$ or a twice punctured sphere. Let $P=\{x=\infty\}$ and
$Q=\{y=\infty\}=\{x=0\}$ denote the north and south pole respectively.
Now consider the action of the $\cW$-current $y=\del\f$. More
precisely, consider the charge $\oint_P x^{n} y$. If we move the contour
to the south pole $Q$ we obtain the identity
$$
\oint_P x^{n} y=\oint_Q {1\over n+1}x^{n+1} y.
$$
In other words, we have
\eqn\clw{
{\p \cF \over \p t_{n}} =\oint_Q {1\over n+1}(\del \tf(y))^{n+1}.  }
Note that the bosons $\phi$ and $\tf$, in the language of section 3
are related by the $S$ transformation that acts on the canonically
conjugated variables $x$ and $y$ as
$$
S:\ (x,y) \to (y,-x).
$$
Therefore the fields $\f(x)$ and $\tf(y)$ are (semi-classically)
related by a Legendre transform. This is highly non-linear
transformation, as the above Ward identity clearly shows --- the
``in-coming'' modes of $\del\f$ are polynomials in the ``out-going''
modes of $\del\tf$.

In accord with our general philosophy, the corresponding quantum
relations are most natural written in terms of branes, which
fermionize the boson fields in the usual fashion
$$
\psi = e^{\f/g_s},\qquad {\widetilde \psi}= e^{\tf/g_s}.
$$
The fermions of the two patches are related by Fourier transform
\eqn\ft{
\widetilde\psi(y) = S\psi(y)=
{1 \over \sqrt{ 2 \pi}}\int dx\; e^{-xy/g_s}\; \psi(x).
}
In terms of the action of the $S$ matrix on the second-quantized fermion
Fock space, the partition function can be written as
$$
Z(t,\tt) = \langle t | S | \tt\rangle
$$

The quantum parts of the fermions, defined by $\psi =
e^{\phi_{cl}/g_s} \psi_{qu}$ have mode expansions
$$
\psi_{qu}(x) = \sum_{p \in \Z + {1\over 2}} \psi_p x^{-p-{1\over 2}},
\qquad
\psi^*_{qu}(x) = \sum_{p \in \Z + {1\over 2}} \psi^*_p x^{-p-{1\over 2}},
$$
with canonical anti-commutation relations $\{ \psi_p,\psi^*_q\} =
\delta_{p+q,0}$.

In terms of the fermions, we get quantum analogue of \clw.  Namely the
operator $\oint_P x^{n}\del \phi(x)$ corresponds to the
$\cW_{1+\infty}$ generator that shifts
$$
(x,y) \to (x,y+nx^{n-1})
$$
so we have
\eqn\quw{
\oint_{x\rightarrow \infty}\psi^{*}(x) x^{n} \psi(x) |V\rangle=
\oint_{y\rightarrow\infty} \psi^{*}(y) (g_s \del_y)^n \psi(y)|V\rangle.
}
This Ward identity for the $c=1$ string amplitude at selfdual radius
was derived in \dmp\ in this form, where the starting point was the
well-known statement that in terms of fermions, the $c=1$ string is
free. Namely, the $c=1$ string scattering matrix which relates the
incoming and the outgoing modes of the fermions consists just of a
phase factor
$$
\widetilde\psi_{-n-{1\over 2}}
= R_{n+{1\over 2}}
\psi_{-n-{1\over 2}}.
$$
Here the reflection factor $R_q$, for general momentum $q$, is given
by
$$
R_{q} = i\left({1+ ie^{-\pi(\mu+iq)}\over
1-i e^{-\pi(\mu+iq)}}\right)^{1\over 2}
\left({\G(\hf-i\mu+q)\over \G(\hf+i\m-q)}\right)^{1\over 2}
$$
In fact, we can derive the fermion scattering amplitude $R_q$ directly from
\ft\ as follows.

If we turn on one set of times only, the partition
function is trivial by momentum conservation, and ${\cal
F}(t,0)=0$. Consequently, in the $x$ patch, the fermions are free,
and the one-point functions are exact at the classical level
$$
\Psi(x) = \langle 0| \psi(x)|V\rangle = x^{-\mu},
$$
corresponding to contribution of the classical piece $e^{\phi_{cl}}$,
in $\phi(x)=e^{\phi_{cl}(x)}\;\psi_{qu}(x)$.
Since fermions transform as wave functions,
this further implies that the fermion
scattering state $|V\rangle$ is of general form
$$
|V\rangle = \exp( \sum_{m,n\geq 0} a_{mn} \psi_{-m-\hf} {\tilde
 \psi}^{*}_{-n-\hf} + {\tilde a}_{mn} {\tilde \psi}_{-m-\hf}
 \psi^{*}_{-n-\hf})|0\rangle,
$$
and we will now show that $a_{mn}= R_{n+\hf} \delta_{mn}$, $\tilde
a_{mn}= R^*_{n+\hf} \delta_{mn}$, which
will reproduce the result we quoted.

Consider $\langle 0|\psi(y)\psi^*(x)|V\rangle$. On the one hand, this
equals
$$
(xy)^{-\mu}\sum_{m,n\geq 0} a_{m,n} x^{-m-1} y^{-m-1}.
$$
On the other hand, we can compute this knowing the correlators in the
$x$-patch alone.
For this, note that similarly to what we had in the $x$ patch,
$$
{\tilde \Psi}(y) = \langle 0| \tilde{\psi}(y)|V\rangle = y^{\mu}.
$$
Moreover, since in the $x-$patch, we have
$$
\langle 0|\psi(\tx)\psi^*(x)|V\rangle = {\tx^{-\mu}\; x^{\mu}\over x
-\tx},
$$
then using \ft\ it follows,
$$
\langle 0|{\tilde \psi}(y)\psi^*(x)|V\rangle =
{1 \over \sqrt{ 2 \pi i}}
\int \;d\tx\; e^{i \tx y}\;{\tx^{-\mu}\; x^{\mu}\over x -\tx}.
$$
which equals
$$
\int d\tx\; e^{i{\tx}y}\;{\tx^{-\mu}\; x^{\mu}\over x -\tx}= (xy)^{\mu}\;
\sum_{n\geq 0}\;(xy)^{-n-1} \;{\widehat R}_{n+{1\over 2}}
$$
where
$$
{\widehat R}_{n+\hf}
= {1 \over \sqrt{ 2 \pi i}}
\oint d\tx \;\tx^{-\mu} {\tx}^{n}\; e^{i\tx}.
$$
Similarly, one finds that $\tilde a_{mn}= {\widehat R}_{n+\hf}^*\delta_{mn}$.
It is easy to show that (see appendix A)
$$
{\widehat R}_{n+{1\over 2}} = R_{n+{1\over 2}},
$$
up to a constant multiple,
and correspondingly
$$
|V\rangle = \exp(\sum_{n\geq 0} {\widehat R}_{n+\hf}\psi_{-n-\hf}
 \tilde{\psi}^*_{-n-\hf} + {\widehat R}^*_{n+\hf}{\tilde \psi}_{-n-\hf}
 {\psi}^*_{-n-\hf})|0\rangle.
$$

When the times are turned on, the curve $xy=\mu$ gets deformed.
The precise form of the deformation can be derived from the dispersionless
solution by using that
$xy= y \partial_y \widetilde \phi(y)=-x\partial_x \phi(x)$.
For example, if we turn on
 $t_1$, $\widetilde t_1$, the curve gets deformed to
$(x-\widetilde t_1)(y+ t_1) =\mu$. A detailed description of the
deformed curve using the Toda hierarchy can be found for example in \kostov.
\subsec{Kontsevich-like model for $c=1$}

It has been known that there is a Kontsevich-like model also for $c=1$
string at self-dual radius \refs{\dmp ,\im}.  We will
now rederive this from the perspective of the present paper.
 Let us consider turning on
only the couplings $t_n$, and set ${\tt_n}=0$ momentarily.  Then all
the amplitudes are trivial, because of momentum conservation.  Thus we
obtain a simple deformation of the geometry
$$
xy-\mu=0\ \longrightarrow \ xy+\sum_{n>0}nt_n x^n -\mu=0.
$$
Now consider adding non-compact probes to this geometry frozen at
positions $y=y_i$, as captured by the eigenvalues of a matrix
$\Lambda$. As we discussed before, adding $N$ of these B-branes
corresponds to turning on $N$ independent times in the space of
couplings ${\tt}_n$.  To write the action on these branes we start by
considering $N$ B-branes at $x=x_i$. In that coordinate patch this
$N$-point function is simply given by the Vandermonde determinant
plus the classical contribution.  The classical contribution is given by
$$\int^{x_i} y(x) dx =\int^{x_i} \mu {dx\over x}=\mu {\rm log} x_i. $$
We thus end up with
$$
\langle N|\psi(x_1)\dots\psi(x_N)|V\rangle = \prod_{i<j} (x_i
-x_j)\;e^{\sum_{i=1}^{N}\Bigl\{ - (i{\mu\over g_s}) \log(x_i) + {i\over g_s} \sum_{n>0} t_n x_i^{n}\Bigr\}}
$$
Now we can ``move'' the branes to the $y$-patch.
This means that the $c=1$ amplitude with
both sets of times turned on equals
$$
Z(t_n,\tt_n )=\int \prod_{i=1}^N dx_i \;e^{\,i\sum_i{1\over g_s} x_i y_i}\;
\langle N|\psi(x_1)\dots\psi(x_N)|V\rangle
$$
We thus get a Kontsevich-like matrix model:
$$Z(t_n,\tt_n )=\int DX \; ({\rm det}X)^{-i(\mu /g_s)}\; e^{
\Bigl\{{1\over g_s}\,{\rm Tr}(\,i \Lambda X \,+\, i \sum_n\, t_n X^n\,)\Bigr\}}$$
where $\tt_n ={g_s\over n}{\rm Tr} \Lambda^{-n}$.  This is exactly the
Imbimbo-Mukhi matrix model for $c=1$ string \im , up to identification
of parameters.  Namely, recall that
the $c=1$ string amplitudes depend only on the ratio ${\hat \mu}$ of
$\mu$ and $g_s$, ${\hat \mu} = {\mu\over g_s}$ -- this can be seen
above, by rescaling appropriately the times $t_n, {\tt}_n$ and
${\Lambda}$ by factors of $\mu$.  After doing so
and a suitable shift ${\hat \mu}={\mu\over g_s}-iN$
 the agreement with
\im\ is manifest.
The fact that these agree up to a shift in $\mu$ by $g_N$ is
quite natural. This is because fermions shift the flux
of the three-form according to \bacr , and correspondingly
the value of $\mu$ depends on where one measures it.

\subsec{General genus zero surfaces}

We can easily generalize the above formalism to more general Riemann
surfaces, with many punctures. Consider for example the genus zero
Riemann surface
$$
y = {\prod_{i=1}^{N_+} (t_i-x)\over \prod_{j=1}^{N_-}
({s_i}- x)}
$$
We can assume, without loss of generality, that $N_+> N_-$.
There are $N_-+1$ punctures corresponding to
$$
{\eqalign{
&P:\quad x\rightarrow \infty,\quad y \rightarrow
(-1)^{N_+ +N_-} \infty \cr
&S_j: \quad \; x\rightarrow s_j, \qquad y\rightarrow \infty.\cr
}}
$$
At the $S_j$ punctures we use $y$ as coordinates, and at $P$, we can use
$x$, for example.

The Ward identities, corresponding to putting branes at fixed $x$, read
$$
\int_{P} \psi^{*}(x) x^{n} \psi(x)
+ \sum_{j=1}^{N_-}\int_{S_j}\psi^{*}(y) (g_s \del_y)^n \psi(y)=0.
$$
One has, in more detail, around the $S_j$ puncture
$$
\int_{S_j}\psi_{qu}^{*}(y)e^{-\phi_{cl,j}(y)/g_s} (g_s
\del_y)^n e^{\phi_{cl,j}(y)/g_s} \psi_{qu}(y)
$$
where
$$
x =x_j(y)= \del_y \psi_{cl,j}(y) = s_j + O(y^{-1})
$$
is the corresponding classical solution near the $j$-th puncture.
Correspondingly, it is clear that, when no B-branes are placed
at $S_j$ punctures, the classical solution for the B-brane at the
$P$-puncture is exact. Namely, the $S_j$ punctures contribute
identically zero to the Ward identity, whereas the contribution of the
$P$-puncture reads
$$
\del^{P}_n {\cal F}=0
$$
Correspondingly, the free energy is completely trivial in this case,
and the B-brane action is given by its classical value
$$
\Psi(x) = \langle 1|\psi(x)|V\rangle = e^{S(x)/g_s}
$$
where
\eqn\sint{
S(x) = \int^{x} y(x) dx.
}
In the same case, note
that the B-brane is trivially annihilated by the Riemann surface Hamiltonian:
$$
H(x,y=g_s \del_x)\Psi(x)=0.
$$
{}From this we get a Kontsevich-like matrix model describing
general non-normalizable deformations of the Riemann surface
corresponding
to placing B-branes at various punctures.
This is given by
$$
\int dX_1 \ldots dX_{N_-} \prod_{i<j}{\rm det}(X_i\otimes 1_j - 1_i \otimes
X_j)\; e^{\sum_{1=1}^{N_-}[- \Tr \,\Lambda_i X_i +\Tr\,W(X_i)]/g_s}
$$
Where
$$
W(x)=S(x)+\sum_{n} t_n x^n
$$
and $S(x)$ is given by \sint.

\subsec{Example 3: Topological string on ${\bf P}^1$}

As we discussed before, if we consider the topological B-model
coupled to a chiral field with superpotential $W(p)$ then
this is equivalent to considering topological strings
on the Calabi-Yau threefold
$$zw-x-W(p)=0.$$
We can use this idea to solve the topological A-model string
on ${\bf P}^1$: It is known \HV\ that mirror symmetry relates
the topological A-model on ${\bf P}^1$ to a LG theory with
superpotential
$$W(u)=e^u+ q e^{-u},$$
where $q=e^{-t}$ where $t$ is the K\"ahler parameter of ${\bf P}^1$ and
$p=e^u \in {\bf C}^*$. We are thus led to consider topological strings
on a Calabi-Yau threefold
$$zw-x-e^u-q e^{-u}=0$$
and corresponding to this a chiral boson on the Riemann surface
$H(x,u)=0$ with
\eqn\hp{H(x,u)=x+e^u+q e^{-u}.}
Deforming the A-model theory by gravitational descendants of the K\"ahler
class and the identity operator should correspond to inserting
appropriate D-branes in this geometry.

To start with, consider insertion of a D-brane at fixed values of $x$
and the associated one-point function $\Psi(x)=\langle
\psi(x)\rangle$. As we discussed above, as operators acting on the
D-brane, $x$ and $u$ are canonically conjugate $[x,u]=g_s$, where we
can take $ u=-g_s \del/\del x$. Moreover the wave function $\Psi(x)$ is
annihilated by $H= H(x,u)$:
$$H\Psi(x)=0,$$
with generalized Schr\"odinger operator
\eqn\eg{H=e^{g_s \del_x} + x + q e^{-g_s \del_x}}
(equivalently, by shifting $x$ by a constant $z$, $\Psi(x)$ is an
eigenfunction of the Lax operator $H$ with eigenvalue $z$).
According to our philosophy the Hamiltonian $H$ is the Lax operator of
an integrable hierarchy, and the corresponding flows are related to
inserting many D-branes at fixed values of $x$.  As it turns out, this
can be identified with turning on gravitational descendants
$\sigma_n(\omega)$ of the K\"ahler class $\omega$ of ${\bf P}^1$.

It has been known for a while that the ${\bf P}^1$ theory is described
by a Toda lattice hierarchy \ehy.  Moreover, the
Lax operator $L$ of the hierarchy \ehy, in the small phase space,
coincides with $H$ in \eg .  The flows of this hierarchy generated by
$(L^n)_+$, $n>0$ correspond to deforming by the descendants of K\"ahler
class: the coupling constants $t_{n,\omega}$ of $\sigma_n(\omega)$ are
related to Toda times $t_n$ by $t_n = {1\over n}t_{n-1,\omega}$
\ehy. (The flows corresponding to turning on descendants of the
identity operator are not the usual Toda flows, but a generalization
thereof \ehy.)  We thus have shown that inserting D-branes at
$x=x_{i}$, $i=1,\ldots,N$ corresponds to turning on
$$
t_{n-1, \omega} = n \sum_i x_i^{-n}.
$$
Alternatively, by thinking about the ${\bf P}^1$ in bosonic terms, we
can make contact with another description of the ${\bf P}^1$ theory
presented in \op.

Note that the Riemann surface \hp\ has two asymptotic infinities
corresponding to $x\rightarrow \infty$ and either
$u\rightarrow \infty$ or $\widetilde u =
-t-u \rightarrow \infty .$
Correspondingly we have, in principle, two chiral bosons $\phi(u)$ and
$\widetilde \phi(\widetilde u)$ which correspond to deforming $x$ from
the $u$ patch and the ${\widetilde u}$ patch as
$x=\del \phi(u)= \del\phi_{cl}(u) +\del\phi_{qu}(u)$ and $x=\del{\tf}
({\widetilde u})=\del{\tf}_{cl}(\widetilde u)+
\del{\tf}_{qu}(\widetilde u)$, respectively where
$$
\del{\phi}_{cl}(u)=e^{u}+ q e^{- u}
$$
and
$$
\del{\tf}_{cl}(\widetilde u)=q e^{-\widetilde u}+ e^{\widetilde u}.
$$
The bosons have mode expansions
$$\del \phi_{qu}(u) =\sum_{n>0} s_n e^{nu}+n {\del\over \del s_n} e^{-n u}$$
and
$$\del{\tf}_{qu}(\widetilde u)=\sum_{n>0} {\widetilde s}_n
e^{n{\widetilde u}}+ n{\del\over \del s_n} e^{-n {\widetilde u}}.$$

The action of the $\cW_{1+\infty}$ algebra generators fixes the
partition function $Z$, since it relates the modes on the two
asymptotic ends. Consider for example the generator corresponding to
$e^{n u} x^m$ for $m,n>0$. That this generates a symmetry implies
that, acting on $Z=e^{\cal F}$
$$
\int_{u\rightarrow \infty}  e^{n u} e^{\phi(u)}\del^m e^{-\phi(u)}
=  \int_{\widetilde u \rightarrow\infty}q^n e^{-n\widetilde u}
e^{{\tf}({\tilde u})}\del^m e^{-{\tf}({\widetilde u})}
$$
so that one finds
%
$$n \; \del_n {\cal F} = q^{n} \; {\widetilde s}_n.$$
More generally, the $\cW_{1+\infty}$ symmetry and the above Ward
identities imply that
$$\del \phi (-t-{\widetilde u})=\del {\tf}(\widetilde u)$$
holds as an operator equation, {\it i.e.}\ that the theory is that of
a {\it globally} defined free chiral scalar on a cylinder of length
$t$. The free energy is therefore quadratic
$${\cal F}(s,\widetilde s, q,g_s) = \sum_{n>0} {q^{n}\over n} s_n
{\widetilde s}_n,$$
and the corresponding partition function is given by
$$Z = \langle s | q^{L_0} |{\widetilde s}\rangle$$
However, as we discussed above, the times $s_n$, $\widetilde{s}_n$
bear {\it no direct relation} to turning on the descendants of the
K\"ahler class, since the modes of $\phi(u)$ correspond to insertions of
B-branes at fixed $u$, whereas the descendants correspond to inserting
B-branes at fixed $x$.

We can remedy this easily in the following way.  It suffices to
consider placing D-branes near only one of the two punctures. Let $W$
be the operator that corresponds to the transformation
$(x,u)\rightarrow (-u,x)$. The partition function corresponding to
${\bf P}^1$ with arbitrary descendants is now given by
$$Z = \langle s|W^{-1}q^{L_0}W^{*}|{\widetilde s}\rangle$$
By the definition of $W$ we have
$$W \cdot \exp \Bigl[\,\sum_{k>0} s_k \alpha_k\,\Bigr] \cdot W^{-1} =
\exp\left[\sum_{k>0} {t_k \over (k+1)!} {\hat W}_0^{k+2}\right],$$
since $\alpha_k$ generates the transformation $(x,u)
\rightarrow (x+e^{ku},u)$, and ${\hat W}_0^{k+2}$ generates
$(x,u)\rightarrow(x,u+x^k)$.

More precisely, ${\hat W}$ generates the corresponding transformation,
in the {\it background} corresponding to the Riemann surface,
so that it is the dressed operator,
$$
{\hat W}_0^{k+1} = \int\;du
\;\psi_{qu}^{*}(u)\,e^{-\phi_{cl}(u)}\,\del^{k}\,e^{\phi_{cl}(u)}\,\psi_{qu}(u)
$$
or equivalently,
$$
{\hat W}_0^{k+1} =e^{-q \alpha_{-1}}\;e^{\alpha_1}\; W_{0}^{k+1}\;
e^{-\alpha_1}\; e^{q \alpha_{-1}}
$$
where $W_{0}^{k+1}= \int\;du
\;\psi_{qu}^{*}(u)\;\del^{k}\;\psi_{qu}(u)$ is the standard
generator.

The operator ${\hat W}^{*}$ is the corresponding operator coming from
the ${\tu}$ puncture which gives rise to
$$
{{{\hat W}}_{0}}^{*,k+1} =e^{q \alpha_{1}}\;e^{- \alpha_{-1}} \;W_{0}^{*,k+1}
\;e^{\alpha_{-1}}\; e^{-q \alpha_{1}}
$$
where
$$
W^{*} \cdot \exp \Bigl[\,\sum_{k>0} {\ts}_k \alpha_{-k}\,\Bigr]
\cdot W^{*-1} =
\exp\left[ \sum_{k>0} {{\tt}_k \over (k+1)!} {{\hat W}}_0^{*,k+1}\right].
$$

Correspondingly, the partition function of ${\bf P}^1$, deformed by the
descendants of the K\"ahler class is, up to a constant factor given by
$$Z = \langle 0| e^{\alpha_1}\; e^{\sum_{k>0} {t_k\over (k+1)!}
W_0^{k+1}}\; q^{L _0}\;e^{\sum_{k>0} {{\tt}_k\over (k+1)!}
W_0^{k+1}}\;e^{\alpha_{-1} }|0\rangle,$$
where we have used that
$$
e^{-\alpha_1}e^{q \alpha_{-1}}q^{L_0}e^{q
\alpha_1} e^{-\alpha_{-1}} = q^{L_0}e^{q}.
$$
Since $L_0$ and $W_0^{k+1}$ commute, there is effectively only one set
of times, {\it i.e.}\ without loss of generality, we can set
${\widetilde t}_k=0$, so we can write
\eqn\parp{Z_{\bf P}^1 =
\langle 0| e^{\alpha_1}\; e^{\sum_{k>0} {t_k\over (k+1)!} W_0^{k+1}}\;
q^{L _0}\;e^{\alpha_{-1} }|0\rangle,}
Note that the partition function is identical to the partition
function for the ${\bf P}^1$ obtained in \op, eqn. 3.13, which is a
nice verification of our
formalism.

We can also conjecture a matrix model that describes this.  Very
analogous to the Kontsevich matrix model of the $(p,1)$ bosonic
minimal models, we can give a simple matrix model description for the
${\bf P}^1$ A-model topological string by making use of the simplicity
of the D-brane amplitudes in the $u$-patch.  There, $x=g_s \del/\del
u$ and the corresponding one-point function $\Psi(u)$ satisfies
$H\Psi(u)=0$ with
$$
H = g_s {\del\over \del u} -e^{u} - qe^{-u}.
$$
Since the Hamiltonian is linear in the momentum, the semi-classical
approximation is {\it exact}, and
$$
\Psi(u) = e^{(e^{u}-q e^{-u})/g_s}.
$$
The matrix model action corresponding to inserting $N$ D-branes at
fixed positions $x_i$ (now considered as the eigenvalues of a matrix
$X$) is
$$
Z_{{\bf P}^1}(X) = \int d_H U \; \exp\Tr\left(XU - e^U-q e^{-U}\right)/g_s]$$
where $U$ is a Hermitian matrix and $d_H U$ is the measure induced
from the Haar measure on the $U(N)$ group manifold on its tangent
space (respecting the periodicity $U \to U + 2\pi i$). Here the traces
${1\over n} \Tr\,X^{-n}=t_n$ correspond to the Toda times.  We
conjecture that this matrix model is describing topological A-model of
the ${\bf P}^1$
when the couplings to the descendants of the
identity operator are turned off.
This should also be equivalent to the matrix model of \ehy .
It is an interesting question to
understand what configuration of B-branes corresponds to turning on
these more general couplings as well.

One can also deform the theory by turning on a twisted mass $\tau$ on
${\bf P}^1$ which gives an equivariant version of the topological
string on ${\bf P}^1$. In this case the superpotential gets deformed
to \HV
$$
W(u)=e^u+ q e^{-u}+\tau u,
$$
and the corresponding Calabi-Yau becomes
$$
zw-x-e^u-q e^{-u}+\tau u=0.
$$
The considerations here should proceed much in the same way as in the
non-equivariant case, and it would be nice to work out the details.

\subsec{Example 4: Integrable structure of the topological vertex}

The topological vertex was introduced in \akmvII\ as the building
block for topological string amplitudes on local, toric Calabi-Yau
threefolds (for further properties and applications of the
topological vertex, see for example
\refs{\ik ,\df ,\orv ,\ek ,\hiv ,\zhouII}).
In terms of the topological A-model, the vertex can be
regarded as an open string amplitude in the ${\bf C}^3$ geometry with
boundaries on certain Lagrangian A-branes.
In \akmvII\ the vertex amplitude is described by the
partition function
\eqn\totalz{
Z(\Lambda_u, \Lambda_v, \Lambda_w)=\sum_{R_i} C_{R_1 R_2 R_3} {\rm Tr}_{R_1} \Lambda_u  {\rm
Tr}_{R_2} \Lambda_v {\rm Tr}_{R_3}\Lambda_w,}
where we are choosing the canonical framing. The entries $C_{R_1 R_2 R_3}$
were obtained
in \akmvII\ from large $N$ duality with Chern-Simons theory, and the result
is
\eqn\csvertex{
C_{R_1 R_2 R_3}= \sum_{R,Q_1,Q_3}N_{Q_1 R}^{R_1
}\; N_{Q_3^t R}^{R_3^{t}}\;
q^{\kappa_{R_2}/2+\kappa_{R_3}/2}\;
{W_{R_2^t Q_1}W_{R_2 Q_3^t}\over W_{R_2}},}
where $W_{R_1 R_2}$ is the large $N$ limit of the CS invariant of the Hopf
link, as
defined in \refs{\amv,\akmvII}, and $W_R=W_{R \cdot}$, where $\cdot$ denotes the
trivial representation,
is the large $N$ limit of the quantum dimension of $R$. In the above
equation, $N_{R_1 R_2}^{R_3}$ are
tensor product coefficients, and $\kappa_R$ is
given by
\eqn\framfac{
\kappa_R= \sum_i \ell_i (\ell_i-2i+1),
}
where $\ell_i$ are the lengths of rows in the Young tableau associated to
$R$. We can consider the
topological vertex in an arbitrary framing (we will discuss framing
from the B-model perspective in the next subsection),
\eqn\carfr{
C^{(n_1, n_2, n_3)}_{R_1 R_2 R_3}=(-1)^{\sum_i n_i \ell(R_i)} q^{\sum_i
n_i \kappa_{R_i}/2}
C_{R_1 R_2 R_3},}
where $\ell(R)$ is the total number of boxes in the representation $R$.
We note that
$\kappa_R$, that implements the change of framing, is the eigenvalue of
the operator $\oint (\partial \phi)^3/3$
acting on the state $| R \rangle$, as it follows from our discussion of
framing in section 3.1. The free energy associated to the vertex is given
by ${\cal F}(g_s, t^i_n)=\log \, Z(g_s, t_n^i),$ where the parameters $t^i_n$ are
related to the three matrices $\Lambda_i$
appearing in \totalz\ by $t^i_n={g_s \over n}{\rm Tr}\, \Lambda_i^n$, $i=u,v,w$. By using
the representation \chiralbosonmodes, it is easy to see that $\oint (\partial \phi)^3/3$
is the ``cut and join" operator considered for example in \llz, and the free energy
associated to the framed vertex satisfies then three ``cut and join" equations (one
for each framing $n_i$).

We want to describe now the topological vertex in the topological
B-model. We first consider the canonical framing $n_i=0$. Mirror symmetry relates
the A-model ${\bf C}^3$ geometry to Kodaira-Spencer theory on the Calabi-Yau
manifold
$$
zw- e^{-u}-e^{v}+1=0 .
$$
As explained in \akmvII, this is described by a chiral boson on the
Riemann surface
\eqn\rv{
e^{-u}+e^{v}-1 =0.
}
In \akmvII\ some evidence for this was presented.  Here we will show
how to recover in the B-model the full free energy of the topological
vertex, following the general philosophy developed in the previous
sections.  Following the discussion of sections 3 and 4 we will show
that we can completely recover the topological A-model amplitudes by
considering the chiral boson on the above Riemann surface. Moreover,
the fermionic perspective provides a remarkably simple description of
the vertex in terms of a wave function quadratic in the fermions. We
will first develop both the bosonic and the fermionic viewpoint and
then we will demonstrate that this indeed agrees with the A-model
amplitudes derived in \akmvII.

The mirror Riemann surface \rv\ is a sphere with three punctures, {\it
i.e.} three asymptotic regions. We will choose the three punctures
have coordinates $u,v$ and $w$ on them, where
$$
u+v+w = i\pi.
$$
The $u$ coordinate parameterizes the asymptote of the Riemann surface
where $u\rightarrow \infty$, and likewise for the $v$- and
$w-$punctures.  These coordinates are defined on a cylinder and have
periodicity $2\pi i$.  With this choice of variables, the Riemann
surface has a ${\bf Z}_3$ symmetry corresponding to cyclic
permutations
$$ u\rightarrow v \rightarrow w \rightarrow u.$$

There are three corresponding chiral scalars deforming the complex
structure. In the $u$-patch, the meromorphic one form is $\lambda=v
du$, and the chiral boson representing the variations of this can be
identified as $v= \del_u \phi^u$. Correspondingly, in terms of the
fermion quantum mechanics,
$$
[v,u] = g_s
$$
with
$$
v = g_s \del_u.
$$
We should emphasize here that in the case of Riemann surfaces which
correspond to mirrors of topological A-model there is always a {\it
preferred} choice of which variable corresponds to momentum $p$ at
each puncture \AV. This is because the classical
value of $\phi(u)$ corresponds to the disk amplitude in the
topological A-model. The physical part of this amplitude is
generated by worldsheet instantons of complexified area $u$ whose
contribution vanishes in the $u\rightarrow \infty$ limit as $e^{-u}$.  This means that
near the $u$-puncture, the classical B-brane action is
$$S(u) = \phi(u)/g_s = {1\over g_s}\int^u p(u) d u.$$
where $p(u)$ is a variable {\it defined} by asking that on the Riemann
surface $p$ vanishes as $u\rightarrow \infty$. In the present case, we
have that in the $u$-patch, the corresponding momentum is
$p_u=\del_u \f^u =v$. Similarly in the $v$
and $w$ patches, we have
$$
w= p_v=\del_v \phi^v, \qquad u=p_w=\del_w \phi^w,
$$
so that, for example
\eqn\relc{u = i \pi - v - g_s \del_v = g_s \del_w,}

Let us now discuss the $\cW_{1+\infty}$ relations.  The chiral boson
defined in the $u$ patch has an expansion
\eqn\ph{\phi(u) = \sum_{n>0} t^u_n \; e^{n u} -
g_s^2{1\over n}{\partial\over \partial t^u_n}\;\; e^{-n u},}
corresponding to turning on a coherent state, and we have a
corresponding fermion%
$$\psi(u) = e^{\phi(u)/g_s}.$$ Consider now the fermion bilinear operator
\eqn\uptwo{\oint_u \psi^{*}(u)e^{n u} \psi(u).}
Moving the contour to the other two patches, and using that
the operator $u$ is represented in the $v-$ and $w-$patches by \relc\
we find the Ward identity
\eqn\twoqb{\oint_u \psi^{*}(u)e^{n u} \psi(u)
+ (-1)^n \oint_v \psi^{*}(v) e^{-n
v - n g_s \partial_v}\psi(v) +\oint_w \psi^{*}(w)e^{n g_s
\partial_w} \psi(w)=0.}
There is in fact
a triple infinity of conserved charges obtained from the above by
the cyclic  ${\bf Z}_3$ symmetry of the problem.
Bosonization of \twoqb\ gives
\eqn\twoqa{\eqalign{\oint_{u} e^{nu}\; \del_u \phi(u)/g_s \;du
= \oint_v {(-1)^n \over [n]}\; e^{-nv - n \Delta_v \phi(v) }\; dv - \oint_w {1
\over [n]} \; e^{n \Delta_w \phi(w)}\; d w}.}
where we have defined, e.g.:
$$n\Delta_v \phi(v): = {1\over g_s}(\phi(v+ ng_s/2) - \phi(v-n
g_s/2)),$$
and introduced the
$q$-number\foot{The appearance of
$1/[n]$ in the r.h.s. of \twoqa\ comes from the normal-ordering factor
which is given by $\exp (\sum_{k>0} [nk]^2/ (2
k))$. This sum contains a finite piece
$$
\exp\Bigl( {1\over 2} \sum_{k>0} {q^{nk} + q^{-nk}\over k} \Bigr)=-{1\over [n]}
$$
which should be included after normal ordering.}
$$
[n]=q^{n\over 2} - q^{-{n\over 2}},
$$
with $q=e^{g_s}$.  These Ward identities, viewed as differential equations
for the free energy ${\cal F}(t^u,t^v,t^w)$,
are strong enough to determine it uniquely.

Let us first discuss the classical limit, $g_s \rightarrow 0$.
In this limit we have that
$$e^{\Delta_u\phi(u)/g_s} \rightarrow e^{\del_u \phi(u)}$$
and only the genus zero piece $F_{0}$ of the free energy ${\cal F}(t)
=\sum_g F_g(t) g_s^{2g -2}$ survives, e.g.
$$
\langle t|e^{\Delta^u \phi^u}|V\rangle_{g_s\rightarrow 0}=\exp{\Bigl\{\sum_{n>0} n t^u_n  \;e^{n u} +
\partial^u_n F_{0}\; e^{-n u} \Bigr\}}.
$$
so that the Ward
identity becomes
\eqn\two{  \oint_{u} e^{nu}\; \del_u \phi^u \;du
= \oint_v {(-1)^n \over n}\; e^{-nv - n \del_v \phi^v }\; dv - \oint_w {1
\over n} \; e^{n \del \phi^w(w)}\; d w .}
The semiclassical limit could have equivalently been obtained by
recalling that, e.g.  $\del^u \phi(u)=v$
and then considering
$$ \oint_{u\rightarrow \infty} e^{n u}\; v\;du,$$
moving the contour to the other patches and integrating by parts.
This classical Ward identity
determines the genus zero free energy.
For example, the leading order solution for $F_{0}$, corresponding
to genus zero with one hole is obtained by setting all the times to
zero to get
\eqn\class{ \del^u_n F_{0} =- {1\over n }.}
Note that this is computing
$$
v = \del_u \phi_{cl}^u = - \sum_n {1\over n} e^{-nu} = \log(1-e^{-u}),
$$
which corresponds to the classical equation of the Riemann surface.
The case of Riemann surfaces which correspond to mirrors of the
toric topological A-model geometries is special, namely
the knowledge of the location of the
punctures in terms of giving the momenta and the
coordinates, suffices to fix the Riemann surface itself. It is thus
not surprising that
the same data fixes the quantum one-point amplitudes as well, as we
will soon see. Expanding about the one-point function, $\phi_{cl}$
it is easy to see that the two point functions precisely correspond to
a global chiral scalar on the Riemann surface \rv\ in accordance with
\akmvII.
The full theory is much more complicated than that, even in the
classical limit, and $F_0$ in general has non-zero coefficients
for all powers of the $t$'s.

We can now use the quantum Ward identity \twoqa\ to solve for the full
free energy. A priori, we should solve the identity
iteratively order by order in the genus expansion.
One can however do better than
this and obtain {\it exact} expressions to all genera, proceeding
iteratively in the number of times turned on. Let us now explain this.

The equation \twoqa, after acting on $Z$, leads to the following identity:
\eqn\fullq{
\eqalign{
g_s  \partial^u_n {\cal F} \;=& {(-1)^n \over [n]}\oint_v \; dv\; e^{-nv}
e^{\Bigl\{\sum_{m>0} -g_s^{-1}[m n] t^v_m e^{m v} -
\Delta_{v,n}{\cal F}(t)\Bigr\}}\cr
&-{1 \over [n]} \oint_w\; dw\; e^{\Bigl\{\sum_{m>0} g_s^{-1}[m n] t^w_m e^{m w} +
\Delta_{w,n}{\cal F}(t)\Bigr\}}
,}
}
where we defined
$$\eqalign{\Delta_{v,n}{\cal F}(t):=&{\cal F}(t^u,\,t^v_m -g_s{[m n] \over m} e^{-m
v},\,t^w)-{\cal F}(t^u,t^v,t^w)\cr
=&\sum_{\ell=1}^{\infty}
\sum_{q_1, \cdots, q_{\ell}}{(-g_s)^{\ell} \over \ell!} \;
\prod_{i=1}^{\ell}{[n q_i]\over q_i} \;\;
\partial^v_{q_1}\cdots \partial^v_{q_{\ell}}{\cal F}(t^u,t^v,t^w)\; e^{-\sum_{i=1}^{\ell} q_i v},}
$$
{\it i.e.} we expand the free energy ${\cal F}(t^u,t^v_m -g_s [m n]e^{-mv}/m,t^w)$
on the r.h.s. in the times around ${\cal F}(t^{u},t^v,t^{w})$,
and similarly for the term in the $w$ integral.
Using \fullq\ and the cyclic property of the vertex it is
straightforward to provide an algorithm to
calculate the exact coefficients in $q$ of
the monomial ${t}^u_{\vec k }\;
{t}^v_{\vec m}\; {t}^w_{\vec n}$ in the free
energy ${\cal F}$ to any given order in the times and total
``winding''.

For example, it is easy to see that the exact amplitude corresponding to
times on the $u-$puncture is given by
\eqn\back{{\cal F}(t^u;t^v=0=t^w) =-\sum_{n=1}^\infty {t_n^u\over  g_s[n]}}
as follows by integrating \fullq\ with respect to $t^u_n$. For $t^v=t^w=0$
the only term that contributes from the righthand side is
the constant $-{1\over [n]}$ coming from the second term.

Similarly the coefficients of
$t^u_{k_1}\ldots t^u_{k_s} \;t^{v}_{m_1} \ldots t^{v}_{m_n}$ and
$t^u_{k_1}\ldots t^u_{k_s}\; {t}^w_{n_1}\ldots t^{w}_{n_r}$
are obtained by integrating   \fullq\ with respect to $t_{k_i}^u$
and iteration over $s$. Each step can be performed
to arbitrary order in the total number of holes
$N_{v,w}=\sum_{i=1}^r k^{v,w}_i$. Explicit
formulas for certain amplitudes  to arbitrary
winding can be obtained, see appendix A. Checks on
the cyclic property are provided by comparison of
the coefficients of
${t}^u_{\vec k} \;{t}^w_{\vec n} \;
{ \longrightarrow \atop {\rm cycl.}}\; {t}^w_{\vec k}\;
{t}^v_{\vec n}$ with the one of
${t}^u_{\vec k}\; {t}^v_{\vec n}\; {\longrightarrow \atop {\rm cycl.}}\;
{t}^v_{\vec k}\; {t}^w_{\vec n}$.
These coefficients can be used to obtain
$t^u_{k_1} \ldots t^u_{k_s}\; {t}^v_{m_1}\ldots t^v_{m_n}\;
{t}^w_{n_1}\ldots t^w_{n_r}$
again by integration (5.23) with respect to $t^u_{k_i}$ and
iteration in $s$. This has been done to order
$(N_u,N_v,N_w)=(4,4,4)$ in complete agreement with the
free energy from the Chern-Simons formula for the vertex \csvertex.

The underlying integrable structure of the topological vertex is better
formulated in the fermionic language, and it implies that albeit the
vertex has a rather complicated structure when written in terms of modes
of a chiral boson, after fermionization it is simply given by the exponential
of a bilinear of fermions acting on the vacuum. We turn to this next.

\subsec{Fermion perspective for the topological vertex}

Let us now consider the fermionic description.
{}From the above, it follows that the fermion one point function
$$
\Psi(u) = \langle 0|e^{\phi(u)/g_s}|V\rangle
$$
is given by the so-called quantum dilogarithm \faddeev\
\eqn\fv{
\Psi(u) =
\exp \sum_{n>0}- {e^{-nu}\over n [n]}.
}
Other useful ways of writing this are
$$
\exp \sum_{n>0}-{e^{-nu}\over n [n]}
= \prod_{n=0}^{\infty}
(1-e^{-u}q^{n+{1\over 2}})^{-1}= \sum_{n\ge 0} (-1)^n {1\over [n]!}
q^{-n(n-1)/4} e^{-n u}.
$$
The one point function is annihilated by the Hamiltonian corresponding
to the Riemann surface \rv , leading to a difference equation
which uniquely fixes it
$$
H(u,v) \Psi(u) =0
$$
where
$$
H(u,v) = q^{-{1\over 2}} e^{-u} +e^{g_s \del_u} - 1.
$$
This reflects, as discussed in section 3, the unbroken $\cW$
symmetries. For the anti-brane a similar reasoning gives
$\prod_{n=0}^{\infty} (1-e^{-u}q^{n+{1\over 2}})$. As we discussed
in section 4, the B-branes transform as wave functions in going
from patch to patch.  With coordinates on the punctures chosen as
in the previous subsection, going from the $u$ to the $v$ patch
the coordinates transform by a symplectic transformation $U$
which, up to a shift, corresponds to the $ST$ element of the
$SL(2,\Z)$.
$$
U \pmatrix{v\cr u} = ST\pmatrix{v\cr u} +\pmatrix{0\cr i\pi}
$$
$$
ST = \pmatrix{0 &  1\cr
              -1 &  -1 \cr}
$$
The element $ST$ generates a ${\bf Z}_3$ subgroup of $SL(2,\bf Z)$,
$(ST)^3=1$, and similarly $U^3={\rm id}$.

In terms of the fermions, these are realized as
$$U\psi(v) = \int du e^{-S_U(v,u)/g_s} \psi(u)$$
$$U^2\psi(w) = \int du e^{-S_{U^2}(w,u)/g_s} \psi(u)$$
where
$$
S_{U}(v,u) = -{v^{2} + 2 uv + 2 i\pi v\over 2}.
$$
Similarly from $w$ to $u$ patches there is a generating function
$$
S_{U}(u,w) = -{u^{2} + 2wu + 2 i\pi u\over 2}.
$$
$$
S_{U^2}(w,u) = {u^{2} + 2 uw + 2 i\pi u\over 2}.
$$
Note that $S_{U^2}(w,u)=-S_{U}(u,w)$, and correspondingly,
it is easy to check that $U^3$ is realized as identity on the fermions.

In this example it is easy to check that the one-point functions $\Psi(u) = \langle
\psi(u)\rangle$
satisfy
\eqn\ff{
\Psi(w)= \int du e^{-S_{U^2}(w,u)/g_s} \Psi(u+g_s/2),
}
by performing a Gaussian integral over $u$, and analytically
continuing in $w$, to $w\rightarrow \infty$ patch. That is, up to
the shift in $u$ we have $\Psi(w) = U^2\Psi(w)$. The shift in $u$
on the right hand side is a quantum shift in the flat coordinate.
While in the case of pure topological gravity, for example, one
could derive very useful statements by considering how fermions
transform from patch to patch, here the one point functions are
complicated in every patch. Incidentally, the statement that the
quantum dialogarithm is its own Fourier transform is well known
\faddeev.

The expression or $|V\rangle$ derived in the previous subsection in
terms of the bosons is hiding the amazing simplicity of the state
$|V\rangle$ --- it is a Bogoliubov transform of the fermionic vacuum!
Namely, we have that
\eqn\vv{
|V \rangle = \exp \left[ \sum_{i,j,n,m} a_{mn}^{ij}
\psi^i_{-m-1/2}
\psi^{j*}_{-n-1/2} \right]|0 \rangle
}
where, due to the cyclicity of the vertex, the coefficients
$a_{mn}^{ij}$ must have the cyclic symmetry
$$
a_{mn}^{ij}=a_{mn}^{i+1,j+1}.
$$
so that there are only three independent matrices of coefficients,
$a^{uu}$, $a^{uv}$ and $a^{vu}$.  That this is true can be shown
as follows. Note that if we consider B-branes in one patch only,
than the corresponding amplitude is manifestly bilinear in the
fermions. This is because the corresponding state in terms of the
bosons is simply a coherent state, which is the exponential of a
bilinear in fermions acting on the vacuum.  Since the fermions
transform linearly from patch to patch, this implies that the
complete state $|V\rangle$ is bilinear in the fermions. Moreover,
the coefficients $a^{ij}$ are easily found if one knows the
amplitudes in one patch.

Since the exact free energy \back\ corresponding to
one puncture ${\cal F}(t^{u},0,0)$ is linear in the times $t^u_n$,
we have that $G(\tu, u)= {1 \over e^{\tu} - e^{u}}$
is the standard fermion Green function on a plane:
\eqn\vc{
\langle 0|\psi(\tu ) \psi^{*}(\tu)|V\rangle = {1 \over e^{\tu} - e^{u}}
\;\Psi(\tu)\;\Psi^{*}(u)={1 \over e^{\tu} - e^{u}} \prod_{n=0}^{\infty} \Biggl(
{1 - e^{-u} q^{n+1/2} \over 1 - e^{-\tu} q^{n+1/2}}\Biggr)
}
This can be used to compute the state $|V \rangle$, as described
in section 4, namely from \vc\ we can compute $a^{uu}$ as:
$$
{\langle 0|\psi(\tu)\psi^{*}(u)
|V\rangle =\sum_{m,n=0}^\infty
a_{mn}^{uu} e^{-(m+1)u}\;e^{-(n+1)\tu} +{1\over e^{\tu} - e^{u}}}.
$$
{}From this we find that:
$$
a^{uu}_{mn}=(-1)^{n}{q^{{m (m+1) \over 4} -{n (n+1) \over 4}} \over
[m+n+1]\; [m]!\; [n]!}.
$$

We can similarly compute $a_{mn}^{vu}$ by evaluating $\langle 0|
\psi(v) \psi^{*}(u)|V\rangle$ and considering how the fermions
transform, namely
\eqn\utv{
\langle 0| \psi(v) \psi^{*}(u)|V\rangle =
\int\; d\tu \;e^{-S_U(v,\tu)/g_s}\;{\Psi(\tu + g_s/2)\;\Psi^{*}(u)\over e^{\tu
+ g_s/2} - e^{u}}
}
and expanding in $e^{-u}$ and $e^{-v}$.
The Gaussian integral in \utv\ gives\foot{For computing this,
it is useful to note that
$$
\int \;du \;e^{-S_{U}(v,u)/g_s}\;e^{nu} \;\Psi(u+g_s/2) = \Psi(v+n g_s)
\;(-1)^n \;e^{-nv} \;q^{-{n^2\over 2}}.
$$
To define this integral and consistently keep track of the quantum
shifts, it is helpful to express it as $\int \;du \int dw
\;e^{-S_{U^2}(v,w)/g_s-S_{U^2}(w,u)/g_s} \;e^{nu} \;\Psi(u+g_s/2)$
using that $U = U^2 \cdot U^2$, and further making use of equation
\ff .}

\eqn\corser{
\langle 0| \psi(v) \psi^{*}(u)|V\rangle=\;e^{-u}\;\sum_{n\geq 0}
\;q^{-{n(n-1)\over 2}}\; (-1)^{n+1}\; e^{-n v} \;e^{-nu}
\;\Psi(v+n g_s)\;\Psi^{*}(u)
}
The coefficient $a_{mn}^{vu}$ of the
$e^{-m v} e^{-(n+1)u}$ term in the expression on
the right hand side is
\eqn\firstcoefa{
a_{mn}^{vu}=
(-1)^{n+1} \; q^{{m(m-1)\over 4}-{n(n-1)\over 4}}\;
 \;\sum_{p=0}^{{\rm min}(m,n)}\; {q^{{p(p-m-n+1)\over 2}}\over [m-p]!\;
[n-p]!}.
}
Similarly, we can compute $a^{uv}$ by considering
$$
\langle 0|
\psi(u) \psi^{*}(v)|V\rangle =
\int\; d\tu \;e^{S_{U}(v,u)/g_s}\;{\Psi(u)\;\Psi^*(\tu - g_s/2)\;\over e^{u}-e^{\tu
- g_s/2}}.
$$
Following similar steps to the ones spelled out above, (or simply
 taking
$g_s \rightarrow -g_s$ which exchanges the branes with the anti-branes) we find
$$
a_{nm}^{uv}=(-1)^m \; q^{-{m(m-1)\over 4}} \;q^{{n(n-1)\over 4}}\;
\;\sum_{p=0}^{{\rm min}(m,n)}\; {q^{-{p(p-m-n+1)\over 2}}\over [m-p]!\;
[n-p]!}
$$
In the next section, we will see that this agrees with the previously
obtained expressions for the vertex. More precisely, this agrees 
up to the factor of $e^{-u}$ and a shift
of $v$ by a $g_s$ factor. The relative factors and shifts are probably related to
the fact that we were not carefuly keeping track of fermion charges.

\subsec{Fermion bilinearity and the equivalence with the
vertex from the topological A-model}

In this section, we show that the vertex amplitude,
as computed above from the topological B-model perspective,
agrees with the expression \csvertex\ for the vertex from the topological
A-model.

As we have seen in the previous section, the vertex amplitude
$|V \rangle$ (which is defined by $Z (t^i) =\langle t^i | V \rangle$)
is of the form
$$
|V \rangle = \exp \biggl[ \sum_{i,j,n,m} a_{mn}^{ij}
\psi^i_{-m-1/2}
\psi^{j*}_{-n-1/2} \biggr]|0 \rangle.
$$
To show that this agrees with \csvertex,
we compute the coefficients $a_{nm}^{ij}$ in terms of the entries
of the vertex in the
representation basis. As it is well-known (see for example \solitons),
the state corresponding to a hook representation $R_{m+1,n+1}$
with $m+1$ boxes in its row
and $n+1$ boxes in its column is given by
$$
| R_{m+1,n+1}\rangle = (-1)^{n+1} \psi^*_{-n-1/2} \psi_{-m-1/2}|0 \rangle.
$$
By taking the inner product with $| V \rangle$ and using
\vv\ we find that
$$
C_{R_{m+1,n+1} \cdot \cdot } =(-1)^{n} a_{m n}^{uu}$$
plus permutations. Similarly, we obtain
\eqn\twof{
C_{R_{p+1,p'+1} R_{q+1,q'+1} \cdot}= (-1)^{p' + q'} (
a^{uu}_{pp'} a^{vv}_{qq'} -a^{uv}_{pq'} a^{vu}_{qp'} ),
}
plus permutations. Finally, the trivalent
vertex with the reps $R_{p+1,p'+1}$, $R_{q+1,q'+1}$, $R_{r+1,r'+1}$
in positions $k$,$l$,$m$, respectively, is given by
$$\eqalign{
(-1)^{p' + q' + r'}
&\Bigl( a_{p q'}^{kl} a_{q r'}^{lm} a_{r p'}^{m
k }+ a_{p r'}^{km} a_{r q'}^{m l} a_{q p'}^{l k }
- a_{p p'}^{kk}a_{q r'}^{lm} a_{r q'}^{ml}\cr &
- a_{q q'}^{ll} a_{p r'}^{km} a_{r p'}^{mk}
- a_{r r'}^{m m } a_{q p'}^{l k } a_{pq'}^{kl}
+ a_{p p'}^{kk} a_{q q '}^{ll} a_{r r'}^{m m }\Bigr)
}
$$
Using these equations one can determine all the coefficients
$a^{ij}_{nm}$ in terms of entries of the trivalent vertex with
hook representations. One first finds
$a_{00}^{uv}= q^{1/6}$, $a_{00}^{vu}= -q^{-1/6}$. Since
$$
C_{R_1 R_2 \cdot} =q^{-\kappa_{R_2^t}/2} W_{R_1 R_2^t}
$$
one finds from \twof\ that
\eqn\asws{
\eqalign{
a^{uv}_{m n}=&(-1)^n q^{{1\over 6}-\kappa_{S_{n+1}}/2}\bigl( W_{ S_{m+1}
S_{n+1}} - W_{S_{m+1}} W_{S_{n+1}}\bigr),\cr
a^{vu}_{m n}=&-(-1)^n q^{-{1\over 6}-\kappa_{A_{m+1}}/2}\bigl( W_{ A_{m+1}
A_{n+1}} - W_{A_{m+1}} W_{A_{n+1}}\bigr),
}}
where $S_n$, $A_n$ are respectively the completely symmetric and
antisymmetric representations with $n$ boxes. The following formula
for $W_{R_1 R_2}$ is a consequence of cyclicity of the vertex (see \zhouIII\ for
a detailed proof):
\eqn\wrr{
W_{R_1 R_2}= q^{(\kappa_{R_1} + \kappa_{R_2})/2}
\sum_{R} W_{R_1^t/R} W_{R_2^t/R}
}
where
$$
W_{R/R'}=\sum_Q N_{Q R'}^{R}W_{Q}
$$
and can be written in terms of skew Schur functions \orv.
Using \wrr\ one easily finds
\eqn\aex{
\eqalign{
a^{uu}_{mn}=&(-1)^{n}{q^{{m (m+1) \over 4} -{n (n+1) \over 4}} \over
[m+n+1]! [m]! [n]!},\cr
a^{uv}_{m n}= &(-1)^{n}q^{ {m (m+1) \over 4} -{n (n+1) \over 4} +{1\over
6}}
\sum_{l=0}^{{\rm min}(m,n)}
{q^{{1\over 2}(l+1) (m+n -l)} \over [m-l]! [n-l]!},\cr
a^{vu}_{m n }= &-(-1)^{n}q^{ {m (m+1) \over 4} -{n (n+1) \over 4}-{1 \over
6} }
\sum_{l=0}^{{\rm min}(m,n)}
{q^{-{1\over 2}(l+1) (m+n -l)} \over [m-l]! [n-l]!}.\cr}}
Notice that the fermionic representation leads to a radical simplification
of the
information encoded in the vertex, and to a full series of consistency
relations between the coefficients $C_{R_1 R_2 R_3}$ that are the analog
of
the Pl\"ucker relations in the KP hierarchy. It is worth mentioning that
the state \vv\ is a tau function of the three-component KP hierarchy
as constructed for example in \kl. It follows in particular that
if we put one set of times to zero we obtain a tau function of the Toda
hierarchy \foot{This fact has also been pointed out in \zhou.}.

It is interesting to note that there is a choice of framing for which at
least some of the amplitudes simplify dramatically.
Suppose we consider the vertex amplitude corresponding to only two stacks of
B-branes, say in the $u$ and the $v$ patches,
with framing $-1$ in the $u-$patch and canonical framing in $v-$patch. The
correlation functions we want to compute are

\eqn\cormix{
\langle 0| \psi^{*} (u) \psi (v) |V^{(-1,0,0)}\rangle=
-\sum_{m,n=0}^{\infty} (-1)^{m+1}q^{-\kappa_{S_{m+1}}/2}
a_{mn}^{uv}e^{-(m+1)u} e^{-(n+1)v},
}
and
$$
\langle 0| \psi (u) \psi^{*} (v) |V^{(-1,0,0)}\rangle=
\sum_{m,n=0}^{\infty} (-1)^{n+1} q^{-\kappa_{A_{n+1}}/2}
a_{mn}^{vu}e^{-(n+1)u} e^{-(m+1)v}.
$$
Using \asws, we find that \cormix\ can be written as
$$
\eqalign{
&\langle 0| \psi (u) \psi^{*} (v) |V^{(-1,0,0)}\rangle = \cr
& q^{1/6} \sum_{m,n=0}^{\infty}(-e^{-u})^{m+1} (-e^{-v})^{n+1}
q^{-\kappa_{S_{m+1}}/2 -
\kappa_{S_{n+1}}/2 }(W_{S_{m+1}  S_{n+1}}-W_{S_{m+1}} W_{S_{n+1}}).}$$
Therefore, we just have to compute the generating function of the $W_{R_1 R_2}$ for
symmetric representations, in the framing $(-1,-1)$. It can be shown by using \wrr\ that
$$
(-1)^{\ell(R_1) + \ell(R_2)} q^{-\kappa_{R_1}/2 -\kappa_{R_2}/2} W_{R_1 R_2} =
\langle R_1| \otimes
\langle R_2 | W\rangle$$
where the state $|W\rangle$ in ${\cal H}^{(1)} \otimes
{\cal H}^{(2)}$ is given by
\eqn\genhopf{
| W \rangle = \exp \Bigl\{ -\sum_{n=1}^{\infty}  {\beta_n \over n}
(\alpha^{(1)}_{-n}
+ \alpha^{(2)}_{-n}) + \sum_{n=1}^{\infty}{1 \over
n}\alpha^{(1)}_{-n}\alpha^{(2)}_{-n}  \Bigr\} |0 \rangle,
}
where $\beta_n = {1\over [n]}$.
Using \genhopf\ one finds
$$
\langle 0| \psi^{*} (u) \psi (v) |V^{(-1,0,0)}\rangle=
-{q^{1/6} \over 1- e^{u +v}} \prod_{n\ge 0}
(1- e^{-u}  q^{n+{1\over 2}})^{-1}(1- e^{-v}  q^{n+{1\over 2}})^{-1}.$$
Similarly,
$$
\langle 0| \psi (u) \psi^{*} (v) |V^{(-1,0,0)}\rangle=
-{q^{-1/6} \over 1- e^{u+v}} \prod_{n\ge 0}
(1- e^{-u}  q^{n+{1\over 2}})(1- e^{-v}  q^{n+{1\over 2}}).$$
%

\subsec{Framing and more general Riemann surfaces}
The fermionic formulation of the vertex in the standard framing is very
easily generalizable to other framings and more general Riemann
surfaces.

Consider a Riemann surface with punctures $P_i$, $i=1,2,\ldots$.
The punctures are defined by
$$p_i = c_i u+ d_i v + t_i \rightarrow 0$$
and come with coordinates
$$q_i = a_i u+b_i v \rightarrow \infty$$
Where $a_i d_i-b_i c_i =1$, for integers $a_i,b_i,c_i,d_i$.
The parameters $t_i$ are complex structure moduli of the Riemann
surface.
According to the discussion above, $[q_i,p_i]=-g_s$, and
$$p_i = g_s {\del \over \del q_i}.$$

The coordinates $(q_i,p_i)$ at different punctures $P_i$ are
related to each other by $SL(2,{\bf Z})$ transformations, up to
constant shifts depending on the $t$'s.
The corresponding Ward identities are given by
\eqn\qca{\oint_{P_i} \psi^{*}(q_i)e^{n q_i} \psi(q_i)
+ \sum_{j\neq i}
\oint_{P_j}\psi^{*}(q_j) e^{-n (A_{ij}q_j + B_{ij} p_j +T_{ij})}
\psi(q_j) =0.}
where
$$q_i = A_{ij}q_j + B_{ij} p_j +T_{ij}(t).$$

For example, consider the topological vertex in arbitrary framing
$(k,l,m)$, i.e. where the coordinates in the $u,v$ and $w$
patches are
$$q_u = u-k v, \quad q_v = v-l w, \quad q_w = w- m u,$$
respectively. The corresponding momenta are unchanged.
We thus have
$${\oint_{q_u} \psi^{*}e^{nq_u } \psi
+
(-1)^n\oint_{q_v}\psi^{*}
e^{n (a_{v}q_v +b_{v}g_s{\del \over \del q_v})}\psi
+(-1)^{nk} \oint_{q_w}\psi^{*}
e^{n (a_{w}q_w+b_{w}g_s{\del \over \del q_w})}\psi
 =0.}$$
where $a_{v} = -k-1$, $b_{v} =-1+l- kl$, $a_{w} = k$, and $b_{w}= 1 +
k + mk$, and similarly for the other charges. The solutions to this
correspond to the framed vertex amplitude, as we have verified in examples.

The Ward identity \qca\ is expected to govern the topological string
amplitudes on general local toric Calabi-Yau geometries. A toric diagram
with $g$ closed meshes is associated to a Riemann surface $H(u,v)=0$ of
genus $g$. In these cases the topological string amplitudes are much
more non-trivial as one has irreducible curves of various genera at
generic degrees. This is reflected in the dependence on the K\"ahler
parameter $t_i$, which enter the periods
of $\lambda$ on $H(u,v)=0$.

As an example we check \qca\ for the ${\cal O}(-3)\rightarrow {\bf P}^2$
A-model geometry whose B-model curve is genus one with three punctures
\eqn\localptwo{
H(e^u,e^v)=r(e^{-t})-e^{-u}-e^v + e^{-t+u-v}=0\ .}
Here $(t,u,v)$ are the flat A-model moduli. The open/closed mirror map is
encoded in $r(e^{-t})=\left(e^{-t}\over z(e^{-t})\right)^{1\over 3}
=1-2 e^{-t}+5 e^{-2t}+\ldots$ as determined by periods of $\lambda$
on $H=0$ \akv . The choice of the coordinate $q_i$ at the punctures
is a choice of framing and the one, which corresponds to framing
$(0,0,0)$ from gluing three vertices \akmvII \foot{Note that we exchanged
$Q_1$ and $Q_2$ w.r.t. \akmvII .}
\eqn\partpII{
\eqalign{
&Z_{\bf P^2}(V_1, V_2, V_3)=\cr
& \sum_{R_i, Q_i}
 C_{Q_3 R_2 R_3^t} C_{Q_2 R_1 R_2^t}
 C_{Q_1 R_3 R_1^t}(-1)^{\sum_i \ell(R_i)}e^{-\sum_i \ell(R_i) t}
q^{\sum_i \kappa_{R_i}} {\rm Tr}_{Q_1}V_1 \, {\rm Tr}_{Q_2} V_2 \,
{\rm Tr}_{Q_3} V_3}
}
is given by
\eqn\ptwocoords{
\eqalign{
(q_1,p_1)&=(-u,u+v+i\pi)\cr
(q_2,p_2)&=(u-v-t,2 u -v -t)\cr
(q_3,p_3)&=(v, 2 v-u +t)}}
The relation between the boundary coordinates is
\eqn\ptworel{
(q_1,p_1)=(q_2-p_2, -3 q_2+2 p_2-t)=(-2 q_3+p_3-t,3 q_3-p_3+t)\ }
yielding the Ward identity
\eqn\ptworel{
\oint_{P_1} \psi^*(q_1) e^{n q_1}\psi(q_1)+
\oint_{P_2} \psi^*(q_2) e^{n q_2-n p_2}\psi(q_2)+
\oint_{P_3} \psi^*(q_3) e^{-2 n  q_3+n p_3-nt}\psi(q_3)\ .}
The simplest check is performed by choosing $\hat q_2=q_2-p_2$,
which corresponds to framing $(0,-1,0)$. In particular the
action of \ptworel\ on $Z$ in the $q_1,\hat q_2,q_3$ coordinates
yields for this choice of framing
\eqn\simplecheck{
{\partial \over \partial t^1_n} {\cal F}|_{t_i^3=0}+
{\partial \over \partial t^2_n} {\cal F}|_{t_i^3=0}=0\ .}
The identity \ptworel\ does not fix the amplitude
as \fullq\ does for the vertex, because of the positive power
in the second patch. However it can be used to reconstruct
the amplitude for three stacks of branes from the
one for one stack.
This is related to the fact that the identities hold for any
 fermion number flux, and thus any superposition
 of fermion number flux in the loop will solve the constraints.

The expression \partpII\ yields for framing $(0,0,0)$ up to order $Q=e^{-t}$ and two
holes after transforming in the winding basis
\eqn\freepII{
\eqalign{F = \sum_n &{1\over [n]} (t^{(1)}_n+t_n^{(2)}+t_n^{(3)}) + Q\;\biggl(- {3\over [1]^2} -
{2\over [1]} (t^{(1)}_1+t_1^{(2)}+t_1^{(3)}) \cr &-
{1\over 2}(t_1^{(1)})^2 - {[2]\over [1]^2}(t^{(1)}_2+t_2^{(2)}+t_2^{(3)}) -
t^{(1)}_1 t^{(2)}_1- t^{(1)}_1 t^{(3)}_1 - t^{(2)}_1 t^{(3)}_1\biggr)+\ldots \ . }}
Using the normal ordering description \fullq\ we can use this to
make a first check on \ptworel . We have checked \ptworel\ up to
order $Q^5$ and $5$ holes involving all three stacks of branes.
Consistency of \fullq\ follows also from the sewing  prescription
of the amplitudes and the fact that the constraints commute with
this operation.

\subsec{Fermion number flux and the topological vertex}

We can illustrate the considerations in section 4 on the fermion number
flux also by considering the topological string amplitudes in
${\cal O}(-3)\rightarrow {\bf P}^2$.

We can then introduce a new variable $\theta$ corresponding to
turning on flux through the nontrivial cycle, and consider the partition function $Z_N$
where $N$ units of flux have been turned on. The computation of these partition functions
can be done by using the topological vertex and the gluing constructions, as follows.
As explained in \akmvII, the propagator associated to an edge is in general of the form
\eqn\prop{
\delta_{R_1 R_2^t}(-1)^{\ell(R_1)(f+1)} q^{f\kappa_{R_1}/2} e^{-\ell(R) t},
}
where $f$ is a framing which
depends on the details of the geometry, and $t$ is the K\"ahler parameter.
This propagator corresponds to setting the momenta
in the loops to zero, but it is easy to construct a state in  ${\cal H}^{(1)}
\otimes {\cal H}^{(2)}$, $\langle P^{(f)}|$, which includes all possible fermion numbers
and is such that $\langle P^{(f)}| R_1 \rangle \otimes
|R_2 \rangle$ equals \prop. This propagator has the form
\eqn\propc{
\langle P^{(f)}|=
\langle 0_{12}| \exp \Biggl\{ \sum_{n=0}^{\infty} e^{-nt-t/2} \Bigl(e ^{i\theta} \epsilon_n \psi^{1*}_{n+1/2} \psi^{2}_{n+1/2}
+  e^{-i\theta} \epsilon'_n \psi^{1}_{n+1/2} \psi^{2*}_{n+1/2} \Bigr) \Biggr\}
}
and it is a bilinear in fermions. The coefficients $\epsilon_n$, $\epsilon'_n$ are given by:
\eqn\epis{
\eqalign{
\epsilon_n=&i^{1+f} (-1)^{(1+f)n}q^{{f\over 2} n(n+1)},\cr
\epsilon'_n=&i^{1+f} (-1)^{(1+f)n}q^{-{f\over 2} n(n+1)}.\cr}
}
Notice as well that the propagator \propc\ is of the form
$$
\langle P^{(f)}|= \sum_{N \in {\bf Z}}e^{i N \theta}{\cal O}_N \langle N, -N |$$
where $N$ is the fermion charge, and ${\cal O}_N$ are operators in the fermion modes, so it includes
all possible fermion numbers.

Let us now consider the computation of the partition function with zero fermion
number flux associated to local ${\bf P}^2$. The topological vertex
computation is \akmvII\
$$
Z_0= \sum_{R_1,R_2,R_3}
(-1)^{\sum_i \ell(R_i)}e^{-\sum_i \ell(R_i) t}
q^{\sum_i \kappa_{R_i}} C_{\cdot R_2 R_3^t} C_{\cdot R_1 R_2^t}
C_{\cdot R_3 R_1^t}.
$$
and it is easy to see that the full answer is given by
$$
Z =\sum_{N \in {\bf Z}} e^{i N \theta} Z_{N}=
\langle 0_{159}| \otimes \langle P^{(2)}_{27}| \otimes \langle P^{(-2)}_{34}| \otimes  \langle P^{(2)}_{86}|
V_{123} \rangle\otimes |V_{546} \rangle \otimes |V_{987} \rangle
$$
where $| V\rangle$ is given by \vv, with the coefficients of the bilinear in fermions given in \aex.
An explicit computation up to order 2 in $e^{-t}$ shows that indeed, up to an overall constant,
\eqn\ptflux{
Z_{N=\pm 1}(e^{-t})=Z_0 (-q^{\pm 3}e^{-t}).
}
This is in agreement with \changeflux, since the intersection number is precisely $3$ for
this geometry (the sign in the r.h.s. of \ptflux\ is due to a quantum shift).

Notice that for example the open string amplitudes for local ${\bf P}^2$ with outer branes, computed in \akmvII,
do not correspond to tau functions of the n-KP hierarchy since they only involve the zero fermion number
sector. In order to obtain amplitudes which are Bogoliubov transforms of the vacuum, we have to use
in the gluing rules the propagator \propc, which includes all sectors.

\newsec{Connections with Non-critical Strings}

It was already shown in \goshv\ that $c=1$ non-critical bosonic string
at self-dual radius is equivalent to topological B-model of the
conifold.  The rationale for the appearance of the conifold can be
viewed as the fact that the holomorphic functions of this CY can be
viewed as the ground ring of $c=1$ at self-dual radius \witteng .  It
was further conjectured in \dvI\ that non-critical bosonic string
coupled to the $(r,s)$ minimal model is equivalent to the topological
string on the $(r,s)$ generalization of the conifold, as reviewed in
this paper.  The fact that we can directly solve for the amplitudes of
these models using the B-model symmetries is in line with the fact
that the symmetries of the ground ring of these theories and their
deformation encode all the scattering amplitudes.  Thus our approach
can be viewed as a method to solve non-critical bosonic string by
quantizing the deformations of the ground ring.  One should identify
the B-branes we have considered with the branes of non-critical
bosonic strings.  This would be interesting to do in detail because of
the fact that we have found fermions/branes to play a key role in
computing the amplitudes.

In the context of topological B-models, we have seen that there
are far more cases that do correspond to a string theory than
would be naively expected based on the experience with
non-critical bosonic strings.  In particular {\it any} matrix
model in the `t Hooft sense is a string theory equivalent to a
topological B-model on some Calabi-Yau \dvI.  Moreover also these
backgrounds do admit `scattering states' that correspond to
deformations at infinity and that are captured by the free energy
${\cal F}(t_n^i)$.  This broadens the class of interesting string
theories to study, and frees us of having to take a double scaling
limit to discuss a string theory\foot{ It would be interesting to
go backwards and see if one can give an interpretation of all
these local B-models in terms of non-critical bosonic strings.
For example it is natural to speculate that the topological vertex
corresponds to non-critical bosonic strings propagating on a
geometry of dimension 1 consisting of three half circles connected
at two points to one another.}.

More recently there has been some study of non-critical superstring
backgrounds (see, for example, \res).  All these models
have been related to certain limits
of matrix models.  Since, as we noted above any matrix model is
equivalent to a topological B-models, so does any limit of them.  Thus
we find, at least at the level of perturbative string theory, a way to
go from non-critical bosonic strings to non-critical superstring: One
simply considers a given matrix model, which is equivalent to B-model
topological string and takes various limits to get one or the other
background.

One main recent motivation to study non-critical superstring has been
to find a non-perturbatively complete string theory.  In this context
it is natural to ask what a non-perturbative definition of topological
string may be.  This would be interesting to study further.

\newsec{Open Questions}

We have seen that the Ward identities leading to the solution of the
B-model can be formulated in terms of fermions whose wavefunctions
transform with suitable Fourier transforms from patch to patch,
encoded by the Riemann surface $H(p,x)=0$ where we view $x,p$ as
conjugate variables.  It would be important to better understand the
meaning of this fact: Why are fermions, which correspond to branes,
not geometrical objects of the usual kind on the Riemann surface?
Moreover why is it that the fermions are free and that the amplitudes
can be given by suitable Wick contraction between patches?

Naively one would think that as we move branes from one position to
the other the partition function is a smooth function of the position.
Let $Z(x)$ denote the partition function of the topological string
with a brane at position $x$.  Let $x=x_i$ denote the asymptotic
infinities of the geometry.  Let us start with a brane at $x$ near a
given $x_i$ and move to another one where $x$ approaches $x_j$. The
statement we have found is that $Z$ transforms as a Fourier transform
instead of analytic continuation as we go from $x_i\rightarrow x_j$.
We believe this may be related to an open string holomorphic anomaly.
Namely, the actual partition function not only depends on $x$ but also
on the complex conjugated variable ${\overline x}$.  Thus each brane
is associated to a particular asymptotic infinity where the
corresponding ${\overline x}\rightarrow {\overline x_i}$.  Let us
define
$$Z_i(x)=Z(x, {\overline x_i}) \qquad Z_j (x)=Z(x,{\overline x_j})$$
 There is
no reason that these are the same functions, in other words
$$Z_i(x)\not= Z_j(x)$$
thus analytic continuation does not need to work.  To find the
relation between these two function we need to study
$$Z_i(x)-Z_j(x)=\int_{\overline x_i}^{\overline x_j}
 d{\overline x}\; {\overline \partial }Z(x,{\overline x})$$
There is an analog of holomorphic anomaly for open string \bcov\ which
has not been fully developed.  In the closed string
case an interpretation of it \wittenanomaly\ led to
the formulation of the closed string amplitudes as quantum
mechanical wavefunctions on the moduli space of Calabi-Yau.
 Here we want to speculate that
there is a similar story in the open string context leading
to the amplitudes being wave functions on the Calabi-Yau itself; that there
could be roughly a heat-kernel-like equation of the form
$$
{\overline \partial Z}=(ag_s^2\partial^2 +bg_s x \partial +c x^2)Z.
$$
Viewing $p$ as conjugate to $x$ this is the same as the action
$$
{\overline \partial Z}=(ap^2+bxp+cx^2) Z
$$
This would imply that an infinitesimal change in the antiholomorphic
coordinates generates a linear symplectic diffeomorphism
$$
\eqalign{
x & \to x+(2ap +bx)\cr
p & \to p-(2cx+bp).\cr
}
$$
Exponentiating this action would then generate the metaplectic action
on $Z$.  It would be important to develop this picture further, if
only because it suggests an interesting way to generate the quantum
mechanical transformation rules from string theory.  It would also
lead to a justification of the procedure followed in this paper, of
treating branes/fermions as wavefunctions.

\appendix{A}{$c=1$ scattering amplitude}

Here we show in detail how the interpretation of the $c=1$ scattering
matrix as a Fourier transform corresponds to the familiar scattering
amplitudes.  Starting from the well-known answer
$$
R_{q} = i\left({1+ ie^{-\pi(\mu+iq)}\over
1-i e^{-\pi(\mu+iq)}}\right)^{1\over 2}
\left({\G(\hf-i\mu+q)\over \G(\hf+i\m-q)}\right)^{1\over 2}
$$
we can use the identity
$$
\G(x)\G(1-x)= {\pi \over \sin(\pi x)}
$$
to write $R_q$ for $q=n+\hf$ as
$$
\eqalign{
R_{n+\hf}^2 & = -{1\over \pi} {1-e^{-i\pi(n-i\m+1)} \over 1+
e^{-i\pi(n-i\m+1)}}\cdot  \sin\left(\pi(n-i\m+1)\right) \cdot
\G(n-i\m+1)^2 \cr & =
-{1\over 2\pi i}
\left(e^{i\pi(n-i\m+1)/2}-e^{-i\pi(n-i\m+1)/2}\right)^2 \G(n+i\m+1)^2
\cr}
$$
So that
$$
R_{n+\hf} = {1\over \sqrt{2\pi}} e^{i\pi/4}
\left(e^{i\pi(n-i\m+1)/2}-e^{-i\pi(n-i\m+1)/2}\right) \G(n+i\m+1)
$$
On the other hand, the expression
$$
{1\over \sqrt{2\pi}} \int_{-\infty}^{+\infty}
 dx \cdot x^{n} \cdot x^{-i\m} e^{ix}
$$
can be evaluated by deforming the contour, by $x \to ix$ for $x>0$ and
to $x \to -i x$ for $x<0$. (This depends on the sign of $n$, or more
generally on where the poles are.) This will turn the Fourier
transform into two Laplace transforms, and gives the result
$$
{1\over \sqrt{2\pi}} \left(e^{i\pi(n-i\mu+1)/2}
- e^{-i\pi(n-i\mu+1)/2}\right) \G(n-i\m+1)  = e^{-i\pi/4}
R_{n+\hf}
$$

\appendix{B}{Topological vertex: Solutions to $W_\infty$ identities}

Here we give explicit formulas for the free energy
of the vertex, obtained by using the ${\cal W}_{\infty}$ identities, up
to -- and including-- terms which are cubic in the couplings $t^i_n$ ({\it i.e.}
up to three holes). The resulting expressions are exact in $q$ and include the
contributions to all genera.

The starting point of the iteration in $s$ is given by \back .
In order to proceed, we note the  following two formulae:
\eqn\firsteq{
\exp\left\{\sum_{k>0} -{[n k] \over k [k]} z^k\right\} = \sum_{k=0}^{n} (-1)^{k}
\biggl[{n \atop k}\biggr] z^k,}
\eqn\secondeq{
z^n \exp \left\{\sum_{k>0} {[n k] \over k [k]} z^k\right\} =
\sum_{m=n}^{\infty} {[n]\over [m]}\biggl[{m \atop n}\biggr]z^m.}
These can be easily proved by considering symmetric polynomials in $n$
variables $x_i$, $i=1, \cdots, n$, with the specialization
$x_i=q^{i-(n+1)/2}$.  The generating function of elementary symmetric
functions $E(z) =\prod_{i=1}^n (1+x_i z)$ can be computed by induction
to be
$$
E(z)=\sum_{k=0}^n \biggl[{n \atop k}\biggr] z^k,$$
which is the r.h.s. of the first equation. It is a well-known fact in
the theory of symmetric functions that $E(-z) = e^{-P(z)}$, with
$P(z)=\sum_{k>0} p_k z^k/k$, and $p_k =\sum_{i=1}^n x_i^k$ (see for
example \macdonald).  With the above specialization, we find $p_k =[n
k]/[k]$, and the first equality \firsteq\ follows. The second equality
follows in a similar way by considering the generating function of
complete symmetric functions $H(z)=\prod_{i=1}^n (1 - x_i z)^{-1}$.

We will now consider amplitudes with $h>1$. The first thing
to notice is that the vanishing of $\CF_{h>1}$
with a single stack of D-branes
allows us to solve for the part of $\CF$ of the form $t^w \cdots t^w t^u$ and
$t^v \cdots t^v t^u$ to all genera. This involves in principle all terms of the
form $\partial^v_{q_1}\cdots \partial^v_{q_{\ell}}\CF|_{t=0}$ and
$\partial^w_{q_1}\cdots \partial^w_{q_{\ell}}\CF|_{t=0}$, but
from \back\ we know them all and only $\del^w_q \CF|_{t=0}$ is non-vanishing.
Using now \firsteq\ and the
value \back, we find that the $w$ integral in \fullq\ leads to
\eqn\expan{
\eqalign{ -&\oint_w {dw \over [n]} e^{-\sum_{k>0} {[n k ]\over k
[k]} e^{- k w} + g_s^{-1}[nk] {t_k^w }e^{k w} + g_s {[n k ] \over k} \del_k^{w}
\CF +...}=\cr =& \sum_{n,m, r}{g_s^{-r} \over r!} \sum_{\sum_i q_i=m}
{(-1)^{m+1}\over[n]}\biggl[{n \atop m}\biggr]\prod_{i=1}^r {[n q_i] t^w_{q_i} } + \cdots}
}
It then follows that the part of $\CF$ of the form $t^w \cdots t^w t^u$ is
\eqn\wsu{
\sum_{n,m, r}{g_s^{-r-1} \over r!} \sum_{\sum_i q_i=m}
{(-1)^{m+1}\over[n]}\biggl[{n \atop m}\biggr]\prod_{i=1}^r
{[n q_i] t^w_{q_i} } {t^{u}_n }+ {\rm  cyclic},
}
where ``cyclic'' denotes the terms obtained by cyclic permutations of the
times $t^u \rightarrow t^v \rightarrow t^w$.
Similarly one finds in the $v$-patch the part of $\CF$ of the form
 $t^v \cdots t^v t^u$
\eqn\vsu{
\sum_{n,m, r}{g_s^{-r-1} \over r!} \sum_{\sum_i q_i=m}
{(-1)^{n+r}\over[m]}\biggl[{m \atop n}\biggr]\prod_{i=1}^r
{[n q_i] t^v_{q_i} } {t^{u}_n }+ {\rm  cyclic}.
}

Notice that, in particular, the free energy at two holes is given by
\eqn\twoholes{
{\cal F}_{h=2} = g_s^{-2}\sum_{n\ge k}{(-1)^{k+1} } {[nk]\over [n]} \biggl[{n \atop k}\biggr]
(t_n^u t_k^w + {\rm cyclic}).}

We can now compute cubic terms of the form $t^w_p t^u_m t^u_n$.
In that case, the only contribution to $\partial_n^u\C F$ that we need to consider
comes from products
of $t^w_k$ with arbitrary derivatives of $\CF$,
$$
-\sum_{p,k} {[np]}(-1)^k\biggl[ {n \atop k}\biggr]  t_p^w \sum_r {g_s^{r-1} \over r!}
\sum_{\sum_i q_i =p-k} \prod_{i=1}^r
{[nq_i] \over q_i}\partial^w_{q_1} \cdots \partial^w_{q_r}\CF, $$
where we consider the contribution to the multiple derivative of \wsu.
The final expression for the coefficient of
$t^w_p t^u_m t^u_n$ is:
\eqn\wwu{
g_s^{-3}(-1)^{p} {[np]\over [n][m]} \sum_{r=1}^p \sum_{k=0}^{p-r} {1\over r!}
\biggl[{n \atop k}\biggr] \biggl[{m \atop p-k} \biggr]\sum_{\sum_{i=1}^r q_i=p-k} \prod_{i=1}^r
 {[n q_i] [m q_i] \over q_i}{t^u_n } {t^u_m } {t^w_p} .}
Of course, if $n=m$ there is an extra $1/2$.
We have checked in many cases
that this complicated albeit explicit expression adds up to a simple
result which is manifestly symmetric in $n$, $m$:
$$
(-1)^p g_s^{-3}{ [np] [m p ] \over [n+m]}\biggl[ {n+m \atop p}\biggr]
{t^u_n } {t^u_m } {t^w_p}
$$
whose genus zero limit is
$$
(-1)^p g_s^{-2} {n m p^2\over n+m} {n+m \choose p}.
$$

Let us now consider the coupling $t^w_p t^v_m t^u_n$. This has two contributions, one
coming from the $w$ patch, and the other from the $v$ patch. For the $w$ patch, we
start again from \expan, but now have to consider the piece of the multiple derivative
that is proportional to $t^v$. This is obtained from \vsu, and we find
for the coefficient of ${t^u_n }{t^v_m}{t^w_p}$
$$
a(n,m,p)=-g_s^{-3}{ [np]\over [n]} \sum_{r=1}^p \sum_{k=0}^{p-r} {1\over r!}
{(-1)^{k+r+m}\over [p-k]}\biggl[{n \atop k}\biggr]
\biggl[{p-k \atop m} \biggr]\sum_{\sum_{i=1}^r q_i=p-k} \prod_{i=1}^r
 {[n q_i] [m q_i] \over q_i}.
$$

The contribution from the $v$ patch is obtained in a similar way,
and one finds
\eqn\bbcoeff{
b(n,m,p)=g_s^{-3}{ [n m ]\over [p]} \sum_{r=1}^{m-1} \sum_{k=n}^{m-r} {1\over r!}
{(-1)^{k+n+ m + r}\over [k]}
\biggl[{k \atop n} \biggr]\biggl[{p \atop m-k}\biggr]
\sum_{\sum_{i=1}^r q_i=m-k}
\prod_{i=1}^r
 {[p q_i] [n q_i] \over q_i}.
}
The coefficient of $ t^u_n t^v_m t^w_p$ in ${\cal F}(g_s,t_n^i)$ is then
given by
\eqn\totalc{
F^{u v w}_{nmp}=a(n,m,p)+ b(n,m,p).
}
Explicit evaluations show that the above coefficient is cyclically symmetric in $n,m,p$,
as it should be. All the results obtained here for $h=2,3$ holes and arbitrary winding numbers $n,m,p$
are in perfect agreement with the free energy as computed from \csvertex.

\vskip 1.3in
\bigskip
\leftline{\bf
   Acknowledgments}

We would like to thank N.~Arkani-Hamed, J.~de Boer, S. ~ Cherkis,
S.~Gukov, C.~Lazaroiu,
W.~Lerche, I.~Low, J. ~McGreevy, G.~Moore, A.~Neitzke,
 L.~Randall, T.~Takayanagi and E.~Verlinde for useful
conversations.

The research of M.A. was supported in part by DOE grant
DE-FG02-96ER40956
and by a DOE OJI Award. The research of
C.V.~was supported in part by NSF grants
PHY-0244821 and DMS-0244464.  The research of R.D.~was partly
supported by FOM and the CMPA grant of the University of Amsterdam. In
addition, CV thanks the hospitality of the theory group at Caltech,
where he was a Gordon Moore Distinguished Scholar.
A.K.~is supported in part by the DFG grant KL-1070/2-1. M.M. would like to thank
the High Energy Theory group of the University of Washington at Seattle for
hospitality during part of this work.

\listrefs
\end

A second important transformation is
$$
T' = \pmatrix{ 1 &  0 \cr
              1 &  1 \cr}
$$
that implements a framing ambiguity
$$
\eqalign{\tp & =  p, \cr
         \tx & =  p + x. \cr}
$$
The corresponding quantum operator is
$$
U(T') \Psi(\tx) = \int {dx \over \sqrt{2\pi g_s}} e^{-(x-\tx)^2/2g_s}
\Psi(x).
$$
Note that operator
$$
T = \pmatrix{ 1 &  1\cr
              0 &  1 \cr}
$$
was excluded in the above formulas because $c=0$. It acts as
$$
\eqalign{\tp & =  p + x,\cr
         \tx & =  x. \cr}
$$
Since we have the shift $\p_x \to \p_x -x/g_s$, this action is
therefore simply given by multiplication of the wave function by a
Gaussian prefactor
$$
U(T)\Psi(x) = e^{-x^2/2g_s} \Psi(x).
$$
In these formulas one easily observes the well-known relations
$T'=TST$.

{}Finally we have the element
$$
ST = \pmatrix {0 &  -1\cr
              1 &  1 \cr}
$$
that is implemented as
$$
U(ST) \Psi(\tx) = \int {dx \over \sqrt{2\pi g_s}}
e^{(-{1\over 2}x^2+x\tx)/g_s} \Psi(x).
$$
Note that this satisfies $(ST)^3=1$. It is reassuring to see this explicitly
in the integral representation using the identity
$$
\eqalign{
& \int dy dz \exp \left[-\hf(x^2 + y^2 +z^2) +2xy +2yz +2z\tx\right]  \cr
& \quad =  \int dy dz \exp \left[-\hf (x-y-z)^2+z(\tx-x)\right] \cr
& \quad = \int dz \exp[z(\tx-x)]=\delta(x-\tx). \cr
}
$$